\documentclass[aps,prb,twocolumn,superscriptaddress,longbibliography]{revtex4-2}
\usepackage{amsmath,amssymb}
\usepackage[pdftex]{hyperref,graphicx}
\hypersetup{colorlinks = true, urlcolor = blue, linkcolor = blue, citecolor = blue}
\usepackage{physics}
\usepackage{caption}
\usepackage{subcaption}
\usepackage{xcolor}
\usepackage[percent]{overpic}

\newcommand{\HPadd}[1]{\textcolor{black}{#1}}

\begin{document}
\title{Avoiding Dilution: Using Diffusion and Vision Transformers to resolve Majorana Features in Nanowires at High Temperature}
\author{Jacob R. Taylor}
\affiliation{Condensed Matter Theory Center and Joint Quantum Institute, Department of Physics, University of Maryland, College Park, Maryland 20742, USA}
\author{Haining Pan}
\affiliation{Department of Physics and Astronomy, Center for Materials Theory, Rutgers University, Piscataway, New Jersey 08854, USA}
\affiliation{Department of Physics, University of Florida, Gainesville, Florida 32611, USA}

\author{Jay D. Sau}
\affiliation{Condensed Matter Theory Center and Joint Quantum Institute, Department of Physics, University of Maryland, College Park, Maryland 20742, USA}
\author{Sankar Das Sarma}
\affiliation{Condensed Matter Theory Center and Joint Quantum Institute, Department of Physics, University of Maryland, College Park, Maryland 20742, USA}

\begin{abstract}
Identifying Majorana zero modes in semiconductor--superconductor nanowires requires ultra-low temperature transport measurements in dilution refrigerators, making device screening slow and resource-intensive. Here, we investigate whether high-temperature conductance data can be used to infer low-temperature Majorana nanowire properties before committing devices to dilution-refrigerator characterization. We generate paired high- and low-temperature conductance simulations for disordered Majorana nanowires and train neural networks to perform two related tasks. First, we use a Shifted Window U-Net Transformer diffusion-inspired architecture to reconstruct low-temperature conductance from thermally broadened high-temperature measurements, achieving high-fidelity recovery with $R^2 \approx {0.95}$ for local conductance and $R^2 \approx {0.91}$ for nonlocal conductance. Second, we train a Video Vision Transformer-based network to predict the low-temperature topological visibility directly from high-temperature conductance, obtaining $R^2 \approx {0.80}$. These results demonstrate that machine-learning models can recover and infer low-temperature Majorana features from experimentally easier high-temperature data, providing a practical route for rejecting poor devices early thus avoiding slow and resource-intensive dilution refrigeration for non-promising devices. This high-temperature screening approach could substantially accelerate the experimental feedback loop for Majorana nanowire device development. 
\end{abstract}

\maketitle

\section{Introduction}
{Topological quantum computing relies on the manipulation of non-Abelian anyons, such as Majorana zero modes (MZMs) in topological superconductors, to achieve fault-tolerant quantum computation~\cite{kitaev2003faulttolerant,nayak2008nonabelian,sarma2015majorana}. Following the theoretical prediction that MZMs could be realized in superconductor-semiconductor hybrid nanowires~\cite{lutchyn2010majorana,sau2010generic,sau2010robustness,oreg2010helical}, intense experimental efforts have been devoted to this platform~\cite{dassarma2023search,kouwenhoven2025perspective}, including a major industrial program by Microsoft~\cite{microsoftquantum2023inasal,aghaee2025interferometric,aghaee2025distinct}. Despite this progress, unambiguous identification of MZMs remains elusive, primarily because disorder suppresses the topological gap and finite wire lengths can lead to MZM overlap, while trivial Andreev bound states generically mimic Majorana signatures in transport measurements~\cite{pan2020physical,pan2020generic,pan2026majorana}. Developing reliable tools to distinguish topological from trivial states in realistic devices remains the central open problem in the field.}
There has been substantial work on using neural networks to solve problems in Majorana nanowires, including the prediction of disorder~\cite{taylor2024machine,pawlowski2026learning}, the mitigation of disorder effects~\cite{taylor2025mitigating,krawczyk2026ai}, and the classification of devices ~\cite{taylor2025vision,taylor2025unreasonable,cheng2024machine}. These networks often use data that is difficult or even uninterpretable for humans, but which can nonetheless contain useful information for achieving the desired task. 

One hurdle that significantly slows the production and experimental characterization of Majorana nanowire devices is the need to cool candidate devices to dilution-refrigerator temperatures, typically $T\approx 50$~mK~\cite{microsoftquantum2023inasal}, before the most relevant Majorana indicators can be meaningfully assessed. The dilution cost is not only financial, but also practical: each device requires sample mounting, wiring, cooldown, thermal stabilization, and measurement time. As a result, a large fraction of experimental throughput can be spent on devices that ultimately show no useful Majorana (or any) signatures. At present, there is no reliable way to determine in advance whether a candidate device is likely to exhibit useful topological behavior before committing it to the dilution-refrigerator cycle.
A high-temperature screening method would therefore be valuable even if it were not perfectly diagnostic, since it could rapidly eliminate devices that are clearly unlikely to succeed and reserve low-temperature resources for the most promising candidates, thus substantially enhancing the throughput for devices to focus on.

The conventional route for relating high-temperature and low-temperature conductance is to treat finite temperature as a thermal convolution and attempt to undo this operation by deconvolution, e.g., through Fourier-space inversion. In principle, this can recover sharper low-temperature structure from thermally broadened measurements. In practice, however, the thermal kernel suppresses energy features on scales smaller than roughly $k_B T$, and direct inversion strongly amplifies noise and experimental imperfections. At pre-dilution temperatures, this limitation is severe: the relevant Majorana features are often far narrower than the thermal resolution, making standard deconvolution useless for either validating or rejecting the presence of MZMs.

The main weakness is that the inverse convolutional method simply tries to undo the operation and is physics-agnostic. It does not know what an MZM, an Andreev bound state, or other states present in the wire are, nor what they should look like; in fact, it does not even know about the superconducting gap. In computer vision, many methods already exist for denoising; diffusion methods in particular provide a robust way to reconstruct high-resolution images from low-resolution, noisy inputs \cite{saharia2022image}. In practice, the high-temperature data can be viewed as a lower-resolution, corrupted version of the low-temperature data, so it should be feasible to reconstruct the low-temperature data from the high-temperature data in a similar way. In standard computer-vision diffusion tasks, the neural network can successfully reconstruct missing or corrupted parts of an image because it has been trained to recognize what similar objects, people, animals, environments, and other features typically look like \cite{rombach2022high,saharia2022image}. As a result, it can generate reasonable approximations of the uncorrupted data.

The objective for the conductance data is similar: by training the neural network on a very large number of Majorana nanowire devices, spanning a wide range of device realizations, the network may learn the relevant physics of these systems. In doing so, it can learn what low-temperature conductance data should look like and use that knowledge to produce highly accurate approximations of the corresponding low-temperature conductance. Moreover, it should be able to do so in a robust manner that is resilient to measurement noise, potentially even suppressing noise present in the low-temperature data.

Doing so would allow topological screening methods to be applied before committing a device to dilution-refrigerator measurements. Once an approximate low-temperature conductance map is reconstructed from high-temperature data, one could apply existing conductance-based assessments, such as topological gap protocol~\cite{pikulin2021protocol}, simple threshold-based checks~\cite{dassarma2016how}, or more advanced neural-network-based methods~\cite{taylor2025vision,taylor2025unreasonable}, to determine whether a candidate device is worthy of further investigation. Prior work has shown that low-temperature conductance contains sufficient information to predict topological properties, motivating the question of whether the full process can be performed starting only from high-temperature conductance data. We therefore further investigate whether the low-temperature topological visibility itself can be predicted directly from high-temperature conductance. This is a domain-to-domain mapping problem in which the network must learn which high-temperature conductance features are predictive of the corresponding low-temperature topological response. Together, conductance reconstruction and direct topological visibility (TV) prediction provide a rapid screening pipeline for identifying devices that have a realistic chance of hosting useful MZM before the most costly low-temperature measurements are performed.

Here we investigate both parts of this screening pipeline using two independent neural networks that each take high-temperature conductance as input but target different outputs. First, we show that modern diffusion-inspired neural-network architectures, specifically a Shifted Window U-Net Transformer (SWIN-UNETR) architecture originally developed for medical image reconstruction \cite{hatamizadeh2021swin}, can reconstruct low-temperature conductance maps from high-temperature conductance measurements with high fidelity; this is essentially denoising in which thermal broadening plays the role of noise. This demonstrates that the network can recover conductance structure beyond what is accessible through direct thermal deconvolution alone. Second, we ask whether the TV itself can be inferred directly from high-temperature conductance data, which is a fundamentally different, domain-to-domain mapping task rather than denoising. Building on prior work~\cite{taylor2025vision,taylor2025unreasonable} showing that low-temperature conductance can be used to predict topological indicators, we train a separate Video Vision Transformer (ViViT)~\cite{arnab2021vivit} to map high-temperature conductance directly to the corresponding low-temperature topological visibility. The conductance data has a natural three-dimensional structure $G(V_{\text{bias}},\mu,B)$; ViViT exploits this by treating the $(\mu,B)$ plane as spatial dimensions and $V_{\text{bias}}$ as the temporal dimension, a factorization that is physically motivated because TV depends only on $(\mu,B)$ and not on $V_{\text{bias}}$. This is a more demanding domain-to-domain inference problem: the network must not only undo thermal degradation, but also extract the topological information hidden in the broadened conductance response.

Together, these two approaches provide a route toward rapid pre-dilution high-temperature screening of Majorana devices. The reconstruction network enables low-temperature conductance diagnostics to be performed approximately at elevated temperatures, while the topology-prediction network estimates whether a device is likely to exhibit negative TV. This suggests a practical workflow in which high-temperature measurements are used to reject poor devices, prioritize promising candidates, and accelerate the experimental feedback loop required for MZM device production.

\HPadd{The remainder of this paper is organized as follows. In Sec.~\ref{sec:model}, we describe the nanowire model and the simulation of paired high- and low-temperature conductance data. In Sec.~\ref{sec:nn}, we introduce the two neural-network architectures employed for conductance reconstruction and TV prediction. In Sec.~\ref{sec:cond}, we present results for reconstructing low-temperature conductance from high-temperature data using the SWIN-UNETR diffusion process. In Sec.~\ref{sec:tv}, we demonstrate direct prediction of the low-temperature TV from high-temperature conductance using ViViT. We conclude in Sec.~\ref{sec:conclusion}. Additional details on the neural-network architectures are provided in Appendices~\ref{Sec:DiffNN} and~\ref{Sec:TVNN}, with supplementary results in Appendices~\ref{Sec:PureGaussianNoise}--\ref{Sec:LowTempTV}.}

\section{Model}\label{sec:model}
We model the one-dimensional single-band semiconductor-superconductor Majorana nanowire using a Bogoliubov-de Gennes Hamiltonian~\cite{lutchyn2010majorana}:
\begin{multline}\label{eq:ham}
H = \left(-\frac{\hbar^2}{2m^*}\partial_x^2 - i\alpha\partial_x\sigma_y - \mu + V_{\text{dis}}(x)\right)\tau_z \\
+ \frac{1}{2}g\mu_B B \sigma_x + \Sigma(\omega)
\end{multline}
where the self-energy from the proximitized superconductor is
\begin{equation}
    \Sigma(\omega) = -\gamma \frac{\omega + \Delta_0 \tau_x}{\sqrt{\Delta_0^2 - \omega^2}},
\end{equation} as in Ref.~\cite{sau2010robustness} with the positive real part branch cut for $\omega < \Delta$ and positive imaginary part otherwise. Here, $\sigma_{i}$ and $\tau_{i}$ are Pauli matrices for spin and particle-hole degrees of freedom, respectively, $\mu_B$ is the Bohr magneton, and the Hamiltonian is written in the Nambu basis $\psi(x) = (u_\uparrow(x), u_\downarrow(x), v_\downarrow(x), -v_\uparrow(x))^\intercal$. The frequency $\omega$ corresponds to the energy of the Bogoliubov quasiparticle. 

We compute the transport properties of the Majorana nanowire from the scattering matrix using the Blonder-Tinkham-Klapwijk formalism~\cite{blonder1982transition}. The scattering matrix, in turn, is obtained from a discretized version of the Hamiltonian $H$ using  \texttt{KWANT}~\cite{groth2014kwant}. This model in Eq.~\eqref{eq:ham} quantitatively reproduces experimental transport features when parameters are appropriately fitted~\cite{dassarma2023spectral,pan2024disordered}. In particular, we use the following realistic parameter values close to \cite{microsoftquantum2023inasal}: effective mass $m^* = 0.03~m_e$, superconducting pairing potential $\gamma = 0.15$ meV, Land\'e $g$-factor $g = 25$, parent Al superconducting gap $\Delta_0(T=50~\text{mK}) = 0.3$~meV, and low temperature $T_{L} = 50$~mK~\cite{pan2021threeterminal,woods2021chargeimpurity,dassarma2023spectral} and high temperature $T_{H} = 300$~mK. We choose the barrier voltage to be $V_{\text{barrier}}^{L/R}=15$~meV~\cite{setiawan2017electron}. We use a wire length of $3~\mu$m. 
For the high-temperature simulations, we use
$\Delta_0(T_H)=\Delta_0(T=0)\tanh\!\left(1.74\sqrt{\frac{T_c - T_H}{T_H}}\right)$ with $T_c=1.2$~K.
The magnetic field $B$ is varied over the range of $[0,0.8]$ Tesla with 20 steps, and the chemical potential is varied over the range $\mu \in [0.2, 0.4]$~meV with 5 steps. These grid sizes (as well as the $V_{\text{bias}}$ grid) are each padded by one point to satisfy the input-dimension requirements of the neural-network architecture, accounting for the slightly larger values quoted below.

We generate device samples by randomly sampling unknown physical parameters from physically plausible ranges. In particular, the spin-orbit coupling $\alpha$ was sampled from $[0.85,1.25] \times 8$ meV nm.
The disorder $V_{\text{dis}}$ follows a correlated Gaussian distribution with the standard deviation $\sigma_{\text{dis}}$ being varied over $[0.15,4.5]$~meV, with the correlation length varied between $20$~nm and $70$~nm.
All parameters were sampled uniformly within these ranges, and no information about their values was provided to the neural networks.

The differential conductances, $G_{ii}$ for the local reflection and $G_{ij}$ ($i\neq j$) for the nonlocal transmission, where $i,j\in\{L,R\}$ denote the left and right terminals, are first computed at zero temperature for fixed $\omega$.
To incorporate finite-temperature effects, the zero-temperature conductance $G_{ij}(\omega,T=0)$ is convolved with the derivative of the Fermi distribution $f(E)$, following Ref.~\cite{setiawan2017electron}:
\begin{equation}
    G_{ij}(V_{\text{bias}},T)=-\int_{-\infty}^\infty d\omega ~
    G_{ij}(\omega,0)\frac{df(\omega-eV_{\text{bias}},T)}{d\omega}.
    \label{eqn:conv}
\end{equation}
This convolution requires conductance values over a sufficiently broad range of $\omega$. 
In our simulations, we use $V_{\text{bias}} \in [-0.15,0.15]$~mV with 151 points at low temperature and $V_{\text{bias}} \in [-0.6,0.6]$~mV with 301 points at high temperature. The larger high-temperature range is needed to accurately capture finite-temperature broadening, which also improves neural-network fidelity. For the neural-network input, we retain only the $[-0.15,0.15]$~mV subset of the high-temperature conductance data, which contains 75 points, and interpolate it to 151 points. This undersampling is due to computational resource limitations; improved sampling is expected to yield higher fidelities.

\section{Neural networks}\label{sec:nn}
Our method consists of generating high-temperature and low-temperature simulation pairs for many wire realizations. Namely, the conductance pairs $G_{ij}(V_{\text{bias}},\mu, B, T)= (-1+2\delta_{ij})dI_i(T)/dV_j(T)$ 
for specific device parameter realizations of $V_\text{dis}(x)$ and $\alpha$. During each run, we simultaneously calculate the low-temperature topological visibility (TV)~\cite{dassarma2016how} data over $(\mu, B)$. We generate a total of 20,000 device realizations, withholding 5\% for testing.

We break the problem into two separate goals: using high-temperature conductance data as input, first solving for the local and nonlocal conductances at low temperature, and second, determining the low-temperature TV. The neural network varies significantly between these two tasks. For conductance reconstruction, the input and output have the same shape, $G(V_{\text{bias}},\mu,B) \to G(V_{\text{bias}},\mu,B)$, so a U-Net-style encoder-decoder (SWIN-UNETR) with skip connections is natural as it preserves pixel-level correspondence. For TV prediction, the output $\mathrm{TV}(\mu,B)$ has fewer dimensions than the input $G(V_{\text{bias}},\mu,B)$ since $V_{\text{bias}}$ must be aggregated away; ViViT's factorized spatial-then-temporal attention naturally handles this by processing each $(\mu,B)$ slice independently and then collapsing across $V_{\text{bias}}$.

We first reconstruct the low-$T$ conductance from high-$T$ data using a diffusion-like process in which the high-temperature effects are treated as noise that the neural network attempts to remove. We reshape the problem from 3D to 2D by flattening the two axes of $\mu$ and $B$ to construct an image of size $(N_{V_{\text{bias}}}+1, (N_{\mu}+1) \times N_{B})= (152 , 6  \times20=120)$ with four conductance channels $[G_{RR}, G_{LL}, G_{LR}, G_{RL}]$ and minor padding.
A SWIN-UNETR~\cite{hatamizadeh2021swin} is used for the denoising process: 
window-based vision transformers, with window mixing, encode the high-temperature conductance, and transposed convolutions are then used during decoding to recover the low-temperature conductance. We train separate networks for local and nonlocal conductances, though in both cases the input includes all four conductance components to provide the neural network with more physics-specific information. We attempted many different architectures; however, this approach significantly outperformed them. See Appendix~\ref{Sec:DiffNN} for additional details.

Second, for the TV prediction, we use a very different neural-network architecture, since this task is no longer a denoising process but rather a computational mapping from one domain to another. In this case, we use a hybrid 3D {Video Vision Transformer }~\cite{arnab2021vivit}, which consists of hierarchical vision transformers applied first to ``space'' (for our purposes, $\mu$ and $B$) and then to ``time'' (for our purposes, $V_{\text{bias}}$ {, since the predicted TV does not depend on  $V_{\text{bias}}$}). We use ViViT to obtain a coarse-grained, low-dimensional encoded TV output. This output is then interpolated and passed through a $\tanh$ activation and a small number of convolutional layers to perform the expansion and fine-tuning operations over the full phase diagram. Additional details are provided in Appendix~\ref{Sec:TVNN}. 

\section{Predicting low-$T$ conductance from high-$T$ conductance}\label{sec:cond}
In principle, the low-temperature features can be extracted by deconvolving the finite-temperature convolution in Eq.~\eqref{eqn:conv}. In practice, however, this procedure is typically limited to a resolution of approximately $k_B T \approx 0.026$~meV, which is insufficient to resolve any of the intricate features associated with MZMs. This limitation arises because the thermal broadening kernel strongly suppresses high-frequency components in energy, so direct inverse deconvolution requires division by exponentially small Fourier components. As a result, even small experimental noise, finite energy resolution, discretization error, or systematic offsets are dramatically amplified, making the reconstruction ill-conditioned. Our approach avoids this unstable direct inversion by using a physics-informed network to learn the physically allowed structure of the low-temperature conductance. This enables reconstruction fidelities far beyond those achievable with standard inverse-deconvolution methods.

\begin{figure}[h]
    \begin{subfigure}[t]{0.49\linewidth}
    \phantomcaption
    \begin{overpic}[width=\textwidth]{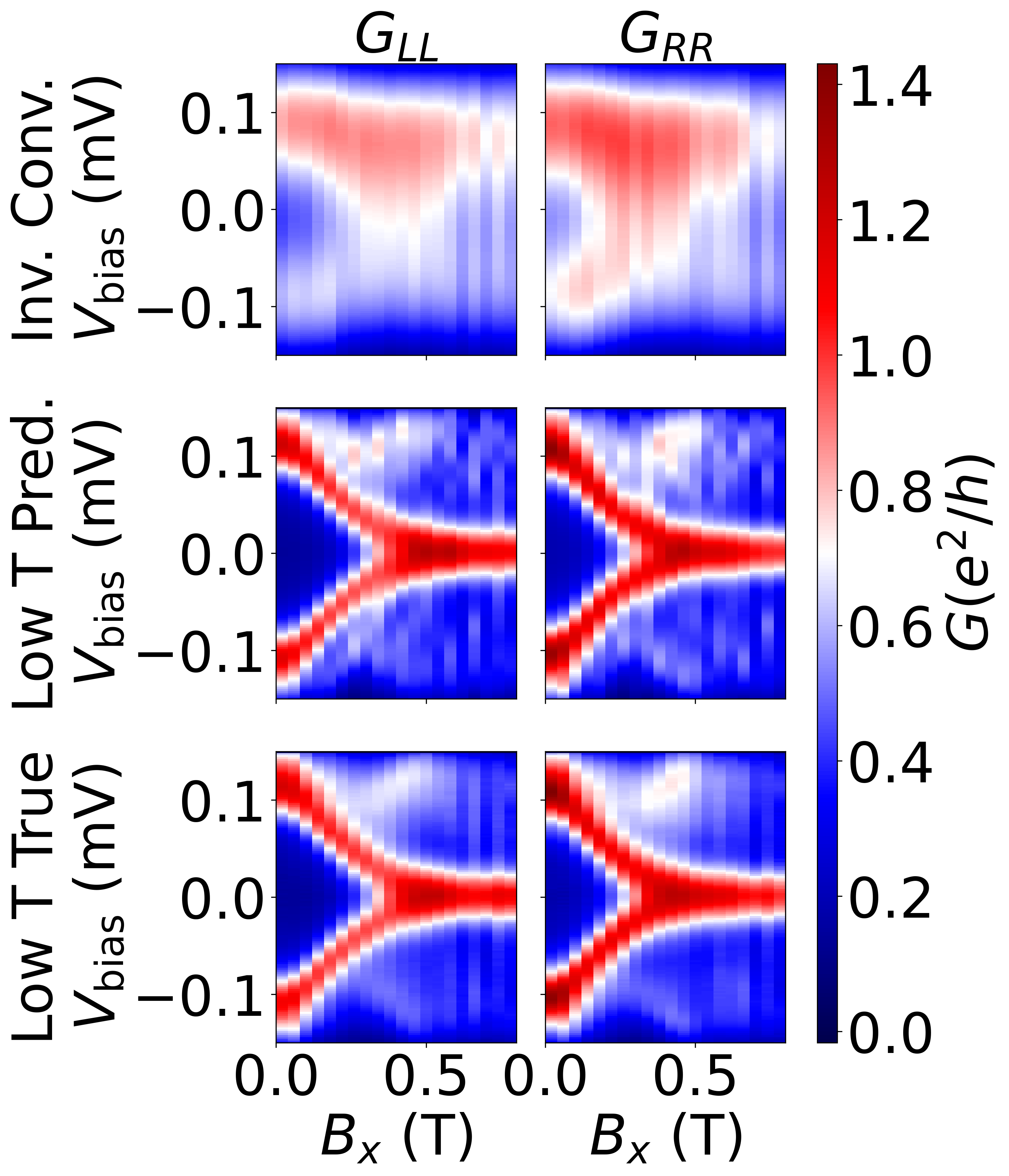}
        \put(0,100){{(\thesubfigure)}}
    \end{overpic}
    \end{subfigure}
    \hfill
    \begin{subfigure}[t]{0.49\linewidth}
        \phantomcaption
    \begin{overpic}[width=\textwidth]{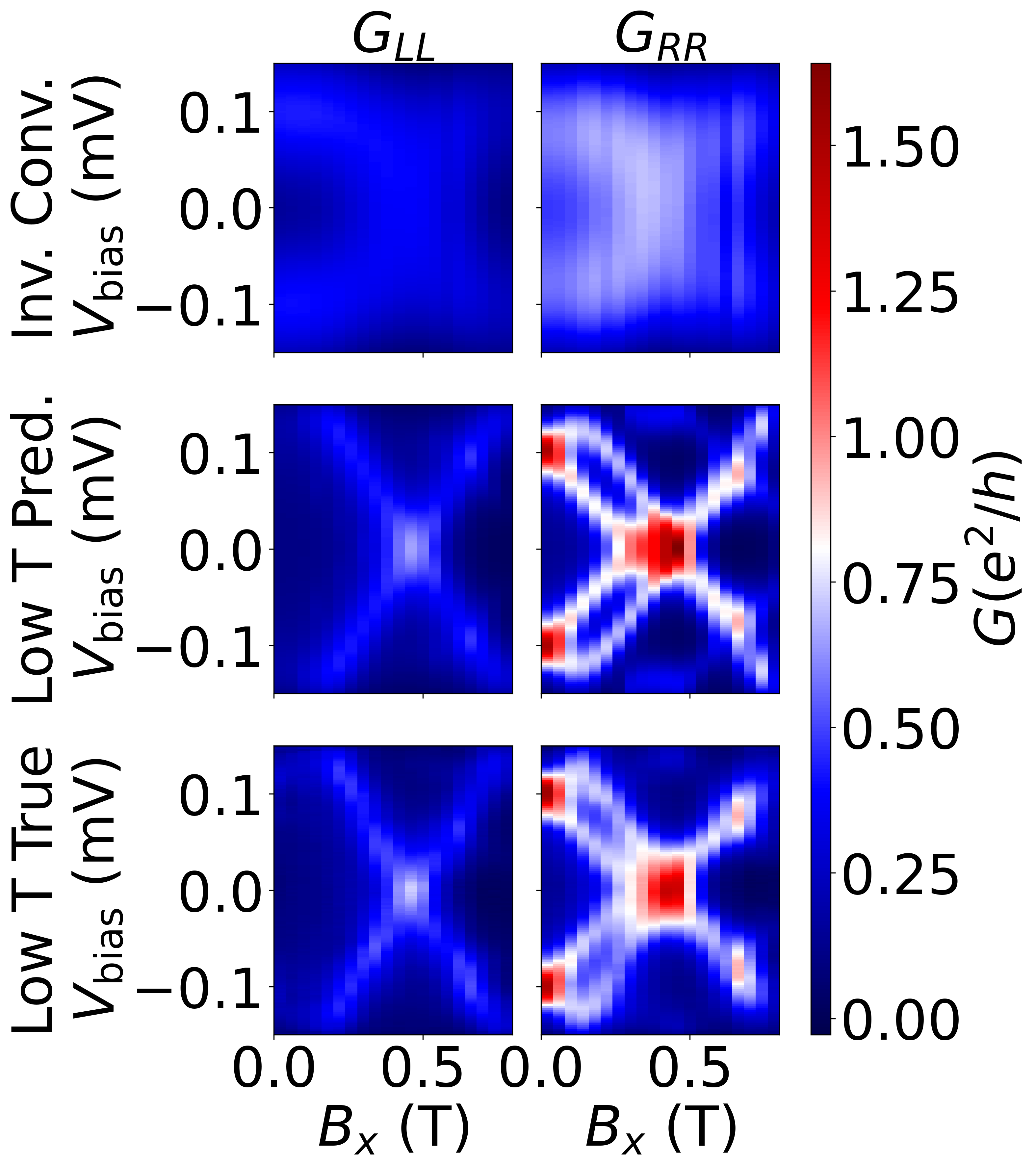}
        \put(0,100){{(\thesubfigure)}}
    \end{overpic}
    \end{subfigure}
    \begin{subfigure}[t]{0.49\linewidth}
        \phantomcaption
    \begin{overpic}[width=\textwidth]{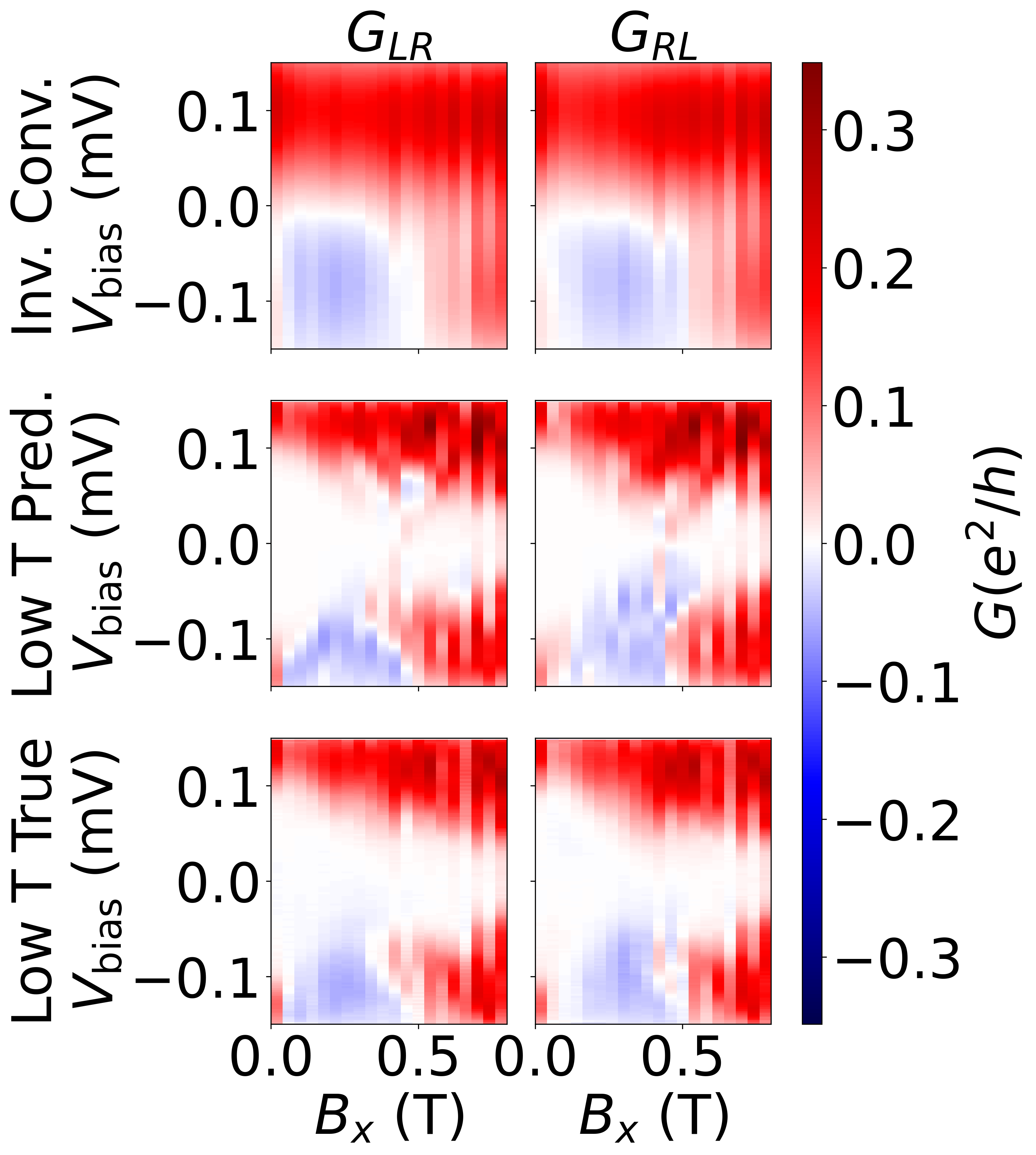}
        \put(0,100){{(\thesubfigure)}}
    \end{overpic}
    \end{subfigure}
    \hfill
    \begin{subfigure}[t]{0.49\linewidth}
    \phantomcaption
    \begin{overpic}[width=\textwidth]{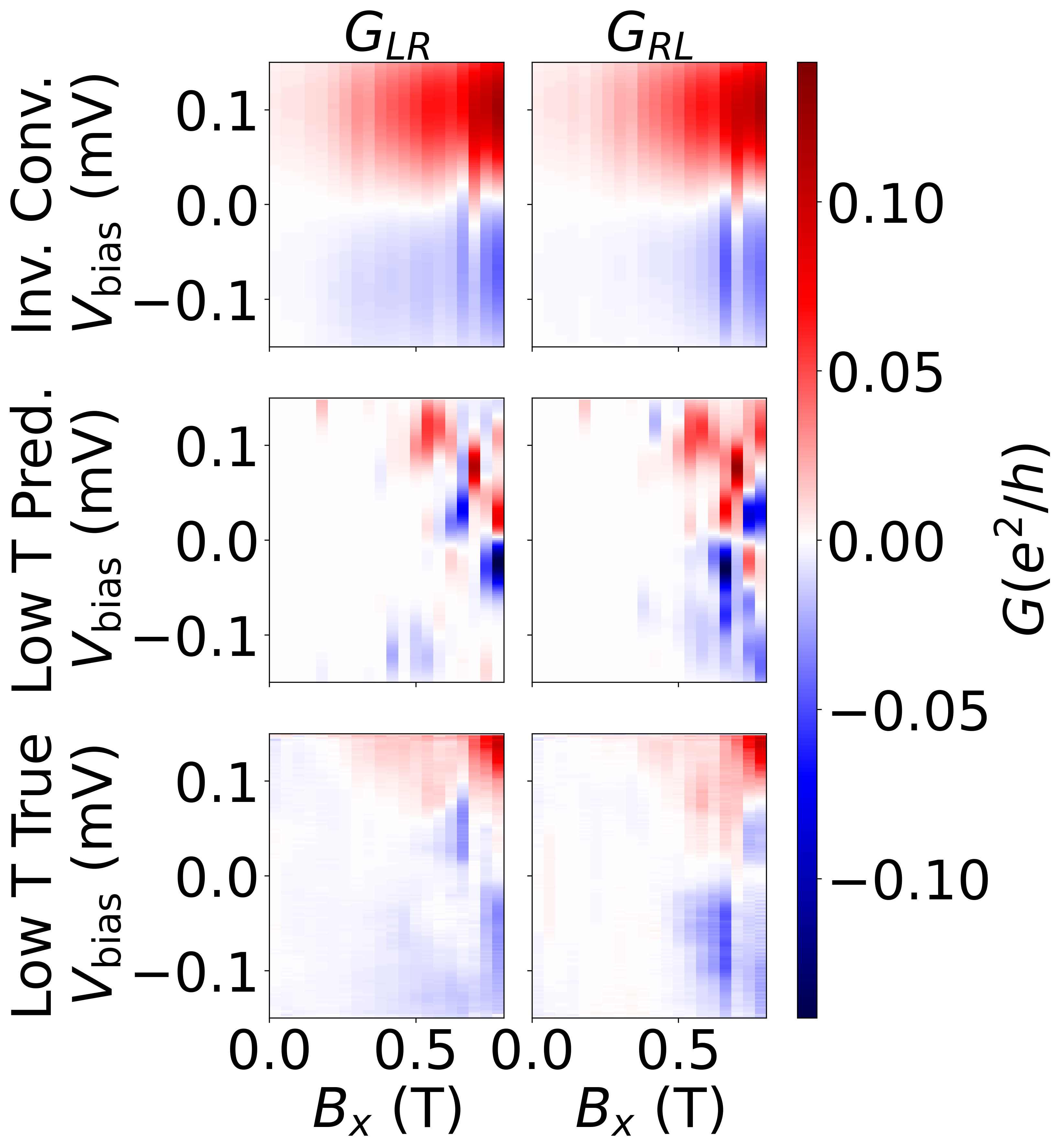}
        \put(0,100){{(\thesubfigure)}}
    \end{overpic}
    \end{subfigure}
    \caption{Low-temperature conductance reconstructed from high-temperature data by reverse convolution (1st row), measured directly (2nd row), and predicted by a neural network using high-temperature data (3rd row). Local conductances are shown in (a,b), and nonlocal conductances in (c,d). Additional maps can be found in Fig. \ref{fig:Extra1LocalCondCompare_a}-\ref{fig:HighTE2NonLocalCondCompare_b}, {including the corresponding high temperature maps.} The chemical potential shown is 0.4 meV.}
    \label{fig:CondCompare}
\end{figure}

We perform a single-step SWIN-UNETR diffusion process, mapping the high-temperature data to low temperature, and find that, for local conductance, this can be done with near-perfect fidelity of $R^2={0.952}$ and root-mean-square error $ \sqrt{\text{MSE}}(G_{RR|LL})={0.060}$. This is achieved despite the very low resolution of the high-temperature input data, with only 75 $V_{\mathrm{bias}}$ points within the relevant parameter range fed to the neural network. Much like diffusion image models, the neural network works to remove data corruption: in denoising diffusion,
this corruption is Gaussian noise, while here it is the high-temperature convolution. By being trained on the actual physics of the devices, the network is able to significantly improve upon the classical Fourier-transform method, for which effectively all relevant information is lost. We present these results in Fig.~\ref{fig:CondCompare}(a)-(b).

The separate neural network trained to predict nonlocal conductance has more difficulties. However, it is still able to predict nonlocal conductance with relatively high fidelity, achieving $R^2={0.9096}$ and $ \sqrt{\text{MSE}}(G_{LR|RL})={0.0097}$. This suggests that, for the neural network, nonlocal conductance is far less predictable than local conductance. We present these results in Fig.~\ref{fig:CondCompare}(c)-(d).

\begin{figure}[h]
    \centering
    \begin{subfigure}[t]{0.49\linewidth}
    \phantomcaption
    \begin{overpic}[width=\textwidth]{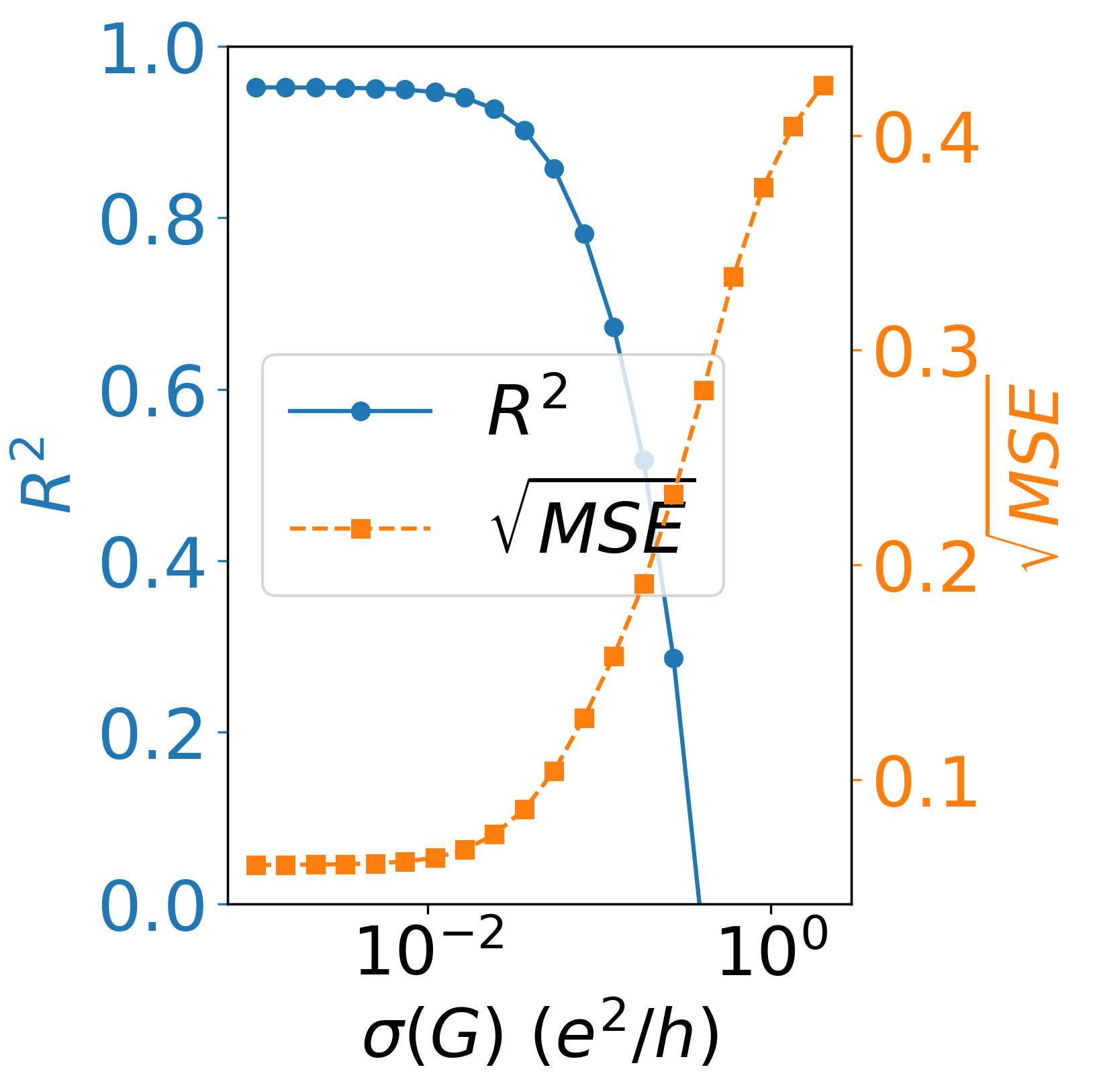}
        \put(0,100){{(\thesubfigure)}}
    \end{overpic}
    \end{subfigure}
    \hfill
    \begin{subfigure}[t]{0.49\linewidth}
        \phantomcaption
    \begin{overpic}[width=\textwidth]{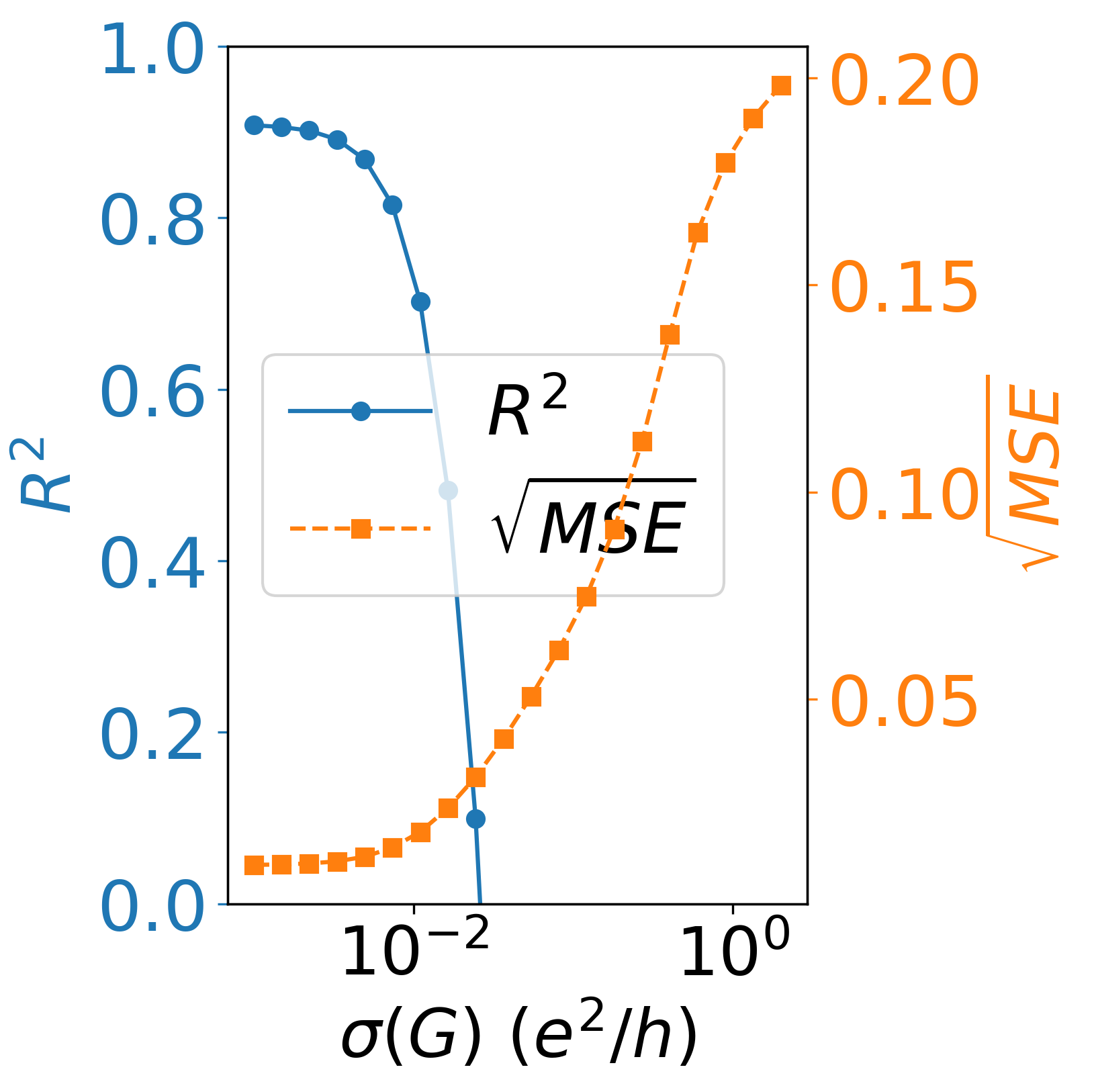}
        \put(0,100){{(\thesubfigure)}}
    \end{overpic}
    \end{subfigure}
    \begin{subfigure}[t]{0.49\linewidth}
    \phantomcaption
    \begin{overpic}[width=\textwidth]{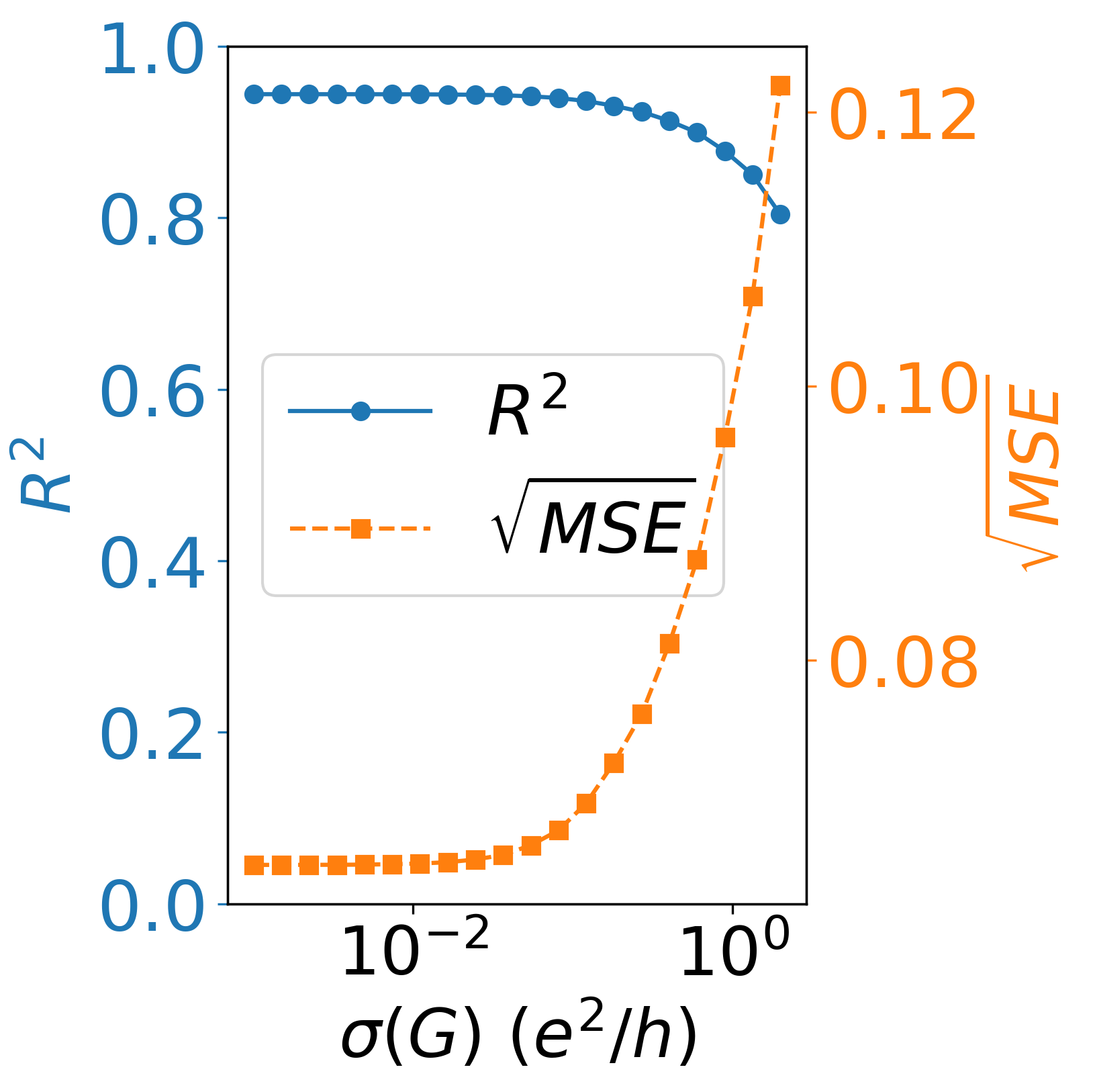}
        \put(0,100){{(\thesubfigure)}}
    \end{overpic}
    \end{subfigure}
    \begin{subfigure}[t]{0.49\linewidth}
    \phantomcaption
    \begin{overpic}[width=\textwidth]{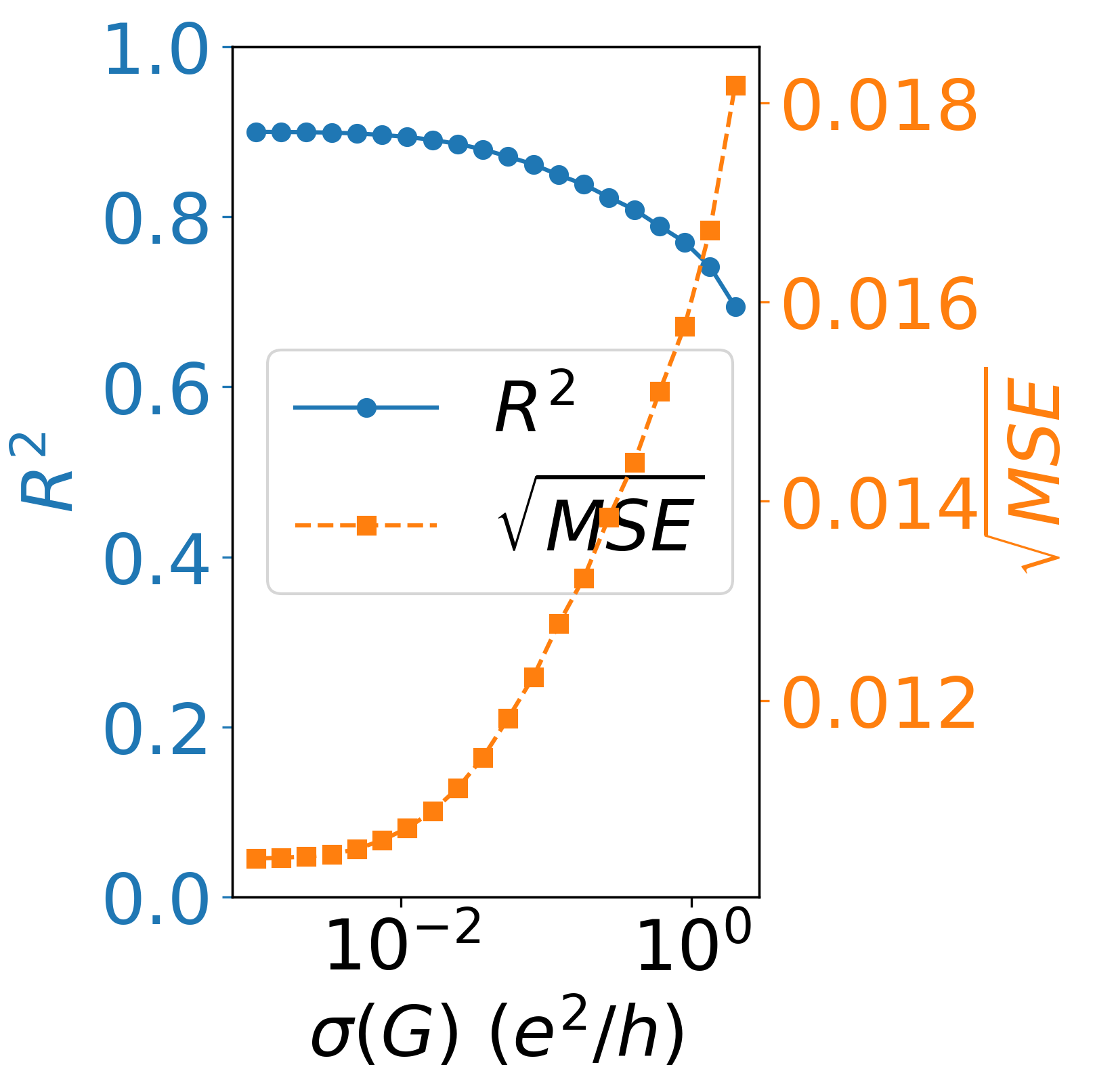}
        \put(0,100){{(\thesubfigure)}}
    \end{overpic}
    \end{subfigure}
    \caption{Local (left) and nonlocal (right) conductance reconstruction fidelity for varying levels of error. The first (second) row is for networks without (with) measurement error during the training stage.}
    \label{fig:Resilience}
\end{figure}

{Following this, we test the network's resilience to measurement imprecision by adding pointwise Gaussian noise $\mathcal{N}(0,\sigma^2)$ directly to the conductance values, where $\sigma(G)$ has units of $e^2/h$. We find that the local conductance prediction collapses around $\sigma(G) \sim 0.1 e^2/h$ and the nonlocal conductance around $\sigma(G) \sim 0.01 e^2/h$ for the base model.} 
To improve robustness, the model is subsequently fine-tuned with noise included during training {(using the same device samples as before)}. This adapts the network to noise and results in approximately 10 times improved resilience, with the local conductance collapsing around $e^2/h$ and the nonlocal conductance around $0.1 e^2/h$. We present this comparison in Fig.~\ref{fig:Resilience}. These results strongly suggest that our diffusion process can enable measurements to be performed at high temperature while still resolving the low-temperature features required for conventional MZM tests~\cite{beenakker2013search,lutchyn2018majorana,rosdahl2018andreev,pan2021threeterminal,microsoftquantum2023inasal,dassarma2016how,pikulin2021protocol}.

{To assess the diffusion method, and in particular where the uncertainty occurs, we test the case of applying diffusion to low-temperature inputs corrupted by additive Gaussian noise in order to recover the underlying true low-temperature result. We find that the fidelity collapses at similar noise levels. This provides support for the interpretation that the neural network works to remove corruption in the high-$T$ case.}
See Sec.~\ref{Sec:PureGaussianNoise} of the Appendix for more details.

\section{Predicting the topology from high-$T$ conductance}\label{sec:tv}
The more useful screening method is to directly predict the topological invariant, e.g., the topological visibility~\cite{dassarma2016how}. Namely, from high-temperature measurements, one would like to determine with high confidence whether a device has a chance of containing MZMs. It is already known from Ref.~\cite{taylor2025vision} that this can be done theoretically at low temperature. Beyond this, it is important to assess how robust such a method is to error, namely to determine how fine-tuned the results are.

\begin{figure}[h]
    \centering
    \begin{subfigure}[t]{0.49\linewidth}
    \phantomcaption
    \begin{overpic}[width=\textwidth]{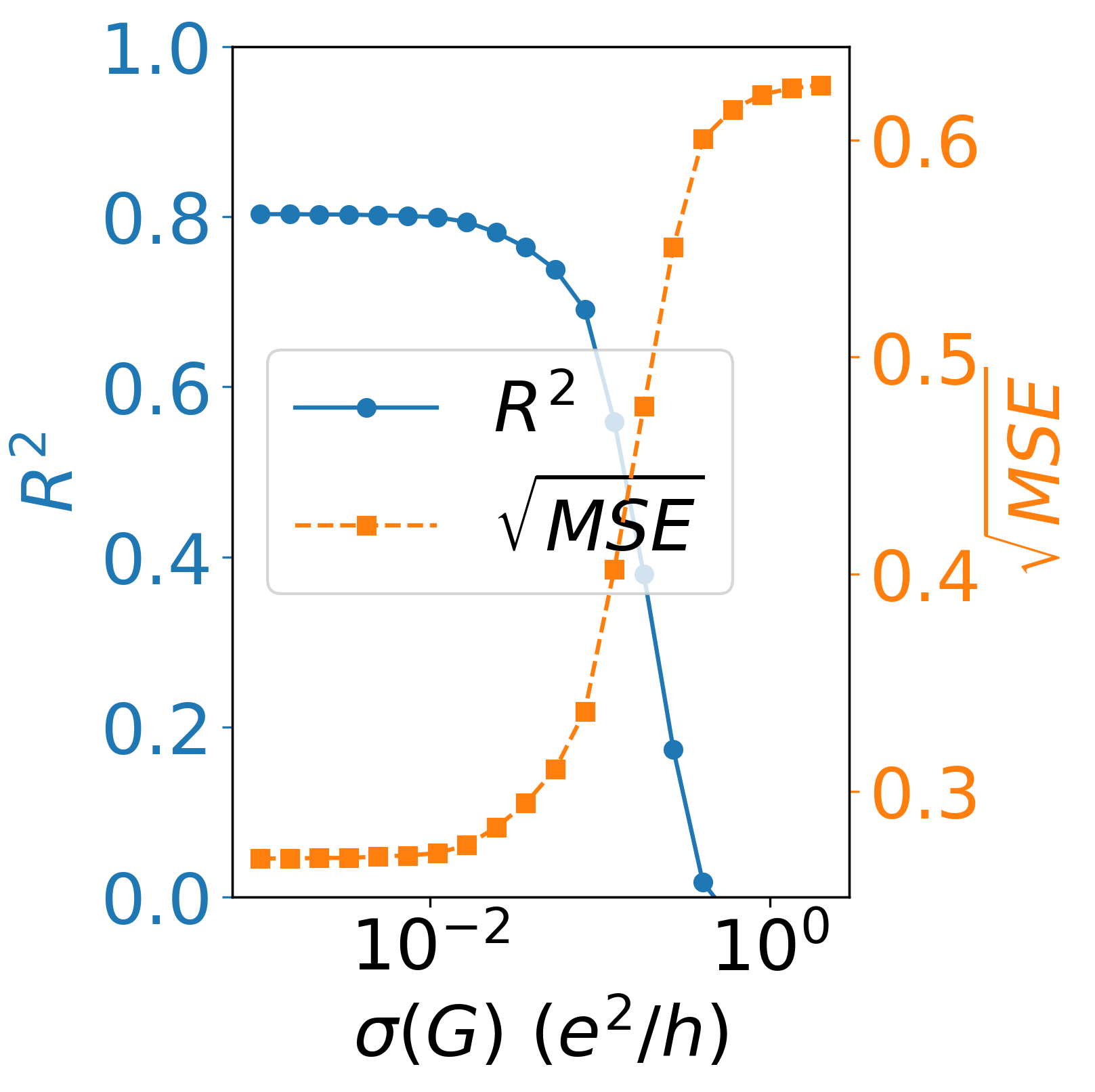}
        \put(0,100){{(\thesubfigure)}}
    \end{overpic}
    \end{subfigure}
    \begin{subfigure}[t]{0.49\linewidth}
    \phantomcaption
    \begin{overpic}[width=\textwidth]{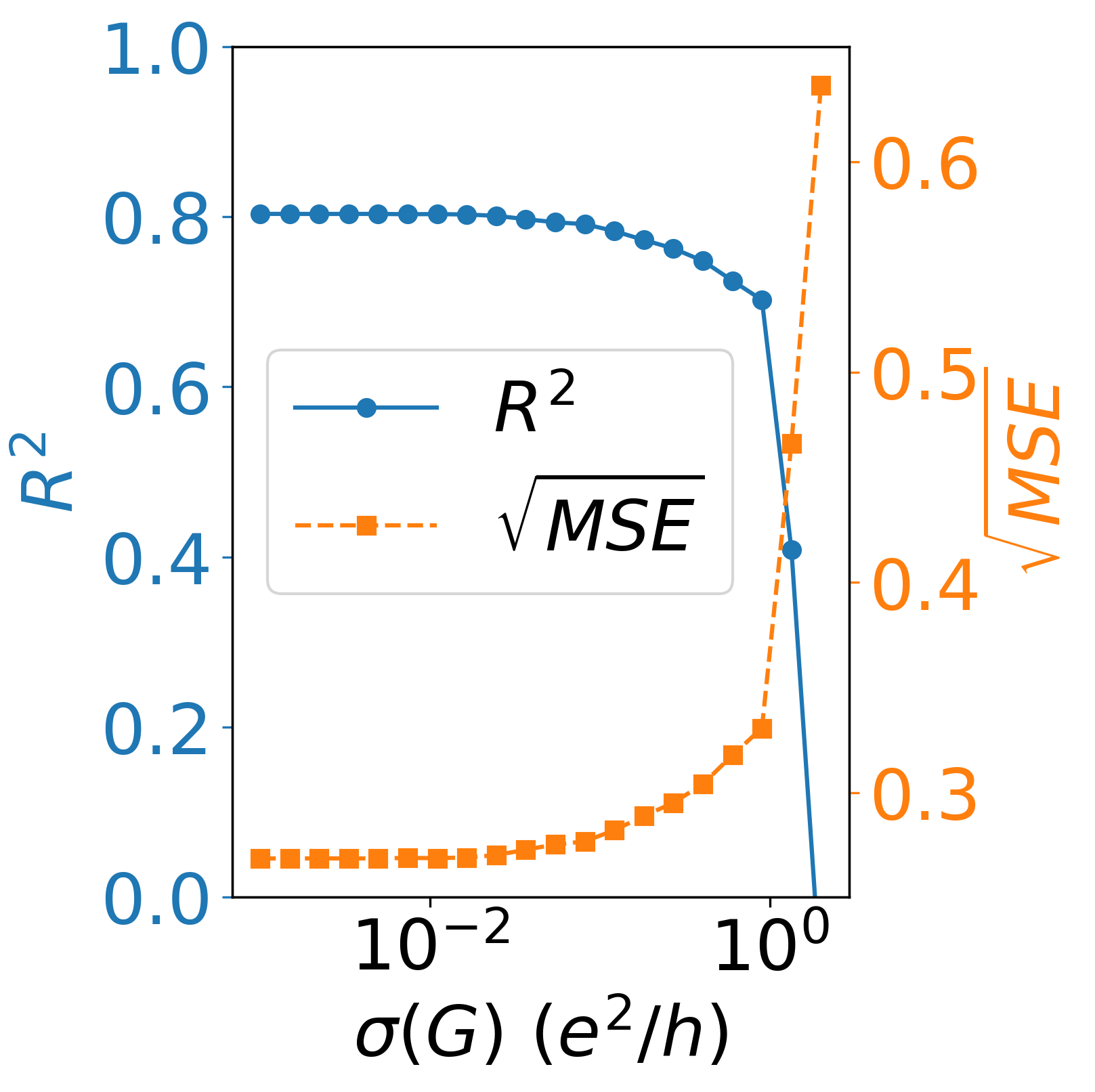}
        \put(0,100){{(\thesubfigure)}}
    \end{overpic}
    \end{subfigure}
    \caption{TV prediction fidelity for varying levels of  Gaussian measurement noise of standard deviation $\sigma(G)$. $R^2 (TV)$ vs.\ $\sigma(G)$ for the neural network without (a) and with (b) noise added during the training stage.}
    \label{fig:TVResilience}
\end{figure}

\begin{figure}[h]
\begin{subfigure}[t]{0.49\linewidth}
    \centering
    \phantomcaption
    \begin{overpic}[width=\textwidth]{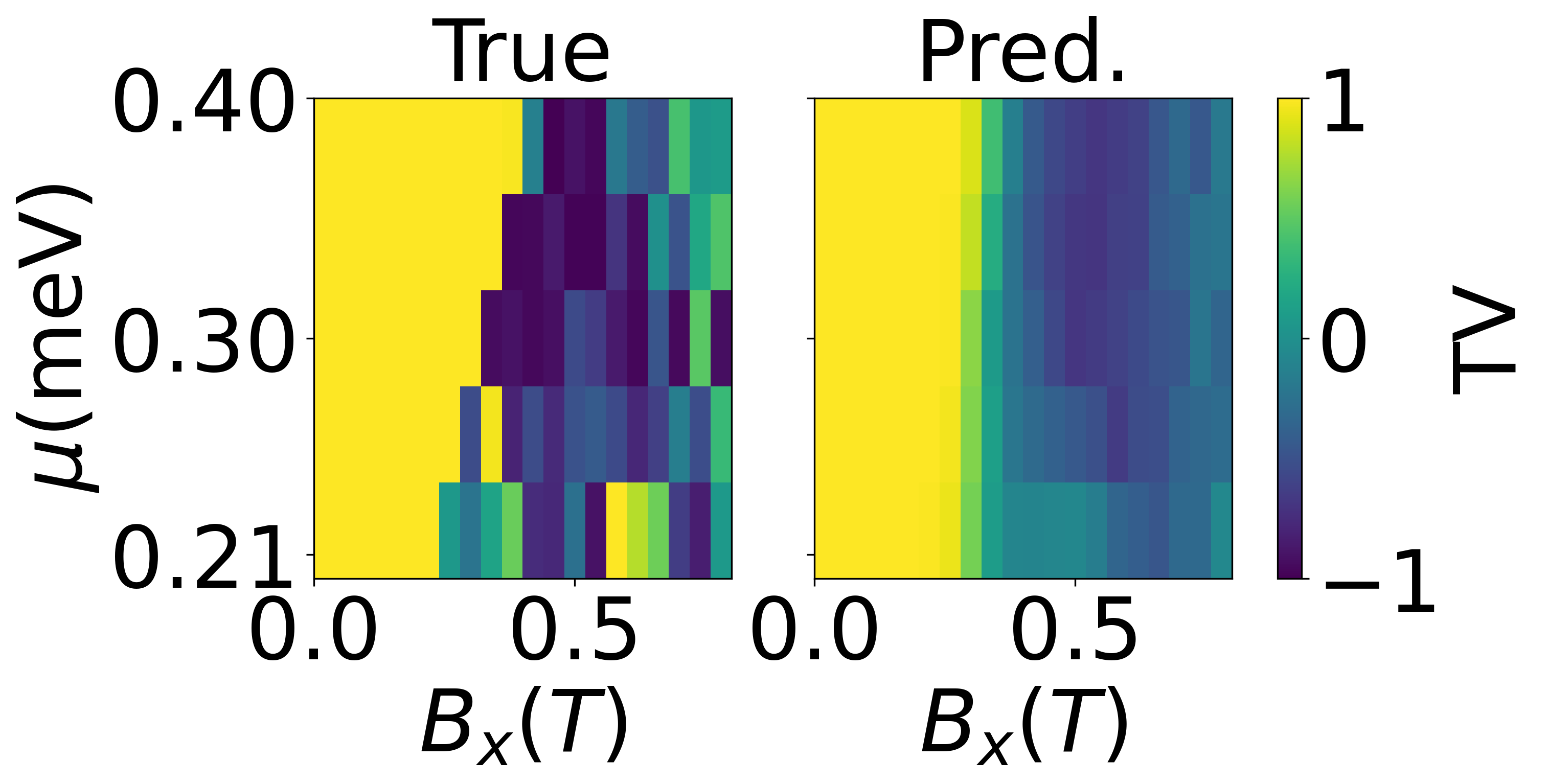}
        \put(0,50){{(\thesubfigure)}}
    \end{overpic}
\end{subfigure}
    \hfill
\begin{subfigure}[t]{0.49\linewidth}
    \centering
    \phantomcaption
    \begin{overpic}[width=\textwidth]{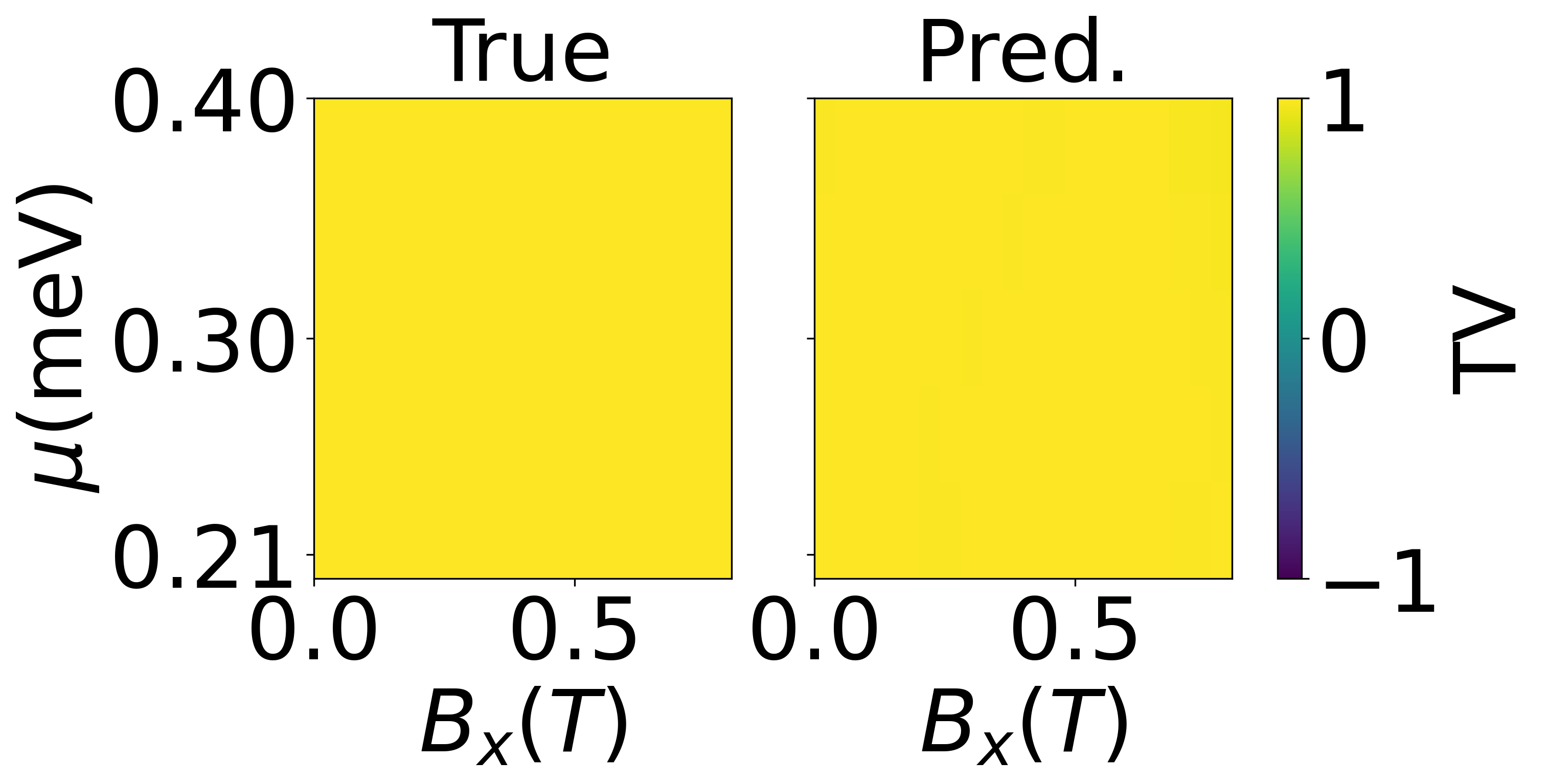}
        \put(0,50){{(\thesubfigure)}}
    \end{overpic}
\end{subfigure}
    \hfill
\begin{subfigure}[t]{0.49\linewidth}
    \centering
    \phantomcaption
    \begin{overpic}[width=\textwidth]{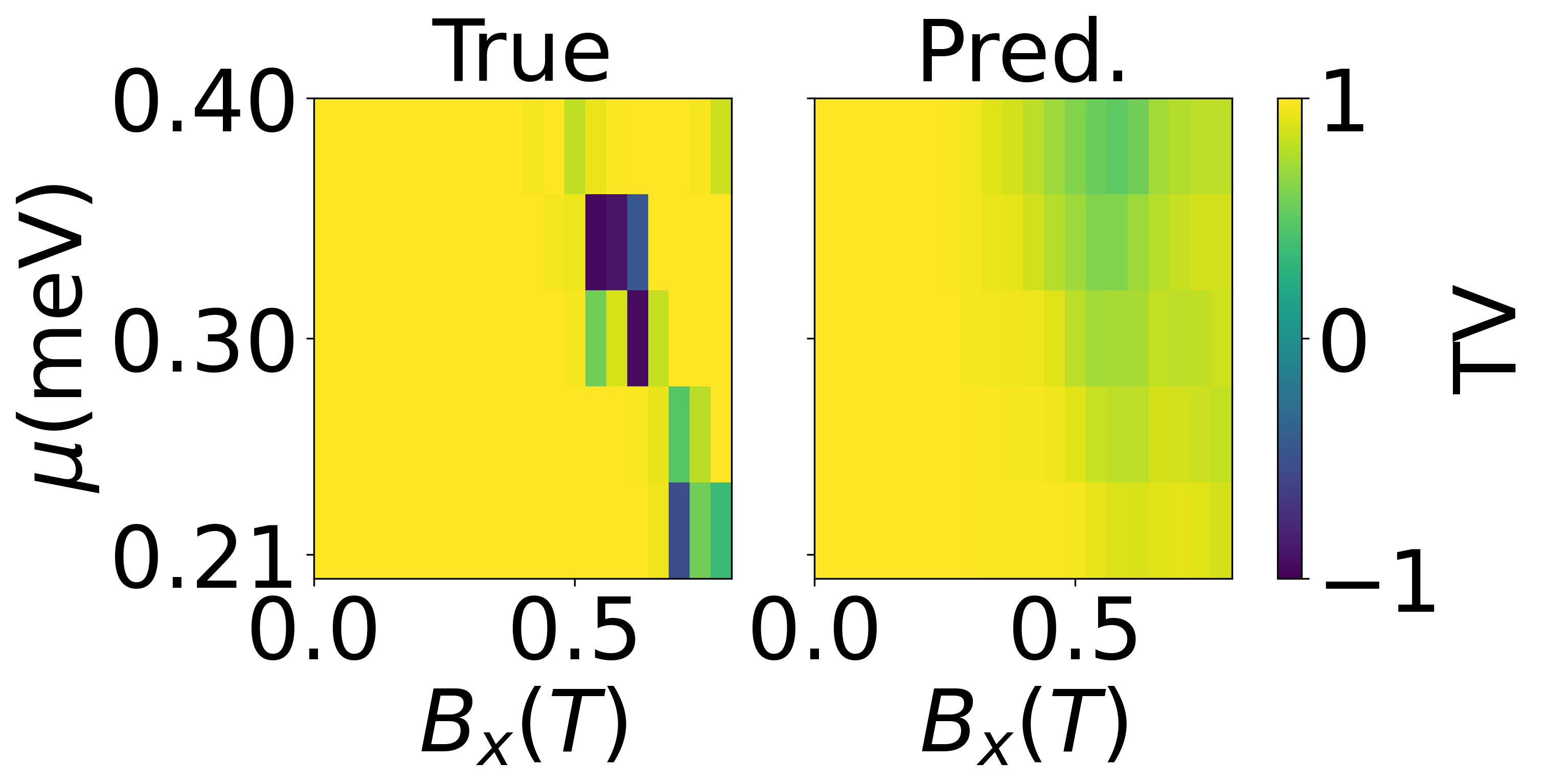}
        \put(0,50){(\thesubfigure)}
    \end{overpic}
\end{subfigure}
    \hfill
\begin{subfigure}[t]{0.49\linewidth}
    \centering
    \phantomcaption
    \begin{overpic}[width=\textwidth]{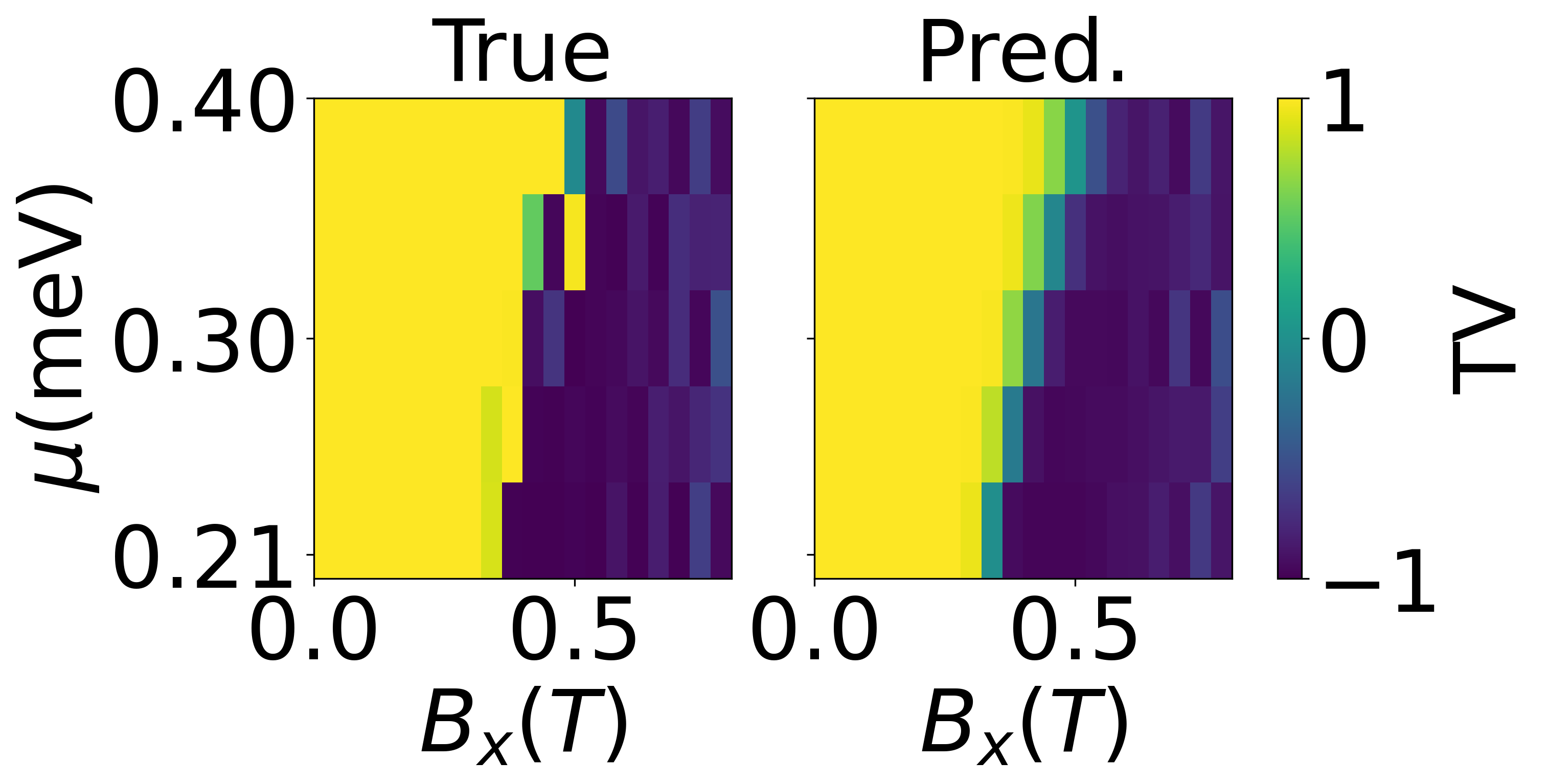}
        \put(0,50){{(\thesubfigure)}}   
    \end{overpic}
\end{subfigure}
    \caption{Comparison of expected and predicted TV from high-$T$ conductance.  Panels (a)-(d) are different device realizations with left being the expected TV and right being the predicted TV. The disorder levels are $\sigma_{\text{dis}}$ = (a) 0.716 meV, (b) 3.55 meV, (c) 0.237 meV, and (d) 1.586 meV,  with correlation lengths of (a) 27 nm, (b) 31.9 nm, (c) 47 nm, and (d) 48 nm respectively.}
    \label{fig:TVCompare}
\end{figure}

Here we show that the TV magnitude can be predicted from only high-temperature conductance measurements with $R^2={0.8017}$ and $ \sqrt{\text{MSE}}(\text{TV})={0.2669}$ {with high resilience (see Fig.~\ref{fig:TVResilience})}. 
Within Fig.~\ref{fig:TVCompare} we can also see that the previous blurring effect of the neural network occurs when the precise boundaries of the complex phase are unknown. The network continues to give signs indicating that it is uncertain about the boundaries within some regions. This is ideal for screening, which these results strongly support, and is similarly useful for experiments because the network provides an indication of its uncertainty. At low temperature, along with a larger $V_{\mathrm{bias}}$ region, this blur can be significantly resolved, as expected from our prior work. Using the same boundary, we find that the low-temperature-to-TV-domain mapping achieves a fidelity of $R^2={0.8317}$ and $\sqrt{\text{MSE}}(\text{TV})={0.246}$. See Fig.~\ref{fig:LowTTVResilience} for more details. This higher fidelity allows far more intricate features of the TV map to be discerned, as shown in Fig.~\ref{fig:LowTTVCompare}.

\section{Conclusion}\label{sec:conclusion} We have shown that high-temperature conductance measurements can be used to infer low-temperature Majorana nanowire features. Using a SWIN-UNETR diffusion process, we reconstruct low-temperature conductance from thermally broadened high-temperature data with high fidelity, enabling conventional low-temperature screening methods to be applied approximately at high temperature, avoiding the need to cool devices to dilution-refrigerator temperatures. Since this reconstruction only requires paired conductance data, the method can, in principle, be fine-tuned directly on experimental measurements without requiring a perfect microscopic model. In general, the method provides much better fidelity for the local conductance compared with the nonlocal conductance for the understandable reason that the nonlocal conductance is much smaller in magnitude and hence more strongly thermally blurred.  It should be possible to improve the fidelity further by using computational resources well beyond our capacity, but we do not believe this is necessary since the main goal is to eliminate ``bad'' devices from dilution refrigerator measurements, which does not necessitate perfect fidelity. We note that it should be straightforward to take the high-temperature data to higher temperatures, perhaps even up to $T> 1$K, but this would require much more training going well beyond our computational resources.

We also demonstrate that low-temperature topological visibility can be predicted directly from high-temperature conductance using a ViViT-based neural network. Together, these two approaches provide a practical high-temperature screening pipeline: poor devices can be rejected early, while promising or uncertain devices can be prioritized for full dilution-refrigerator characterization, massively increasing throughput. Our technique could considerably improve the yield rate for ``good'' devices entering the exhaustive dilution refrigerator measurement protocols, since all devices failing our procedure could be safely discarded from further consideration. Future improvements may come from more realistic training data, additional temperature steps (like in conventional diffusion processes), and direct fine-tuning on experimental high/low-temperature conductance pairs.

\begin{acknowledgments}
This work is supported by the Laboratory for Physical Sciences through the Condensed Matter Theory Center at Maryland. 
J.R.T. thanks the Joint Quantum Institute for additional support. 
H.P. is supported by US-ONR grant No.~N00014-23-1-2357 and startup funds at the University of Florida.
\end{acknowledgments}
\bibliography{ML_HighT,Paper_High_T}

@article{aghaee2025interferometric,
  title = {Interferometric Single-Shot Parity Measurement in {{InAs}}--{{Al}} Hybrid Devices},
  author = {Aghaee, Morteza and Alcaraz Ramirez, Alejandro and Alam, Zulfi and Ali, Rizwan and Andrzejczuk, Mariusz and Antipov, Andrey and Astafev, Mikhail and Barzegar, Amin and Bauer, Bela and Becker, Jonathan and Bhaskar, Umesh Kumar and Bocharov, Alex and Boddapati, Srini and Bohn, David and Bommer, Jouri and Bourdet, Leo and Bousquet, Arnaud and Boutin, Samuel and Casparis, Lucas and Chapman, Benjamin J. and Chatoor, Sohail and Christensen, Anna Wulff and Chua, Cassandra and Codd, Patrick and Cole, William and Cooper, Paul and Corsetti, Fabiano and Cui, Ajuan and Dalpasso, Paolo and Dehollain, Juan Pablo and {de Lange}, Gijs and {de Moor}, Michiel and Ekefj{\"a}rd, Andreas and El Dandachi, Tareq and Estrada Salda{\~n}a, Juan Carlos and Fallahi, Saeed and Galletti, Luca and Gardner, Geoff and Govender, Deshan and Griggio, Flavio and Grigoryan, Ruben and Grijalva, Sebastian and Gronin, Sergei and Gukelberger, Jan and Hamdast, Marzie and Hamze, Firas and Hansen, Esben Bork and Heedt, Sebastian and Heidarnia, Zahra and Herranz Zamorano, Jes{\'u}s and Ho, Samantha and Holgaard, Laurens and Hornibrook, John and Indrapiromkul, Jinnapat and Ingerslev, Henrik and Ivancevic, Lovro and Jensen, Thomas and Jhoja, Jaspreet and Jones, Jeffrey and Kalashnikov, Konstantin V. and Kallaher, Ray and Kalra, Rachpon and Karimi, Farhad and Karzig, Torsten and King, Evelyn and Kloster, Maren Elisabeth and Knapp, Christina and Kocon, Dariusz and Koski, Jonne V. and Kostamo, Pasi and Kumar, Mahesh and Laeven, Tom and Larsen, Thorvald and Lee, Jason and Lee, Kyunghoon and Leum, Grant and Li, Kongyi and Lindemann, Tyler and Looij, Matthew and Love, Julie and Lucas, Marijn and Lutchyn, Roman and Madsen, Morten Hannibal and Madulid, Nash and Malmros, Albert and Manfra, Michael and Mantri, Devashish and Markussen, Signe Brynold and Martinez, Esteban and Mattila, Marco and McNeil, Robert and Mei, Antonio B. and Mishmash, Ryan V. and Mohandas, Gopakumar and Mollgaard, Christian and Morgan, Trevor and Moussa, George and Nayak, Chetan and Nielsen, Jens Hedegaard and Nielsen, Jens Munk and Nielsen, William Hvidtfelt Padkar and Nijholt, Bas and Nystrom, Mike and O'Farrell, Eoin and Ohki, Thomas and Otani, Keita and Paquelet W{\"u}tz, Brian and Pauka, Sebastian and Petersson, Karl and Petit, Luca and Pikulin, Dima and Prawiroatmodjo, Guen and Preiss, Frank and Puchol Morejon, Eduardo and Rajpalke, Mohana and Ranta, Craig and Rasmussen, Katrine and Razmadze, David and Reentila, Outi and Reilly, David J. and Ren, Yuan and Reneris, Ken and Rouse, Richard and Sadovskyy, Ivan and Sainiemi, Lauri and Sanlorenzo, Irene and Schmidgall, Emma and Sfiligoj, Cristina and Shah, Mustafeez Bashir and Simoes, Kevin and Singh, Shilpi and Sinha, Sarat and Soerensen, Thomas and Sohr, Patrick and Stankevic, Tomas and Stek, Lieuwe and Stuppard, Eric and Suominen, Henri and Suter, Judith and Teicher, Sam and Thiyagarajah, Nivetha and Tholapi, Raj and Thomas, Mason and Toomey, Emily and Tracy, Josh and Turley, Michelle and Upadhyay, Shivendra and Urban, Ivan and Van Hoogdalem, Kevin and Van Woerkom, David J. and Viazmitinov, Dmitrii V. and Vogel, Dominik and Watson, John and Webster, Alex and Weston, Joseph and Winkler, Georg W. and Xu, Di and Yang, Chung Kai and Yucelen, Emrah and Zeisel, Roland and Zheng, Guoji and Zilke, Justin},
  year = 2025,
  month = feb,
  journal = {Nature},
  volume = {638},
  number = {8051},
  pages = {651--655},
  publisher = {Nature Publishing Group},
  issn = {1476-4687},
  url = {https://www.nature.com/articles/s41586-024-08445-2},
  urldate = {2025-02-20},
  copyright = {2025 The Author(s)},
  langid = {english}
}

@article{beenakker2013search,
  title = {Search for {{Majorana Fermions}} in {{Superconductors}}},
  author = {Beenakker, C.W.J.},
  year = 2013,
  month = mar,
  journal = {Annu. Rev. Condens. Matter Phys.},
  volume = {4},
  number = {1},
  pages = {113--136},
  issn = {1947-5454},
  url = {https://www.annualreviews.org/doi/10.1146/annurev-conmatphys-030212-184337},
  urldate = {2018-10-22},
  day = 8
}

@article{dassarma2023search,
  title = {In Search of {{Majorana}}},
  author = {Das Sarma, Sankar},
  year = 2023,
  month = feb,
  journal = {Nat. Phys.},
  volume = {19},
  number = {2},
  pages = {165--170},
  publisher = {Nature Publishing Group},
  issn = {1745-2481},
  url = {https://www.nature.com/articles/s41567-022-01900-9},
  urldate = {2023-02-17},
  copyright = {2023 Springer Nature Limited},
  langid = {english},
  day = 6
}

@article{dassarma2023spectral,
  title = {Spectral Properties, Topological Patches, and Effective Phase Diagrams of Finite Disordered {{Majorana}} Nanowires},
  author = {Das Sarma, Sankar and Sau, Jay D. and Stanescu, Tudor D.},
  year = 2023,
  month = aug,
  journal = {Phys. Rev. B},
  volume = {108},
  number = {8},
  pages = {085416},
  publisher = {American Physical Society},
  url = {https://link.aps.org/doi/10.1103/PhysRevB.108.085416},
  urldate = {2023-09-14},
  day = 10
}

@article{groth2014kwant,
  title = {Kwant: A Software Package for Quantum Transport},
  author = {Groth, Christoph W and Wimmer, Michael and Akhmerov, Anton R and Waintal, Xavier},
  year = 2014,
  journal = {New Journal of Physics},
  volume = {16},
  number = {6},
  pages = {063065},
  url = {http://iopscience.iop.org/article/10.1088/1367-2630/16/6/063065/meta}
}

@article{kouwenhoven2025perspective,
  title = {Perspective on {{Majorana}} Bound-States in Hybrid Superconductor-Semiconductor Nanowires},
  author = {Kouwenhoven, Leo},
  year = 2025,
  month = jan,
  journal = {Mod. Phys. Lett. B},
  volume = {39},
  number = {03},
  pages = {2540002},
  issn = {0217-9849, 1793-6640},
  url = {https://www.worldscientific.com/doi/10.1142/S0217984925400020},
  urldate = {2025-04-13},
  langid = {english},
  day = 30
}

@article{blonder1982transition,
  title = {Transition from Metallic to Tunneling Regimes in Superconducting Microconstrictions: {{Excess}} Current, Charge Imbalance, and Supercurrent Conversion},
  shorttitle = {Transition from Metallic to Tunneling Regimes in Superconducting Microconstrictions},
  author = {Blonder, G. E. and Tinkham, M. and Klapwijk, T. M.},
  year = 1982,
  month = apr,
  journal = {Phys. Rev. B},
  volume = {25},
  number = {7},
  pages = {4515--4532},
  url = {https://link.aps.org/doi/10.1103/PhysRevB.25.4515},
  urldate = {2018-10-22},
  day = 1
}

@article{lutchyn2010majorana,
  title = {Majorana {{Fermions}} and a {{Topological Phase Transition}} in {{Semiconductor-Superconductor Heterostructures}}},
  author = {Lutchyn, Roman M. and Sau, Jay D. and Das Sarma, S.},
  year = 2010,
  month = aug,
  journal = {Phys. Rev. Lett.},
  volume = {105},
  number = {7},
  pages = {077001},
  url = {https://link.aps.org/doi/10.1103/PhysRevLett.105.077001},
  urldate = {2018-10-22},
  day = 13
}

@article{lutchyn2018majorana,
  title = {Majorana Zero Modes in Superconductor--Semiconductor Heterostructures},
  author = {Lutchyn, R. M. and Bakkers, E. P. A. M. and Kouwenhoven, L. P. and Krogstrup, P. and Marcus, C. M. and Oreg, Y.},
  year = 2018,
  month = may,
  journal = {Nature Reviews Materials},
  volume = {3},
  number = {5},
  pages = {52--68},
  issn = {2058-8437},
  url = {https://www.nature.com/articles/s41578-018-0003-1},
  urldate = {2018-09-30},
  copyright = {2018 Macmillan Publishers Ltd., part of Springer Nature},
  langid = {english}
}

@article{microsoftquantum2023inasal,
  title = {{{InAs-Al}} Hybrid Devices Passing the Topological Gap Protocol},
  author = {{Microsoft Quantum} and Aghaee, Morteza and Akkala, Arun and Alam, Zulfi and Ali, Rizwan and Alcaraz Ramirez, Alejandro and Andrzejczuk, Mariusz and Antipov, Andrey E. and Aseev, Pavel and Astafev, Mikhail and Bauer, Bela and Becker, Jonathan and Boddapati, Srini and Boekhout, Frenk and Bommer, Jouri and Bosma, Tom and Bourdet, Leo and Boutin, Samuel and Caroff, Philippe and Casparis, Lucas and Cassidy, Maja and Chatoor, Sohail and Christensen, Anna Wulf and Clay, Noah and Cole, William S. and Corsetti, Fabiano and Cui, Ajuan and Dalampiras, Paschalis and Dokania, Anand and {de Lange}, Gijs and {de Moor}, Michiel and Estrada Salda{\~n}a, Juan Carlos and Fallahi, Saeed and Fathabad, Zahra Heidarnia and Gamble, John and Gardner, Geoff and Govender, Deshan and Griggio, Flavio and Grigoryan, Ruben and Gronin, Sergei and Gukelberger, Jan and Hansen, Esben Bork and Heedt, Sebastian and Herranz Zamorano, Jes{\'u}s and Ho, Samantha and Holgaard, Ulrik Laurens and Ingerslev, Henrik and Johansson, Linda and Jones, Jeffrey and Kallaher, Ray and Karimi, Farhad and Karzig, Torsten and King, Cameron and Kloster, Maren Elisabeth and Knapp, Christina and Kocon, Dariusz and Koski, Jonne and Kostamo, Pasi and Krogstrup, Peter and Kumar, Mahesh and Laeven, Tom and Larsen, Thorvald and Li, Kongyi and Lindemann, Tyler and Love, Julie and Lutchyn, Roman and Madsen, Morten Hannibal and Manfra, Michael and Markussen, Signe and Martinez, Esteban and McNeil, Robert and Memisevic, Elvedin and Morgan, Trevor and Mullally, Andrew and Nayak, Chetan and Nielsen, Jens and Nielsen, William Hvidtfelt Padk{\ae}r and Nijholt, Bas and Nurmohamed, Anne and O'Farrell, Eoin and Otani, Keita and Pauka, Sebastian and Petersson, Karl and Petit, Luca and Pikulin, Dmitry I. and Preiss, Frank and {Quintero-Perez}, Marina and Rajpalke, Mohana and Rasmussen, Katrine and Razmadze, Davydas and Reentila, Outi and Reilly, David and Rouse, Richard and Sadovskyy, Ivan and Sainiemi, Lauri and Schreppler, Sydney and Sidorkin, Vadim and Singh, Amrita and Singh, Shilpi and Sinha, Sarat and Sohr, Patrick and Stankevi{\v c}, Toma{\v s} and Stek, Lieuwe and Suominen, Henri and Suter, Judith and Svidenko, Vicky and Teicher, Sam and Temuerhan, Mine and Thiyagarajah, Nivetha and Tholapi, Raj and Thomas, Mason and Toomey, Emily and Upadhyay, Shivendra and Urban, Ivan and Vaitiek{\.e}nas, Saulius and Van Hoogdalem, Kevin and Van Woerkom, David and Viazmitinov, Dmitrii V. and Vogel, Dominik and Waddy, Steven and Watson, John and Weston, Joseph and Winkler, Georg W. and Yang, Chung Kai and Yau, Sean and Yi, Daniel and Yucelen, Emrah and Webster, Alex and Zeisel, Roland and Zhao, Ruichen},
  year = 2023,
  month = jun,
  journal = {Phys. Rev. B},
  volume = {107},
  number = {24},
  pages = {245423},
  publisher = {American Physical Society},
  url = {https://link.aps.org/doi/10.1103/PhysRevB.107.245423},
  urldate = {2023-06-26},
  day = 21
}

@article{nayak2008nonabelian,
  title = {Non-{{Abelian}} Anyons and Topological Quantum Computation},
  author = {Nayak, Chetan and Simon, Steven H. and Stern, Ady and Freedman, Michael and Das Sarma, Sankar},
  year = 2008,
  month = sep,
  journal = {Reviews of Modern Physics},
  volume = {80},
  number = {3},
  pages = {1083--1159},
  issn = {0034-6861, 1539-0756},
  url = {https://link.aps.org/doi/10.1103/RevModPhys.80.1083},
  urldate = {2018-09-19},
  langid = {english},
  day = 12
}

@article{oreg2010helical,
  title = {Helical {{Liquids}} and {{Majorana Bound States}} in {{Quantum Wires}}},
  author = {Oreg, Yuval and Refael, Gil and {von Oppen}, Felix},
  year = 2010,
  month = oct,
  journal = {Phys. Rev. Lett.},
  volume = {105},
  number = {17},
  pages = {177002},
  url = {http://link.aps.org/doi/10.1103/PhysRevLett.105.177002}
}

@article{pan2020physical,
  title = {Physical Mechanisms for Zero-Bias Conductance Peaks in {{Majorana}} Nanowires},
  author = {Pan, Haining and Das Sarma, S.},
  year = 2020,
  month = mar,
  journal = {Phys. Rev. Research},
  volume = {2},
  number = {1},
  pages = {013377},
  publisher = {American Physical Society},
  url = {https://link.aps.org/doi/10.1103/PhysRevResearch.2.013377},
  urldate = {2020-05-12},
  copyright = {All rights reserved},
  day = 30,
  tag = {disordered systems,quantum transport and s-matrix theory,semiconductor-superconductor hybrid nanowire,topological phases of matter,topological quantum computation}
}

@article{pan2021threeterminal,
  ids = {pan2020threeterminal},
  title = {Three-Terminal Nonlocal Conductance in {{Majorana}} Nanowires: {{Distinguishing}} Topological and Trivial in Realistic Systems with Disorder and Inhomogeneous Potential},
  shorttitle = {Three-Terminal Nonlocal Conductance in {{Majorana}} Nanowires},
  author = {Pan, Haining and Sau, Jay D. and Das Sarma, S.},
  year = 2021,
  month = jan,
  journal = {Phys. Rev. B},
  volume = {103},
  number = {1},
  pages = {014513},
  publisher = {American Physical Society},
  url = {https://link.aps.org/doi/10.1103/PhysRevB.103.014513},
  urldate = {2021-01-25},
  archiveprefix = {arXiv},
  copyright = {All rights reserved},
  day = 20,
  tag = {disordered systems,quantum transport and s-matrix theory,semiconductor-superconductor hybrid nanowire,topological phases of matter}
}

@article{pan2024disordered,
  title = {Disordered {{Majorana}} Nanowires: {{Studying}} Disorder without Any Disorder},
  shorttitle = {Disordered {{Majorana}} Nanowires},
  author = {Pan, Haining and Das Sarma, Sankar},
  year = 2024,
  month = aug,
  journal = {Phys. Rev. B},
  volume = {110},
  number = {7},
  pages = {075401},
  issn = {2469-9950, 2469-9969},
  url = {https://link.aps.org/doi/10.1103/PhysRevB.110.075401},
  urldate = {2025-04-13},
  langid = {english},
  day = 2
}

@article{rosdahl2018andreev,
  title = {Andreev Rectifier: {{A}} Nonlocal Conductance Signature of Topological Phase Transitions},
  shorttitle = {Andreev Rectifier},
  author = {Rosdahl, T. {\"O}. and Vuik, A. and Kjaergaard, M. and Akhmerov, A. R.},
  year = 2018,
  month = jan,
  journal = {Phys. Rev. B},
  volume = {97},
  number = {4},
  pages = {045421},
  url = {https://link.aps.org/doi/10.1103/PhysRevB.97.045421},
  urldate = {2019-08-19},
  day = 22
}

@article{sarma2015majorana,
  title = {Majorana Zero Modes and Topological Quantum Computation},
  author = {Sarma, Sankar Das and Freedman, Michael and Nayak, Chetan},
  year = 2015,
  month = oct,
  journal = {npj Quantum Information},
  volume = {1},
  pages = {15001},
  issn = {2056-6387},
  url = {https://www.nature.com/articles/npjqi20151},
  urldate = {2018-10-22},
  copyright = {2015 Nature Publishing Group},
  langid = {english},
  day = 27
}

@article{sau2010generic,
  title = {Generic {{New Platform}} for {{Topological Quantum Computation Using Semiconductor Heterostructures}}},
  author = {Sau, Jay D. and Lutchyn, Roman M. and Tewari, Sumanta and Das Sarma, S.},
  year = 2010,
  month = jan,
  journal = {Phys. Rev. Lett.},
  volume = {104},
  number = {4},
  pages = {040502},
  url = {https://link.aps.org/doi/10.1103/PhysRevLett.104.040502},
  urldate = {2018-10-22},
  day = 27
}

@article{sau2010robustness,
  title = {Robustness of {{Majorana}} Fermions in Proximity-Induced Superconductors},
  author = {Sau, Jay D. and Lutchyn, Roman M. and Tewari, Sumanta and Das Sarma, S.},
  year = 2010,
  month = sep,
  journal = {Phys. Rev. B},
  volume = {82},
  number = {9},
  pages = {094522},
  url = {https://link.aps.org/doi/10.1103/PhysRevB.82.094522},
  urldate = {2018-10-22},
  day = 27
}

@article{taylor2024machine,
  title = {Machine {{Learning}} the {{Disorder Landscape}} of {{Majorana Nanowires}}},
  author = {Taylor, Jacob R. and Sau, Jay D. and Das Sarma, Sankar},
  year = 2024,
  month = may,
  journal = {Phys. Rev. Lett.},
  volume = {132},
  number = {20},
  pages = {206602},
  publisher = {American Physical Society},
  url = {https://link.aps.org/doi/10.1103/PhysRevLett.132.206602},
  urldate = {2025-04-13},
  day = 16
}

@article{taylor2025vision,
  title = {Vision Transformer Based Deep Learning of Topological Indicators in {{Majorana}} Nanowires},
  author = {Taylor, Jacob R. and Das Sarma, Sankar},
  year = 2025,
  month = mar,
  journal = {Phys. Rev. B},
  volume = {111},
  number = {10},
  pages = {104208},
  publisher = {American Physical Society},
  url = {https://link.aps.org/doi/10.1103/PhysRevB.111.104208},
  urldate = {2025-04-13},
  day = 28
}

@article{woods2021chargeimpurity,
  ids = {woods2021charge},
  title = {Charge-{{Impurity Effects}} in {{Hybrid Majorana Nanowires}}},
  author = {Woods, Benjamin D. and Das Sarma, Sankar and Stanescu, Tudor D.},
  year = 2021,
  month = nov,
  journal = {Phys. Rev. Applied},
  volume = {16},
  number = {5},
  pages = {054053},
  publisher = {American Physical Society},
  url = {https://link.aps.org/doi/10.1103/PhysRevApplied.16.054053},
  urldate = {2022-04-25},
  archiveprefix = {arXiv},
  day = 30
}

@inproceedings{arnab2021vivit,
  title={Vivit: A video vision transformer},
  author={Arnab, Anurag and Dehghani, Mostafa and Heigold, Georg and Sun, Chen and Lu{\v{c}}i{\'c}, Mario and Schmid, Cordelia},
  booktitle={Proceedings of the IEEE/CVF international conference on computer vision},
  pages={6836--6846},
  year={2021}
}

@article{dassarma2016how,
  title = {How to Infer Non-{{Abelian}} Statistics and Topological Visibility from Tunneling Conductance Properties of Realistic {{Majorana}} Nanowires},
  author = {Das Sarma, S. and Nag, Amit and Sau, Jay D.},
  year = 2016,
  month = jul,
  journal = {Phys. Rev. B},
  volume = {94},
  number = {3},
  pages = {035143},
  url = {https://link.aps.org/doi/10.1103/PhysRevB.94.035143},
  urldate = {2018-10-22},
  day = 20
}

@inproceedings{hatamizadeh2021swin,
  title={Swin unetr: Swin transformers for semantic segmentation of brain tumors in mri images},
  author={Hatamizadeh, Ali and Nath, Vishwesh and Tang, Yucheng and Yang, Dong and Roth, Holger R and Xu, Daguang},
  booktitle={International MICCAI brainlesion workshop},
  pages={272--284},
  year={2021},
  organization={Springer}
}

@inproceedings{akiba2019optuna,
  title={{O}ptuna: A Next-generation Hyperparameter Optimization Framework},
  author={Akiba, Takuya and Sano, Shotaro and Yanase, Toshihiko and Ohta, Takeru and Koyama, Masanori},
  booktitle={Proceedings of the 25th ACM SIGKDD International Conference on Knowledge Discovery \& Data Mining},
  pages={2623--2631},
  year={2019},
  doi={10.1145/3292500.3330701}
}

@article{taylor2025unreasonable,
  title={Unreasonable effectiveness of unsupervised learning in identifying Majorana topology},
  author={Taylor, Jacob and Pan, Haining and Sarma, Sankar Das},
  journal={arXiv preprint arXiv:2512.13825},
  year={2025}
}

@article{cheng2024machine,
  title={Machine learning detection of Majorana zero modes from zero-bias peak measurements},
  author={Cheng, Mouyang and Okabe, Ryotaro and Chotrattanapituk, Abhijatmedhi and Li, Mingda},
  journal={Matter},
  volume={7},
  number={7},
  pages={2507--2520},
  year={2024},
  publisher={Elsevier}
}

@article{krawczyk2026ai,
  title={AI-enhanced tuning of quantum dot Hamiltonians toward Majorana modes},
  author={Krawczyk, Mateusz and Paw{\l}owski, Jaros{\l}aw},
  journal={arXiv preprint arXiv:2601.02149},
  year={2026}
}

@article{taylor2025mitigating,
  title={Mitigating disorder and optimizing topological indicators with vision-transformer-based neural networks in Majorana nanowires},
  author={Taylor, Jacob R and Das Sarma, Sankar},
  journal={Physical Review B},
  volume={112},
  number={4},
  pages={L041110},
  year={2025},
  publisher={APS}
}

@article{pawlowski2026learning,
  title={Learning Hamiltonians for solid-state quantum simulators},
  author={Paw{\l}owski, Jaros{\l}aw and Krawczyk, Mateusz},
  journal={arXiv preprint arXiv:2603.02889},
  year={2026}
}

@inproceedings{liu2021swin,
  title={Swin transformer: Hierarchical vision transformer using shifted windows},
  author={Liu, Ze and Lin, Yutong and Cao, Yue and Hu, Han and Wei, Yixuan and Zhang, Zheng and Lin, Stephen and Guo, Baining},
  booktitle={Proceedings of the IEEE/CVF international conference on computer vision},
  pages={10012--10022},
  year={2021}
}

@inproceedings{ronneberger2015u,
  title={U-net: Convolutional networks for biomedical image segmentation},
  author={Ronneberger, Olaf and Fischer, Philipp and Brox, Thomas},
  booktitle={International Conference on Medical image computing and computer-assisted intervention},
  pages={234--241},
  year={2015},
  organization={Springer}
}

@article{isensee2021nnu,
  title={nnU-Net: a self-configuring method for deep learning-based biomedical image segmentation},
  author={Isensee, Fabian and Jaeger, Paul F and Kohl, Simon AA and Petersen, Jens and Maier-Hein, Klaus H},
  journal={Nature methods},
  volume={18},
  number={2},
  pages={203--211},
  year={2021},
  publisher={Nature Publishing Group US New York}
}

@article{saharia2022image,
  title={Image super-resolution via iterative refinement},
  author={Saharia, Chitwan and Ho, Jonathan and Chan, William and Salimans, Tim and Fleet, David J and Norouzi, Mohammad},
  journal={IEEE transactions on pattern analysis and machine intelligence},
  volume={45},
  number={4},
  pages={4713--4726},
  year={2022},
  publisher={IEEE}
}

@inproceedings{rombach2022high,
  title={High-resolution image synthesis with latent diffusion models},
  author={Rombach, Robin and Blattmann, Andreas and Lorenz, Dominik and Esser, Patrick and Ommer, Bj{\"o}rn},
  booktitle={Proceedings of the IEEE/CVF conference on computer vision and pattern recognition},
  pages={10684--10695},
  year={2022}
}

@misc{aghaee2025distinct,
  author = {Aghaee, Morteza and Alam, Zulfi and Andersen, Rikke and Andrzejczuk, Mariusz and Antipov, Andrey and Astafev, Mikhail and Avilovas, Lukas and Azizimanesh, Ahmad and Banek, Eric and Bauer, Bela and Becker, Jonathan and Bhaskar, Umesh Kumar and Boa, Andrea G. and Boddapati, Srini and Bohac, Nichlaus and Bommer, Jouri D. S. and Borovsky, Jan and Bourdet, L{\'e}o and Boutin, Samuel and Casparis, Lucas and Chakravarthi, Srivatsa and Chalabi, Hamidreza and Chapman, Benjamin J. and Chatzaras, Nikolaos and Chien, Tzu-Chiao and Cho, Jason and Codd, Patrick and Cole, William and Cooper, Paul W. and Corsetti, Fabiano and Cui, Ajuan and Dandachi, Tareq El and Dinesen, Celine and Ekefj{\"a}rd, Andreas and Fallahi, Saeed and Galletti, Luca and Gardner, Geoffrey C. and Gonzalez, Gonzalo Leon and Govender, Deshan and Griggio, Flavio and Grigoryan, Ruben and Grijalva, Sebastian and Gronin, Sergei and Gukelberger, Jan and Hamdast, Marzie and Hamida, A. Ben and Hansen, Esben Bork and Hansen, Caroline Tynell and Heedt, Sebastian and Ho, Samantha and Holgaard, Laurens and van Hoogdalem, Kevin and Hornibrook, John and Ingerslev, Henrik and Ivancevic, Lovro and Jamo, Sherwan and Jantos, Max and Jensen, Thomas and Jhoja, Jaspreet Singh and Jones, Jeffrey C. and Joshi, Vidul and Kalashnikov, Konstantin V. and Kallaher, Ray and Kalra, Rachpon and Karimi, Farhad and Karzig, Torsten and Kimes, Seth and King, Evelyn and Kloster, Maren Elisabeth and Knapp, Christina and Koski, Jonne V. and Kostamo, Pasi and Laeven, Tom and Lai, Jeffrey and de Lange, Gijs and Larsen, Thorvald W. and Lee, Kyunghoon and Li, Kongyi and Li, Guangze and Liang, Shuang and Lindemann, Tyler and Looij, Matthew and Lucas, Marijn and Lutchyn, Roman and Madsen, Morten Hannibal and Madulid, Nasiari and Manfra, Michael J. and Manjunath, Laveena and Markussen, Signe and Martinez, Esteban and Mattila, Marco and Mattinson, J. R. and McNeil, R. P. G. and Millan, Alba P{\'e}rez and Mishmash, Ryan V. and Mittal, Sarang and M{\o}llgaard, Christian and de Moor, M. W. A. and Morejon, Eduardo Puchol and Morgan, Trevor and Moussa, George and Nabar, B. P. and Narla, Anirudh and Nayak, Chetan and Nielsen, Jens Hedegaard and Nielsen, William Hvidtfelt Padk{\ae}r and Nolet, Fr{\'e}d{\'e}ric and Nystrom, Michael J. and O'Farrell, Eoin and Ohki, Thomas A. and Otani, Keita and Papon, Camille and Petersson, Karl D. and Petit, Luca and Pikulin, Dima and Rajpalke, Mohana and Ramirez, Alejandro Alcaraz and Razmadze, David and Ren, Yuan and Sadovskyy, Ivan and Sainiemi, Lauri and Salda{\~n}a, Juan Carlos Estrada and Sanlorenzo, Irene and dos Santos, Tatiane Pereira and Schaal, Simon and Schack, John and Schmidgall, Emma R. and Sfetsou, Christina and Sfiligoj, Cristina and Sinha, Sarat and Sohr, Patrick and S{\o}rensen, Thomas L. and Spiegelhauer, Kasper and Stankevi{\'c}, Toma{\v s} and Stek, Lieuwe J. and {Str{\o}m-Hansen}, Patrick and Suominen, Henri J. and Suter, Judith and Teicher, Samuel M. L. and Tholapi, Raj and Thomas, Mason and Tom, D. W. and Toomey, Emily and Tracy, Joshua and Turley, Michelle and Turner, Matthew D. and Upadhyay, Shivendra and Urban, Ivan and Viazmitinov, Dmitrii V. and Viazmitinova, Anna Wulff and Viegas, Beatriz and Vogel, Dominik J. and Watson, John and Webster, Alex and Weston, Joseph and Williamson, Timothy and Winkler, Georg W. and van Woerkom, David J. and Wuetz, Brian Paquelet and Yang, Chung-Kai and {Shang-Jyun} and Yu and Yucelen, Emrah and Zamorano, Jes{\'u}s Herranz and Zeisel, Roland and Zheng, Guoji and Zimmerman, A. M.},
  year = 2025,
  month = sep,
  number = {arXiv:2507.08795},
  primaryclass = {cond-mat},
  publisher = {arXiv},
  url = {http://arxiv.org/abs/2507.08795},
  urldate = {2025-12-09},
  archiveprefix = {arXiv},
  title = {Distinct {{Lifetimes}} for {$X$} and {$Z$} {{Loop Measurements}} in a {{Majorana Tetron Device}}},
  day = 4
}

@article{kitaev2003faulttolerant,
  title = {Fault-Tolerant Quantum Computation by Anyons},
  author = {Kitaev, A. {\relax Yu}.},
  year = 2003,
  month = jan,
  journal = {Annals of Physics},
  volume = {303},
  number = {1},
  pages = {2--30},
  issn = {0003-4916},
  url = {http://www.sciencedirect.com/science/article/pii/S0003491602000180},
  urldate = {2018-09-30},
  day = 1
}

@article{pan2020generic,
  title = {Generic Quantized Zero-Bias Conductance Peaks in Superconductor-Semiconductor Hybrid Structures},
  author = {Pan, Haining and Cole, William S. and Sau, Jay D. and Das Sarma, S.},
  year = 2020,
  month = jan,
  journal = {Phys. Rev. B},
  volume = {101},
  number = {2},
  pages = {024506},
  publisher = {American Physical Society},
  url = {https://link.aps.org/doi/10.1103/PhysRevB.101.024506},
  urldate = {2020-09-07},
  copyright = {All rights reserved},
  day = 8,
  tag = {disordered systems,quantum transport and s-matrix theory,random matrix theory,semiconductor-superconductor hybrid nanowire,topological phases of matter,topological quantum computation}
}

@article{pan2026majorana,
  title = {Majorana Zero Modes in Semiconductor-Superconductor Hybrid Structures: {{Defining}} Topology in Short and Disordered Nanowires through {{Majorana}} Splitting},
  shorttitle = {Majorana Zero Modes in Semiconductor-Superconductor Hybrid Structures},
  author = {Pan, Haining and Das Sarma, Sankar},
  year = 2026,
  month = apr,
  journal = {Phys. Rev. B},
  volume = {113},
  number = {16},
  pages = {165420},
  publisher = {American Physical Society},
  url = {https://link.aps.org/doi/10.1103/2v41-yvs1},
  urldate = {2026-05-22},
  day = 20,
  tag = {disordered systems,quantum transport and s-matrix theory,semiconductor-superconductor hybrid nanowire,topological phases of matter,topological quantum computation}
}

@misc{pikulin2021protocol,
  title = {Protocol to Identify a Topological Superconducting Phase in a Three-Terminal Device},
  author = {Pikulin, Dmitry I. and van Heck, Bernard and Karzig, Torsten and Martinez, Esteban A. and Nijholt, Bas and Laeven, Tom and Winkler, Georg W. and Watson, John D. and Heedt, Sebastian and Temurhan, Mine and Svidenko, Vicky and Lutchyn, Roman M. and Thomas, Mason and de Lange, Gijs and Casparis, Lucas and Nayak, Chetan},
  year = 2021,
  month = mar,
  number = {arXiv:2103.12217},
  primaryclass = {cond-mat},
  publisher = {arXiv},
  url = {http://arxiv.org/abs/2103.12217},
  urldate = {2025-04-13},
  archiveprefix = {arXiv},
  day = 22
}

@article{setiawan2017electron,
  title = {Electron Temperature and Tunnel Coupling Dependence of Zero-Bias and Almost-Zero-Bias Conductance Peaks in {{Majorana}} Nanowires},
  author = {Setiawan, F. and Liu, Chun-Xiao and Sau, Jay D. and Das Sarma, S.},
  year = 2017,
  month = nov,
  journal = {Phys. Rev. B},
  volume = {96},
  number = {18},
  pages = {184520},
  url = {https://link.aps.org/doi/10.1103/PhysRevB.96.184520},
  urldate = {2019-09-13},
  day = 22
}
\appendix
\clearpage

\section{Diffusion Neural Network Architectures}
\label{Sec:DiffNN}
\begin{figure}[h]
    \centering
    \includegraphics[width=1\linewidth]{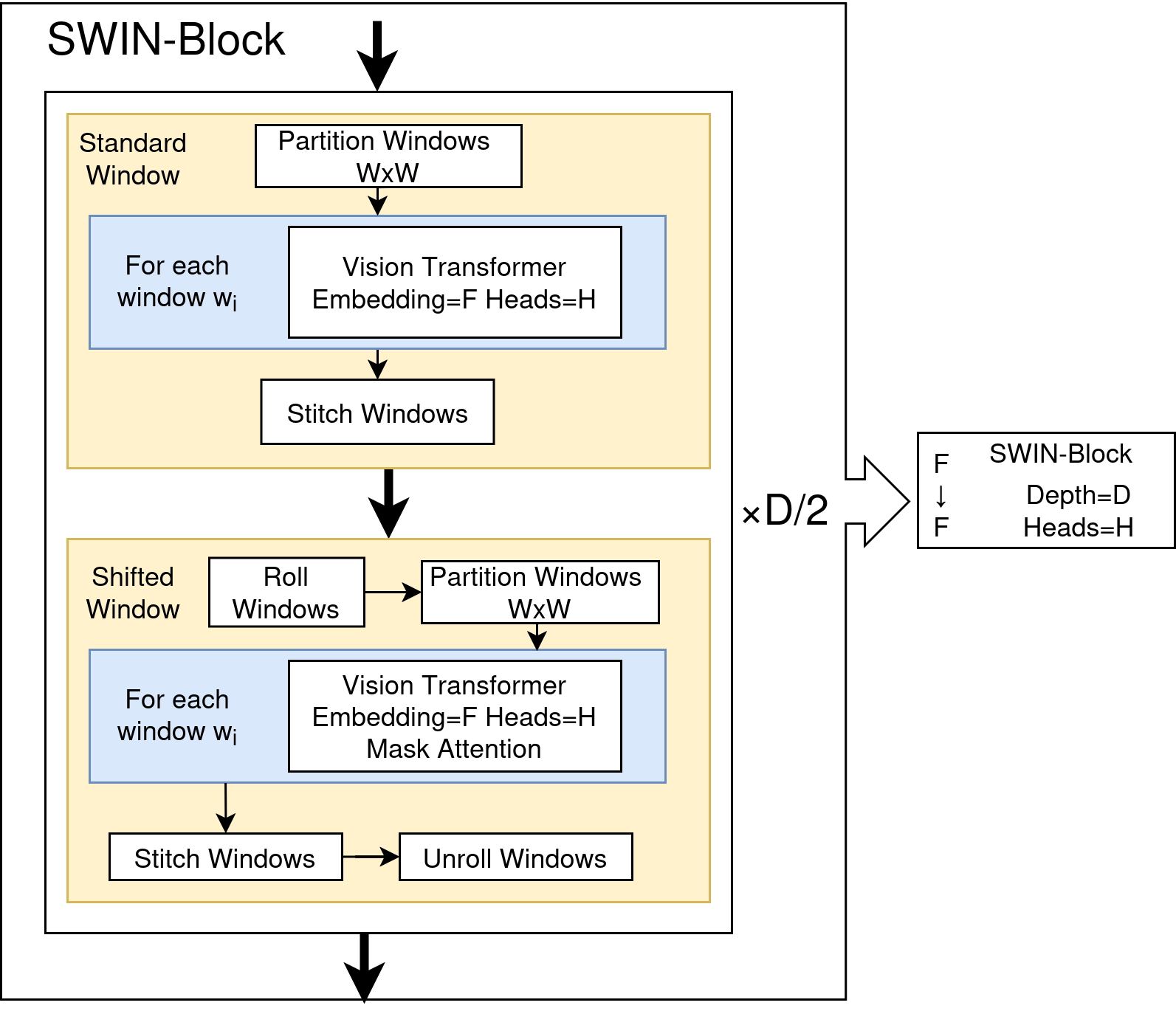}
    \caption{Schematic of the SWIN-Block used in the Conductance U-Net. The block alternates standard and shifted-window attention: features are partitioned into $W \times W$ windows, processed by a window-wise transformer with $H$ heads, stitched back together, and repeated for depth $D$. The shifted-window stage enables information exchange between neighboring windows.
}\label{fig:SWINBlock}
\end{figure}
\begin{figure}[h]
    \centering
    \includegraphics[width=1.0\linewidth]{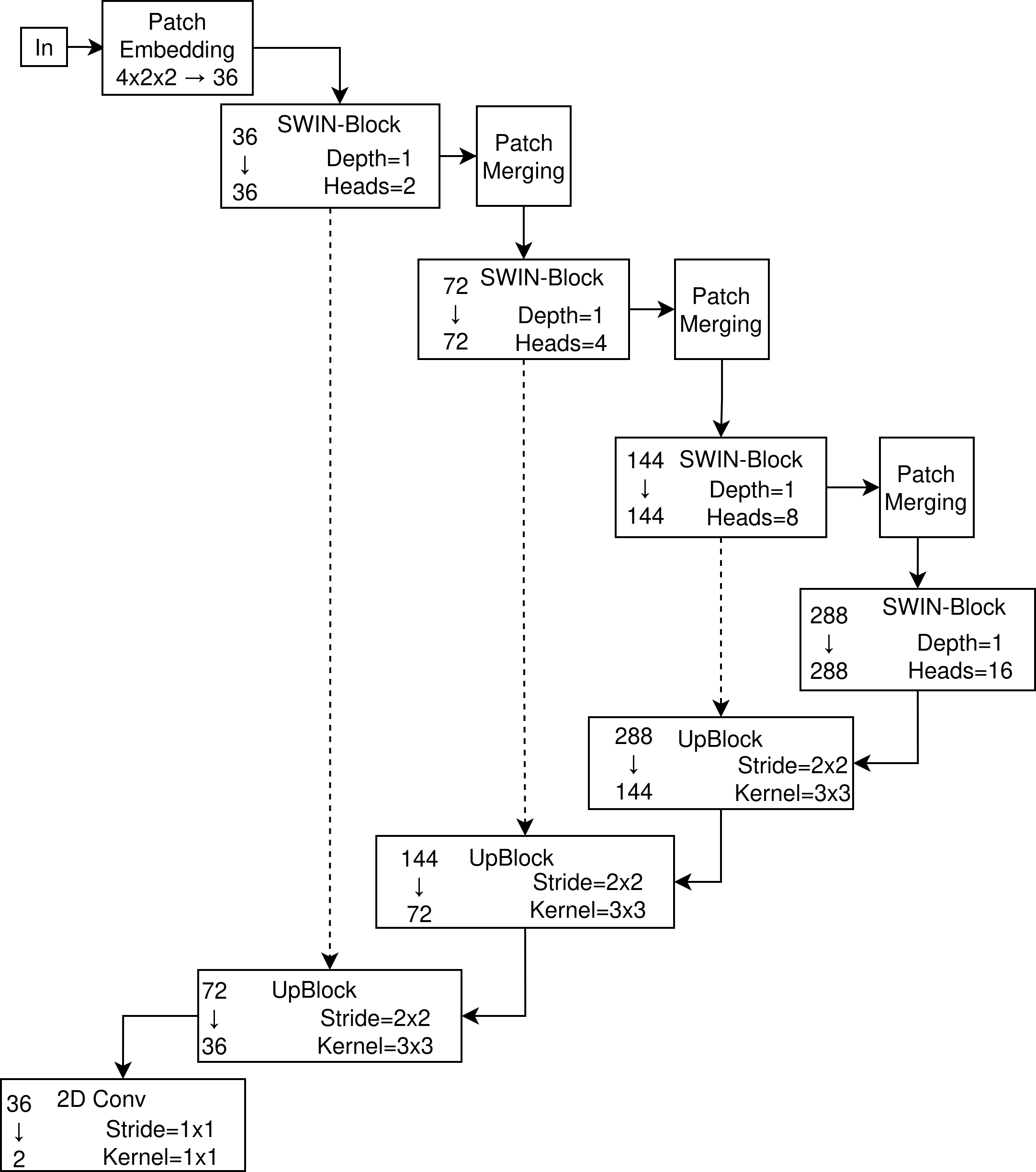}
    \caption{Diagram of Conductance SWIN-UNETR Neural network used to map away the temperature and experimental noise-based errors.}\label{fig:Unet}
\end{figure}

 
We make use of two dramatically different neural network architectures depending on whether we are predicting conductance or TV. In the case of conductance, we are seeking to resolve from other conductance data in a manner similar to denoising diffusion to reverse "noise" or temperature effects that obscure the relevant Majorana features. In this case, we make use of a neural network that is trained on a large number of physical devices and thus knows what physical features to expect when it denoises. The input training data is in practice allowing the conductance neural network to do a physics-informed diffusion process. The conductance neural network consists of a SWIN-UNETR originally proposed in \cite{hatamizadeh2021swin} for the purpose of medical imaging. The network takes the form of a very standard shape for a UNET similar to those outlined in \cite{ronneberger2015u} and so forth, where an encoding half decreases the size of the spatial dimensions while increasing the size of the channel dimension. Following the encoder layers, then an Up-block transverse convolutional process is performed to decode. At each encoder block step, there is a skip to concatenate the partially encoded information to the partially decoded at the equivalent layer, which is then used in the next decoding step. The full neural network can be seen in Fig. \ref{fig:Unet}, where we use 2x2 patch embeddings and 4 layers of SWIN-Blocks. The main innovation with the SWIN-UNETR is to make use of the SWIN-Block~\cite{liu2021swin} transformer-based process for the encoding phase. 
\begin{figure}[h]
    \centering
    \includegraphics[width=0.9\linewidth]{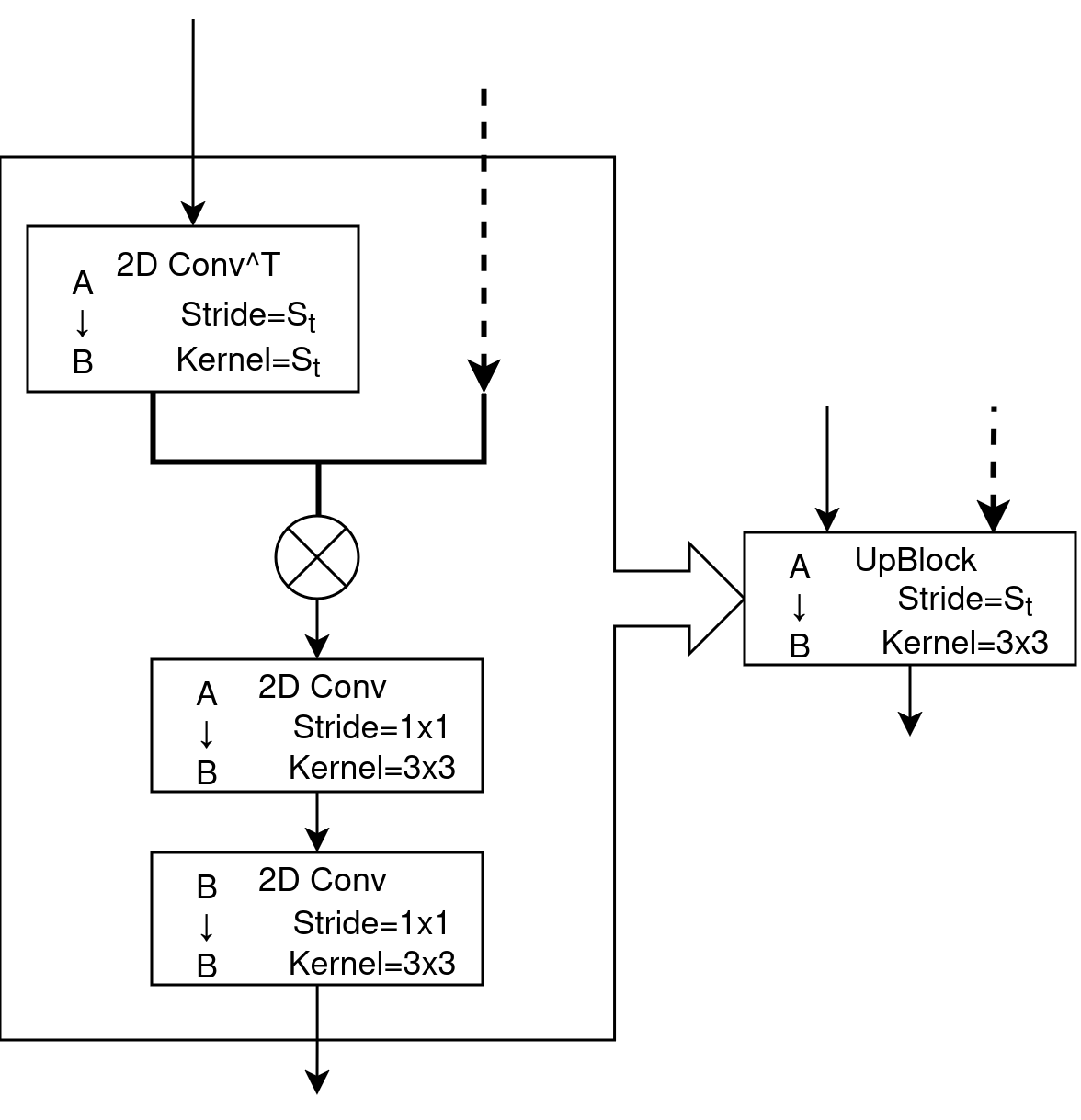}
    \caption{Schematic of the upscaling block used in the Conductance U-Net. The lower-resolution decoder feature map is first upsampled using a 2D transposed convolution with stride $S_t$ and kernel size $S_t$, mapping the channel dimension from $A$ to $B$. The upsampled feature map is then combined with the corresponding encoder skip connection and passed through two $3 \times 3$ convolutional layers with unit stride. This UpBlock increases the spatial resolution while incorporating higher-resolution features from the contracting path.}\label{fig:Upblock}
\end{figure} 

The SWIN-Block works by converting the "image" into many 7x7 windows and then applying a vision transformer (the same vision transformer) independently to each window. The windows are then stitched back together because the same process is applied again, but using a rolling process to shift the data and thus what data is included in each window, though a masking process is applied to prevent periodic long-range attention. Depending on its depth $D$, the SWIN-Block alternates between these two operations $D/2$ times, where $D=1$ implies only the non-shifted layer is applied. After every SWIN-Block is a patch merging whereby a linear operation on a 2x2 window in the spatial dimension is converted to a $2\times \text{Channels}$ in the channel dimension. For a diagram showing the SWIN-Block, see Fig. \ref{fig:SWINBlock}. The decoder UpBlocks take in both the partially decoded and partially encoded images, and after applying a transverse convolutional layer on the partially decoded data to get them the same size, are concatenated along the channel dimension, and after which 2 standard 2D convolutional layers are applied. For more details, see Fig. \ref{fig:Upblock}.  The hyperparameters were tuned with 100 samples using the OPTUNA~\cite{akiba2019optuna} package over 30 epochs. The final model used approximately 200 epochs, where we stopped once the validation data stopped improving. We had tried many different network setups, in particular a more standard convolutional UNET dynunet~\cite{isensee2021nnu}, and a full vision transformer-based encoding setup instead of SWIN. We also tried a 3D version of the SWIN-UNETR, however the 2D version far outperformed all other architectures. 

\section{TV Neural Network Architecture}
\label{Sec:TVNN}
\begin{figure}[h]
    \centering
    \includegraphics[width=0.95\linewidth]{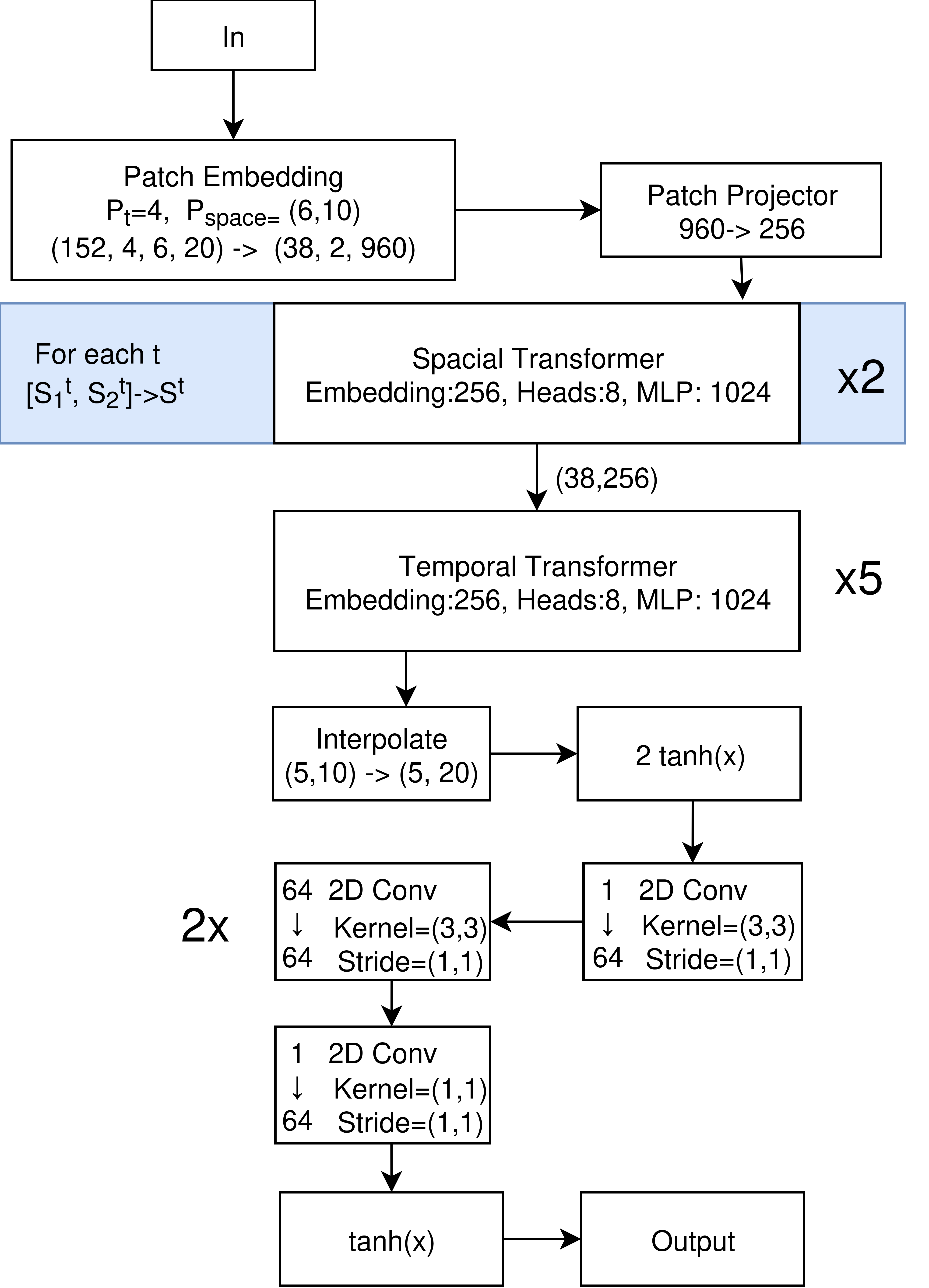}
    \caption{
Architecture of the TV ViViT network used to infer low-temperature topological visibility from high-temperature conductance data. The input is divided into spatiotemporal patches with temporal patch size $P_t$, taken along the bias-voltage direction, and spatial patch size $P_{\mathrm{space}}$, taken over the $(\mu,B)$ plane. The resulting patch vectors are projected into the transformer embedding space. For each bias-voltage patch index $t$, the spatial patch tokens $\{S_i^t\}$ are processed by a shared spatial transformer to produce an encoded spatial representation $S^t$. These encoded representations are then passed through temporal transformer blocks to mix information along the bias-voltage direction and produce a coarse latent representation of TV. The coarse prediction is interpolated to the target resolution and refined using convolutional layers. The $\tanh$ nonlinearities cap the output, with the final activation enforcing TV$\in[-1,1]$.}\label{fig:vivit}
\end{figure}
The TV network is based on a ViViT architecture \cite{arnab2021vivit} designed to map high-temperature conductance data to the corresponding low-temperature topological visibility. This transformer-based network allows the spatial (in our case $\mu$ and $B$) to be processed independently from time (in our case $V_{\text{bias}}$), which makes sense since TV is only over $\mu$ and $B$. The input conductance is first divided into spatiotemporal patches and projected into a 256-dimensional embedding space. The spatial transformer blocks then process the spatial patch structure independently for each temporal slice, allowing the network to learn local spatial correlations in the conductance maps. The same ViT is used for each patch in time. The resulting encoded sequence is passed to temporal transformer blocks, which mix information across the remaining sequence direction and produce a coarse latent representation of TV.

This coarse transformer output is reshaped and interpolated to the target spatial resolution. A convolutional refinement head is then applied to recover sharper local structure and improve the final pixel-level prediction. In this way, the transformer component learns the global and long-range structure of the phase diagram, while the convolutional layers refine the coarse ViViT prediction into a high-resolution TV map. Nonlinear $\tanh$ activations are included to keep the predicted visibility bounded, with the final $\tanh$ enforcing the physical range TV $\in [-1,1]$. See Fig. \ref{fig:vivit} for additional details. 

\section{Diffusion of Pure Gaussian Noise}
\label{Sec:PureGaussianNoise}
\begin{figure}[!h]
    \begin{subfigure}[t]{0.48\linewidth}
    \centering
    \includegraphics[width=\textwidth]{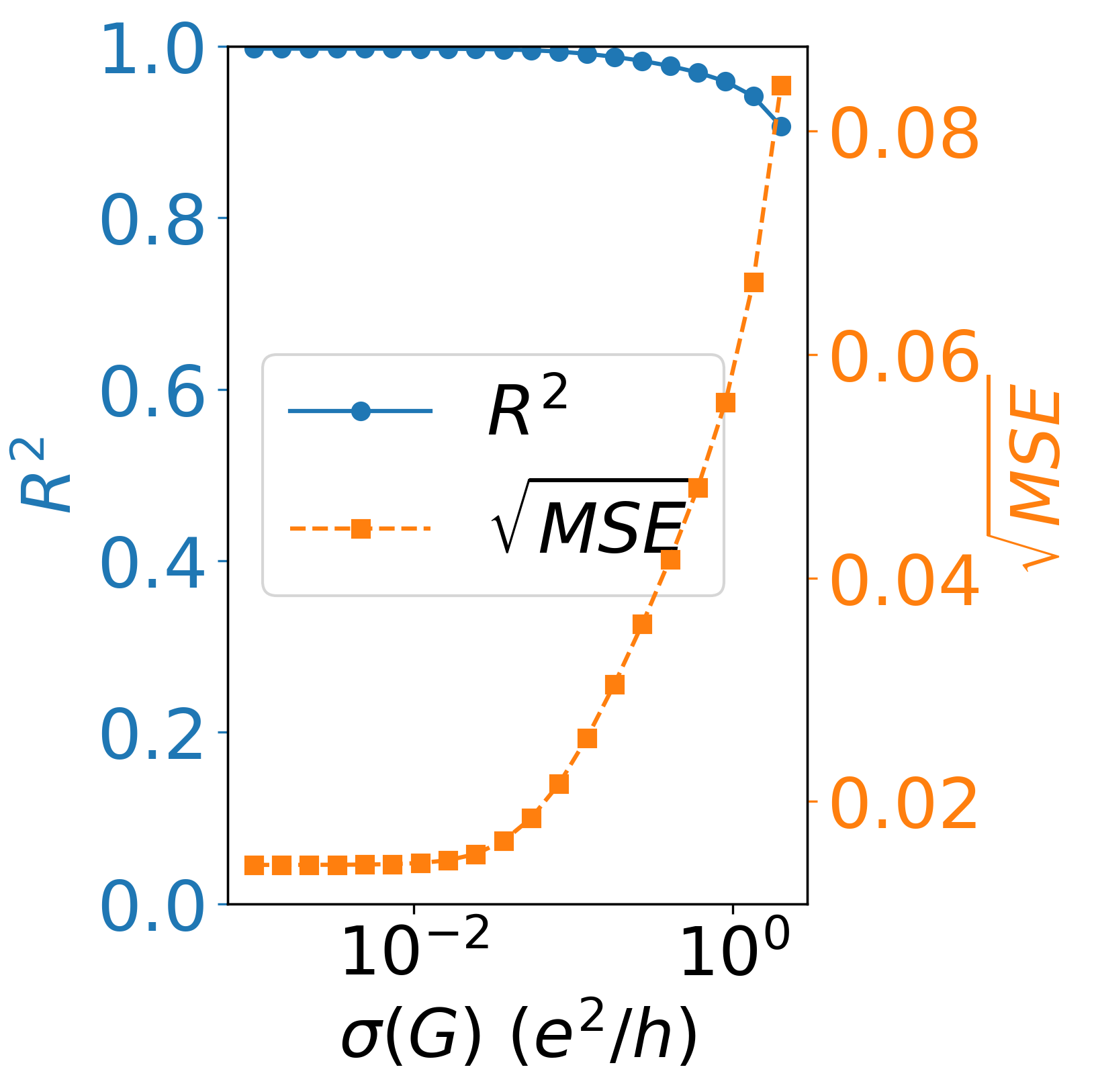}
    \caption{}
    \end{subfigure}
    \begin{subfigure}[t]{0.48\linewidth}
    \centering
    \includegraphics[width=\textwidth]{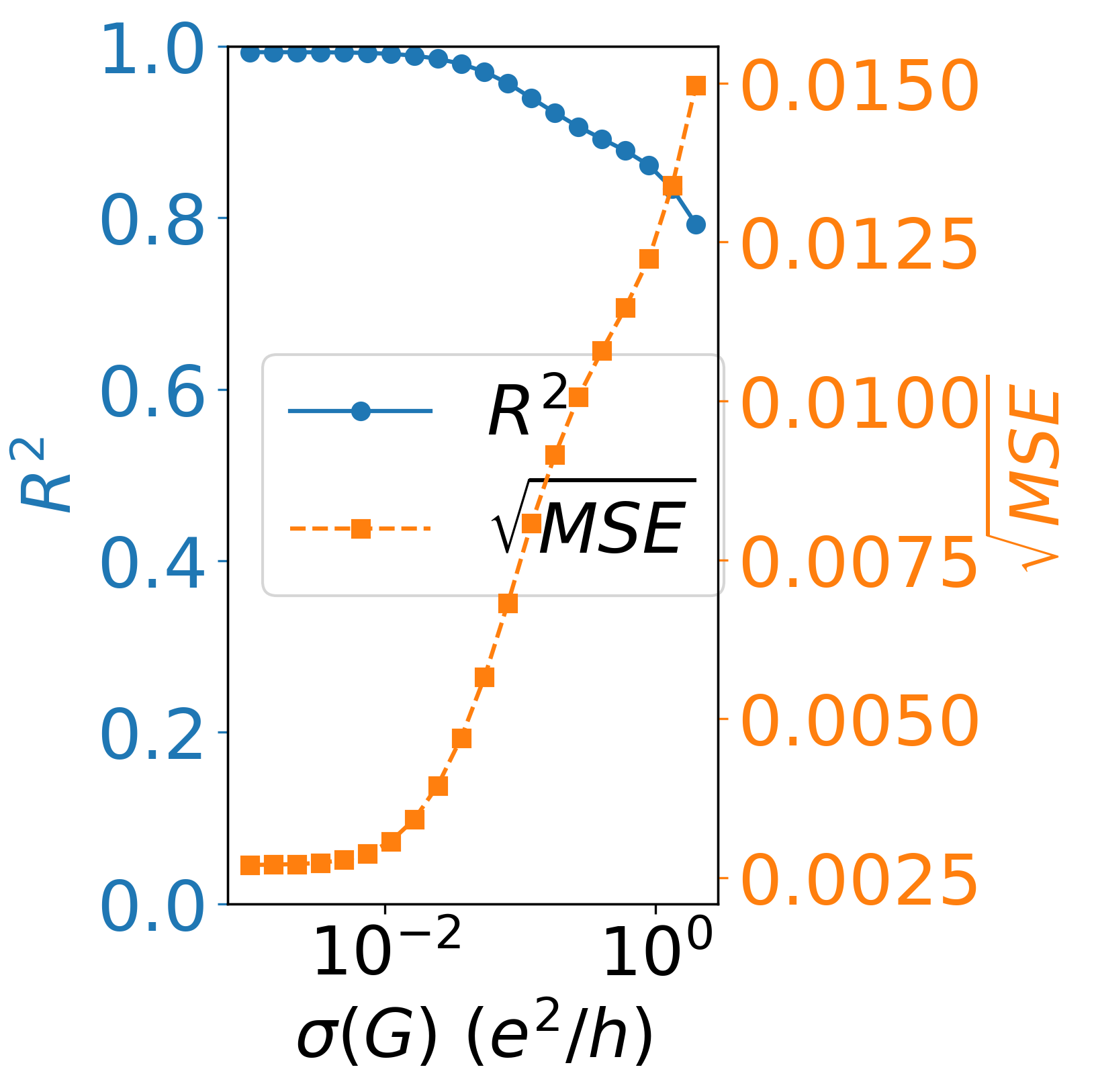}
    \caption{}
    \end{subfigure}
    \hfill
    \caption{{Local (left) and nonlocal (right) conductance reconstruction fidelities for different levels of additive Gaussian measurement noise. This is for the network trained to take in low-temperature conductance and predict low-temperature conductance after removing noise.} }\label{fig:ReconstructionFidelities}
\end{figure} 
Here, we provide results showing the ability to reconstruct the low-temperature conductance using low-temperature conductance with varying amounts of additive Gaussian measurement noise as input. See Fig. \ref{fig:ReconstructionFidelities}.

\section{Low Temperature Results}
\label{Sec:LowTempTV}
\begin{figure}[h]
    \begin{subfigure}[t]{0.48\linewidth}
    \centering
    \includegraphics[width=\textwidth]{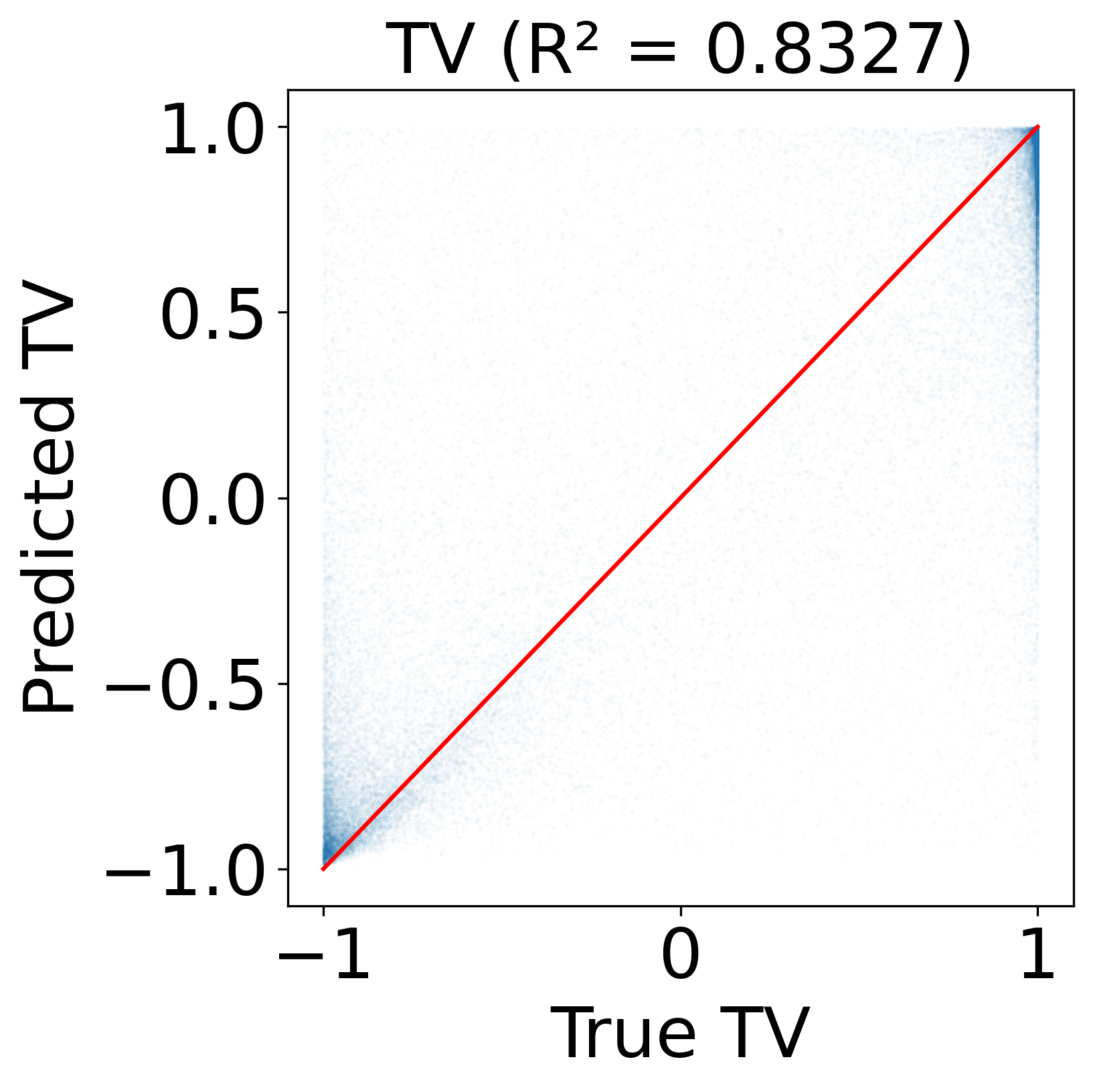}
        \caption{}
    \end{subfigure}
    \hfill
    \begin{subfigure}[t]{0.48\linewidth}
    \centering
        \includegraphics[width=\textwidth]{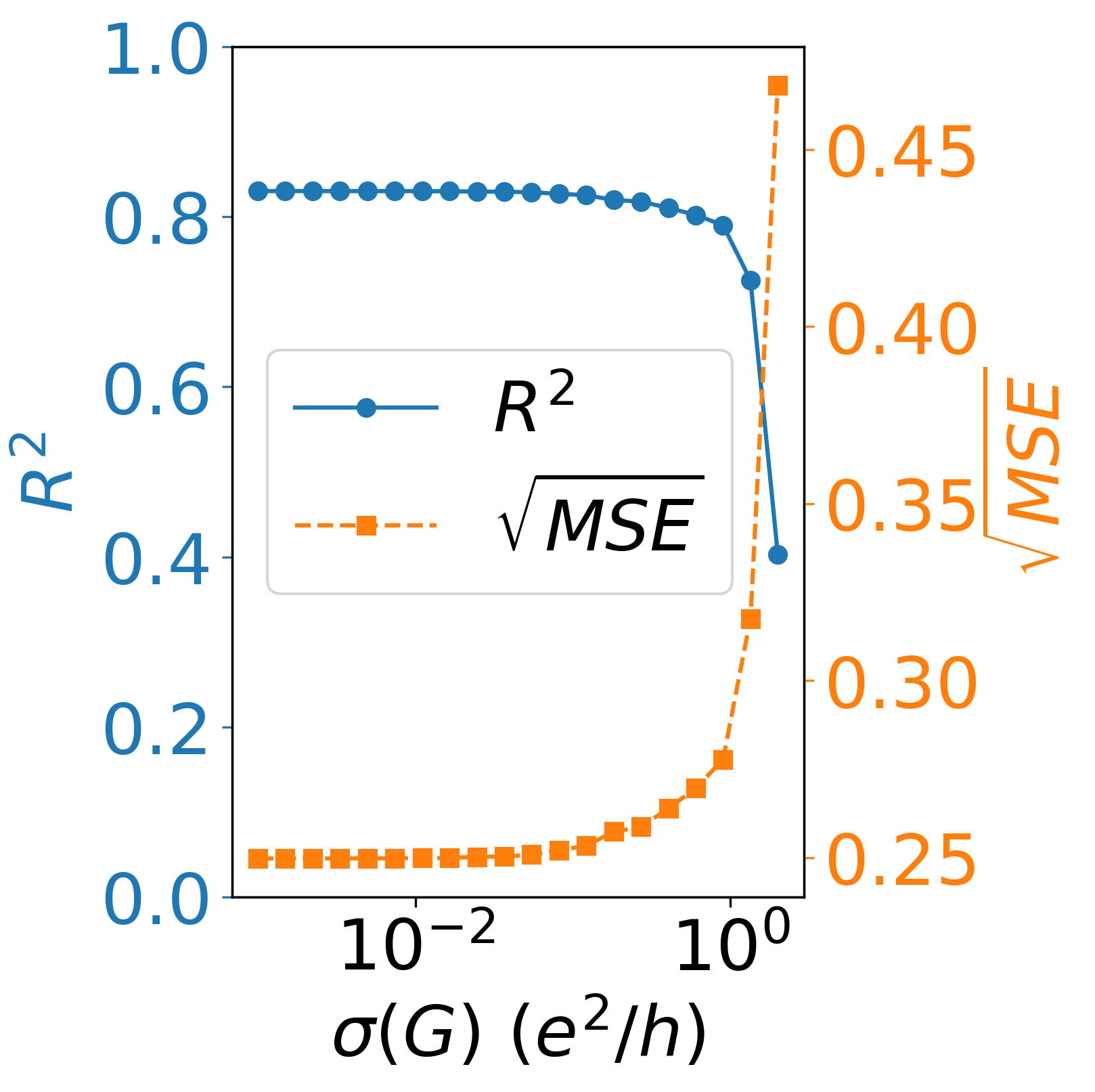}
        \caption{}
    \end{subfigure}
    \caption{Low temperature TV prediction fidelity. (a) Predicted TV vs Expected TV using low-temperature conductance as input. (b) $R^2$ and $ \sqrt{\text{MSE}}$ for varying levels of additive Gaussian noise.  }\label{fig:LowTTVResilience}
\end{figure}

\begin{figure*}[t]
    \begin{subfigure}[t]{0.48\linewidth}
    \centering
    \includegraphics[width=\textwidth]{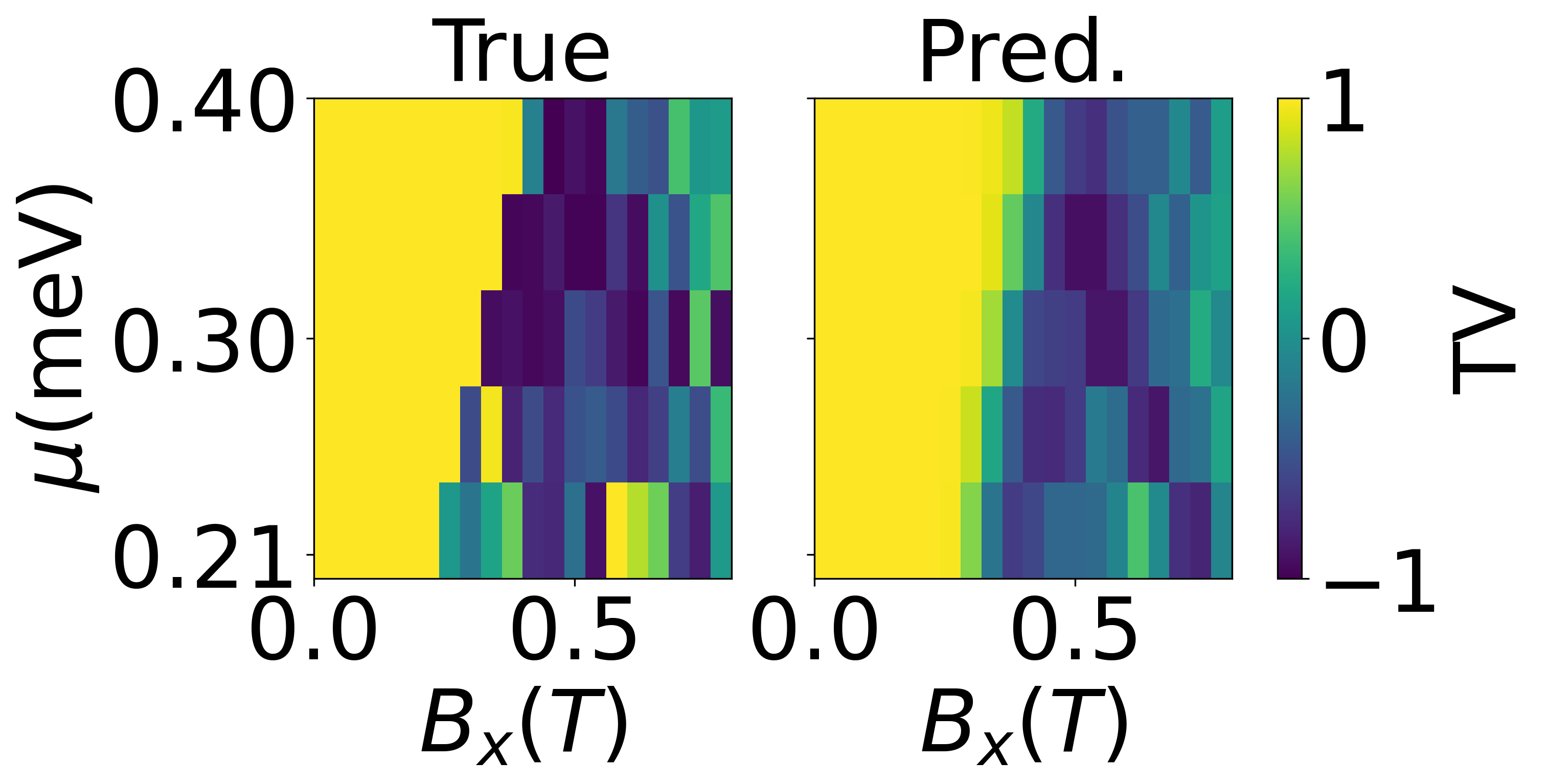}
        \caption{}
    \end{subfigure}
    \hfill
    \begin{subfigure}[t]{0.48\linewidth}
        \centering
        \includegraphics[width=\textwidth]{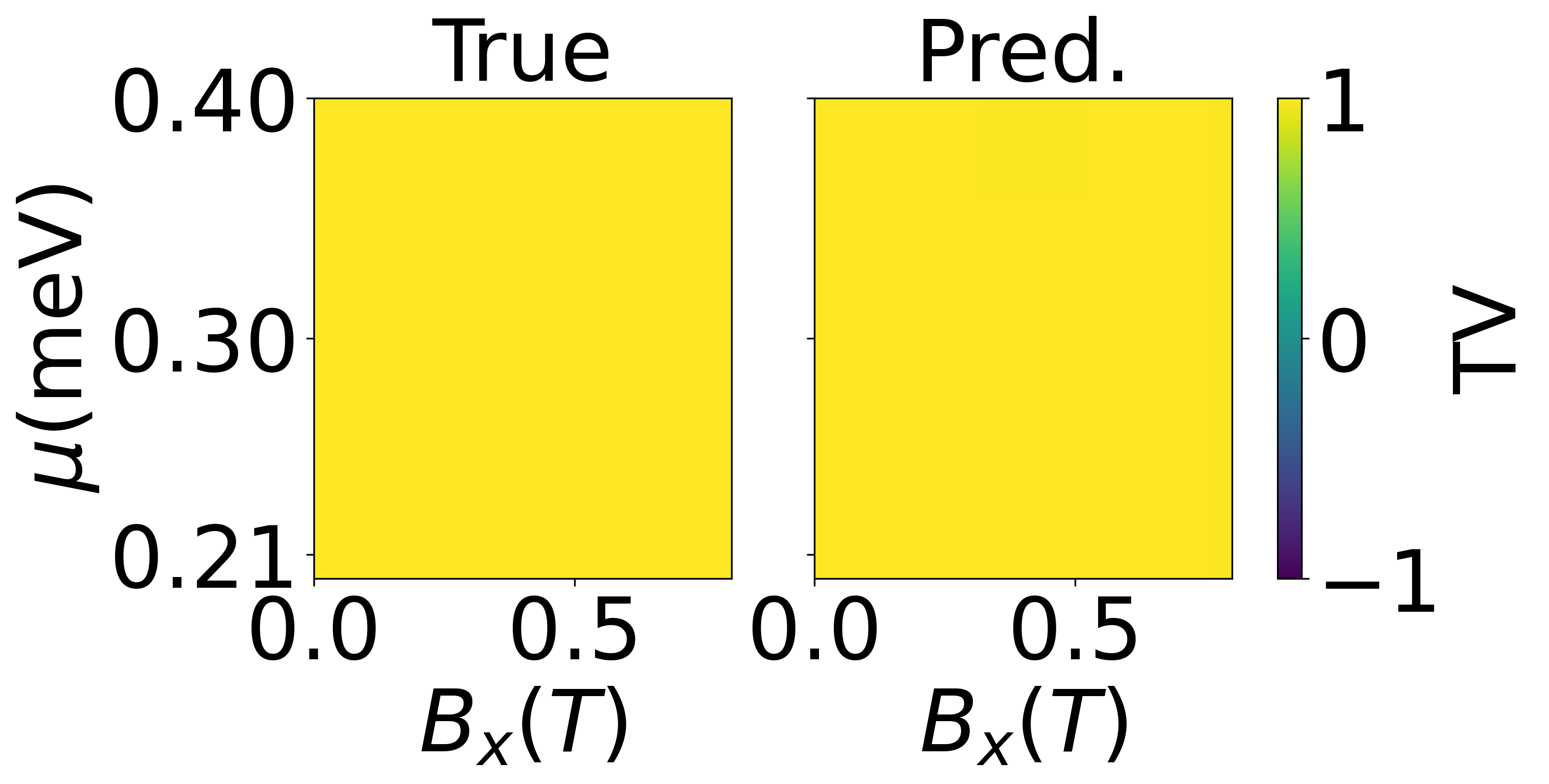}
        \caption{}
    \end{subfigure}
    \hfill
    \begin{subfigure}[t]{0.49\linewidth}
        \centering
        \includegraphics[width=\textwidth]{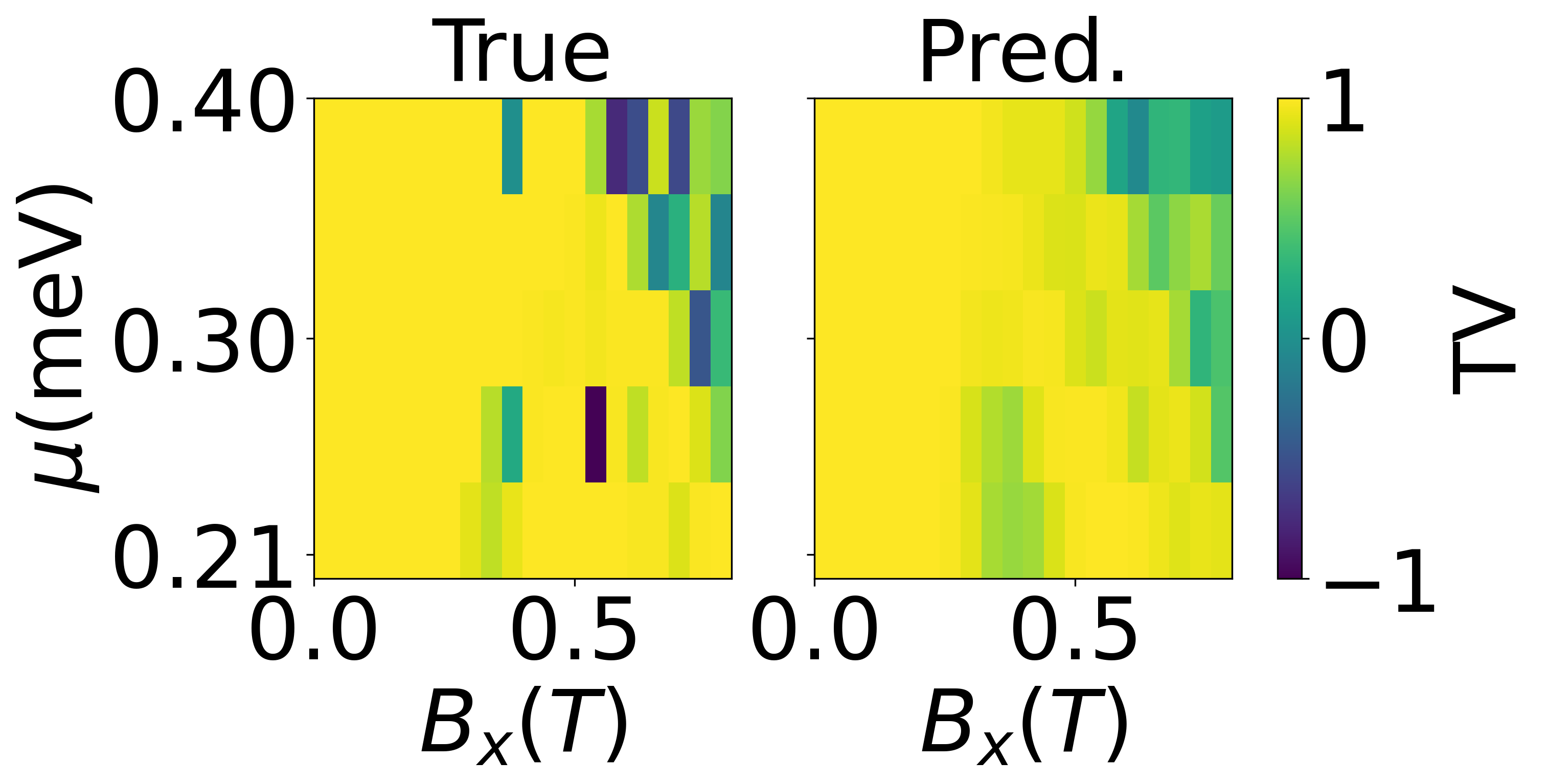}
        \caption{}
    \end{subfigure}
    \hfill
\begin{subfigure}[t]{0.49\linewidth}
        \centering
        \includegraphics[width=\textwidth]{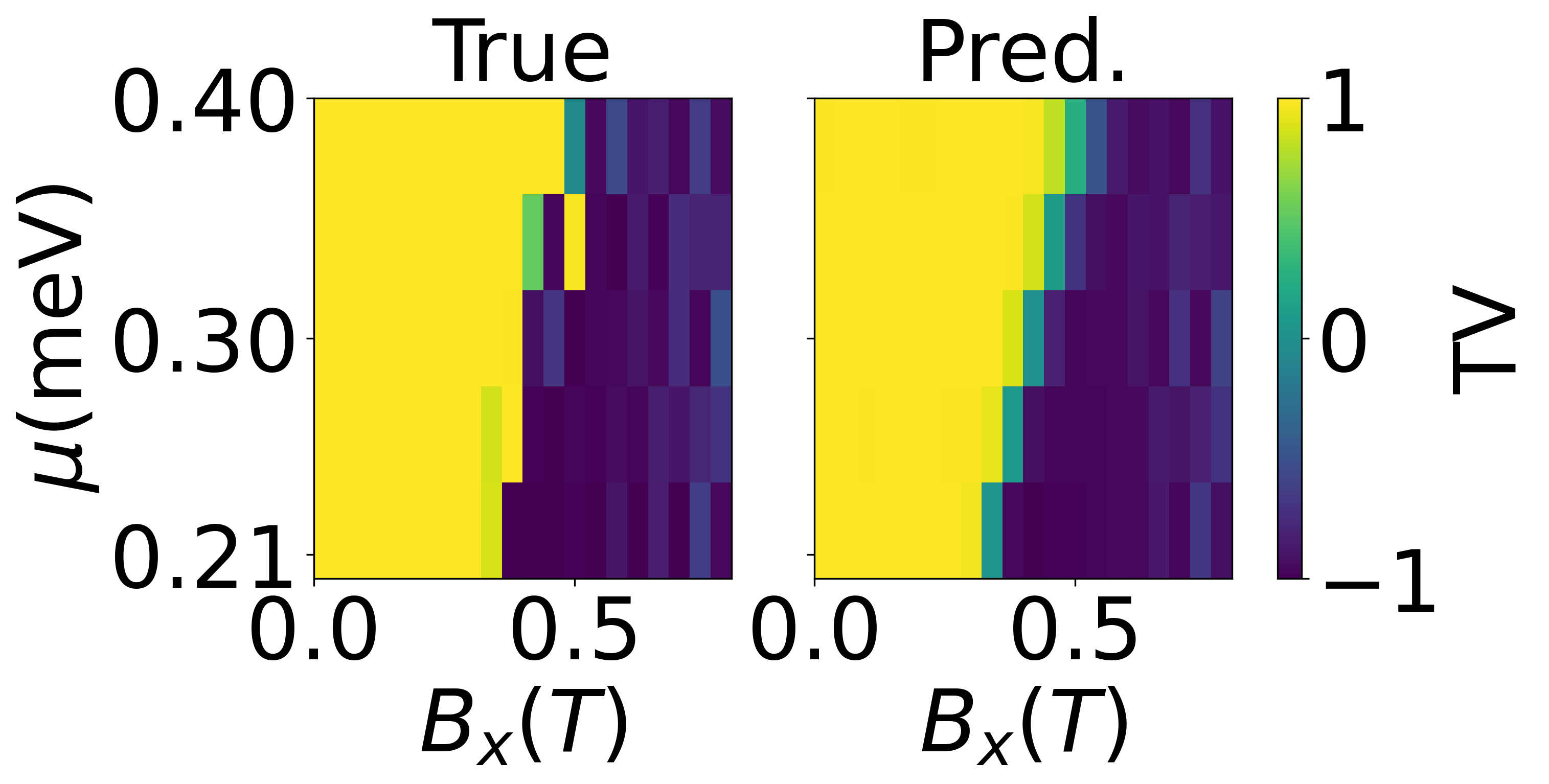}
        \caption{}
    \end{subfigure}
    \caption{Comparison of expected and predicted TV from low-$T$ conductance.  Panels (a)-(d) are different device realizations, with the left being the expected TV and the right being the predicted TV. }\label{fig:LowTTVCompare}
\end{figure*}

The ability to predict the TV from low-temperature conductance was shown in our previous work \cite{taylor2025vision}. We improve upon the neural network used there with our new ViViT/Convolution-based architecture, able to get more intricate details. We show the fidelity outcomes in Fig. \ref{fig:LowTTVResilience} and a comparison between the expected and predict TV in Fig. \ref{fig:LowTTVCompare}. We also found that this can be improved by increasing the $V_{\text{bias}}$ range. 

\section{Inverse Convolution}
\begin{figure*}[!t]
    \centering
    \begin{minipage}[t]{0.47\textwidth}
        \centering
        \includegraphics[width=\linewidth]{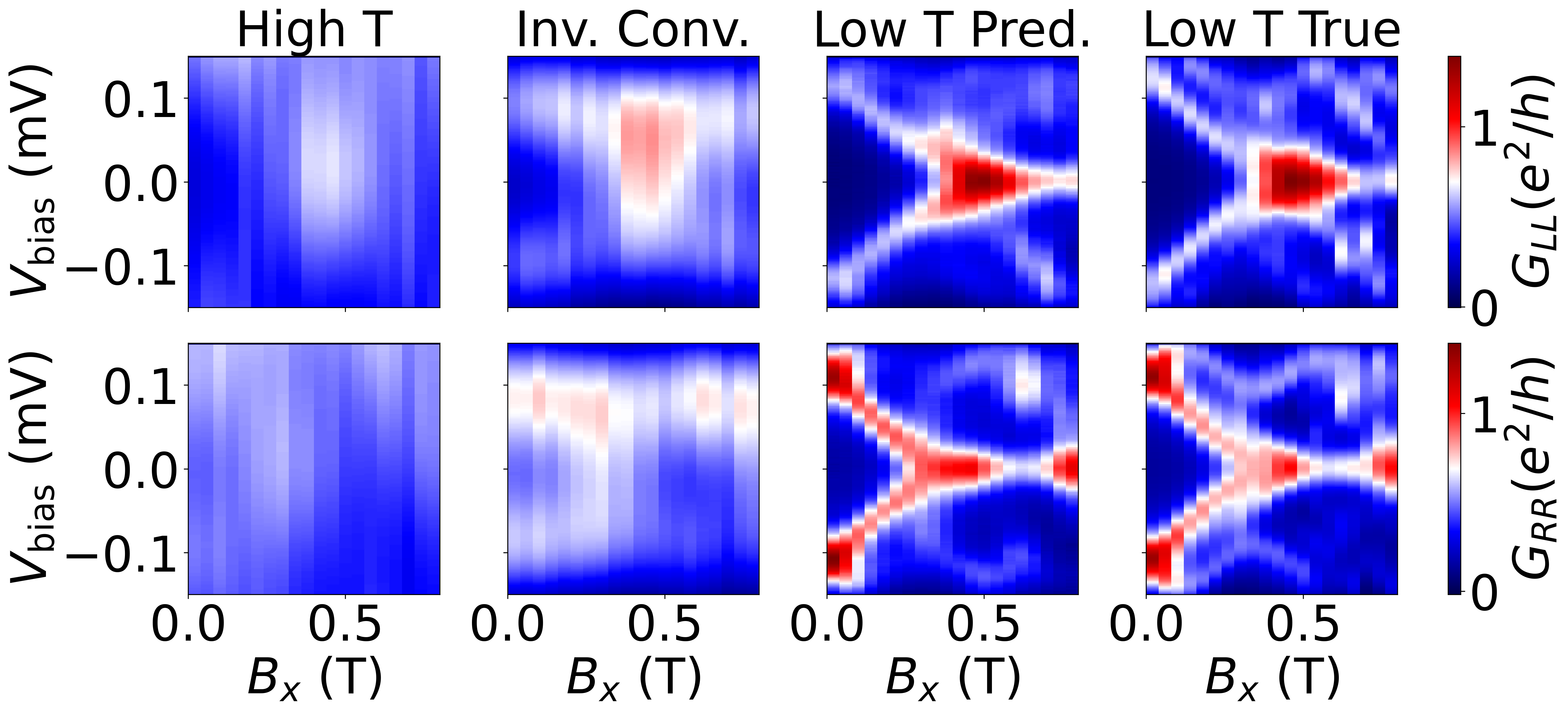}
        \caption{Comparison of expected and predicted local conductance. \HPadd{Columns show, from left to right, the high-temperature input, inverse-convolution result, neural-network prediction, and low-temperature ground truth, for $G_{LL}$ (top row) and $G_{RR}$ (bottom row).}}
        \label{fig:Extra1LocalCondCompare_a}
    \end{minipage}
    \hfill
    \begin{minipage}[t]{0.47\textwidth}
        \centering
        \includegraphics[width=\linewidth]{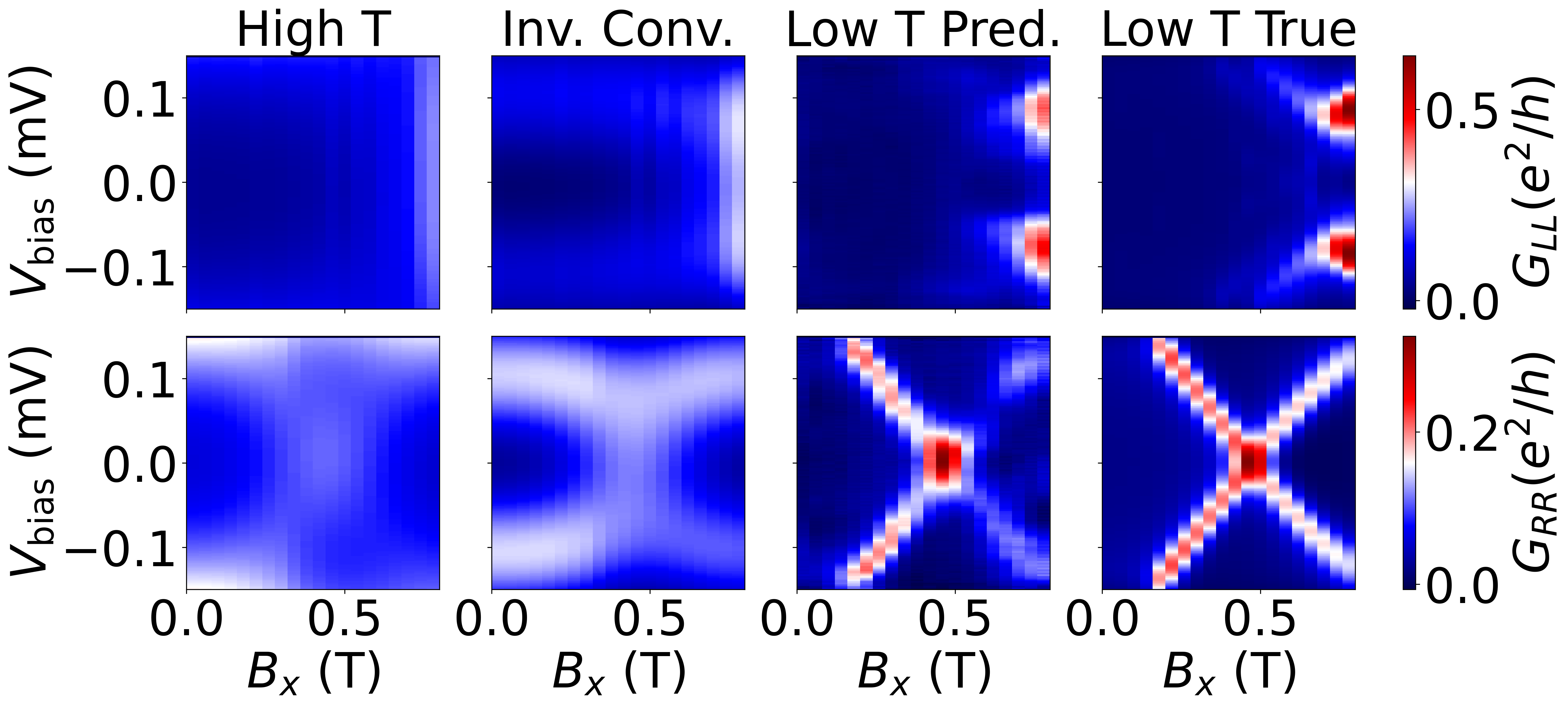}
        \caption{Comparison of expected and predicted local conductance. \HPadd{Columns show, from left to right, the high-temperature input, inverse-convolution result, neural-network prediction, and low-temperature ground truth, for $G_{LL}$ (top row) and $G_{RR}$ (bottom row).}}
        \label{fig:Extra1LocalCondCompare_b}
    \end{minipage}
\end{figure*}

\begin{figure*}[!t]
    \centering
    \begin{minipage}[t]{0.49\textwidth}
        \centering
        \includegraphics[width=\linewidth]{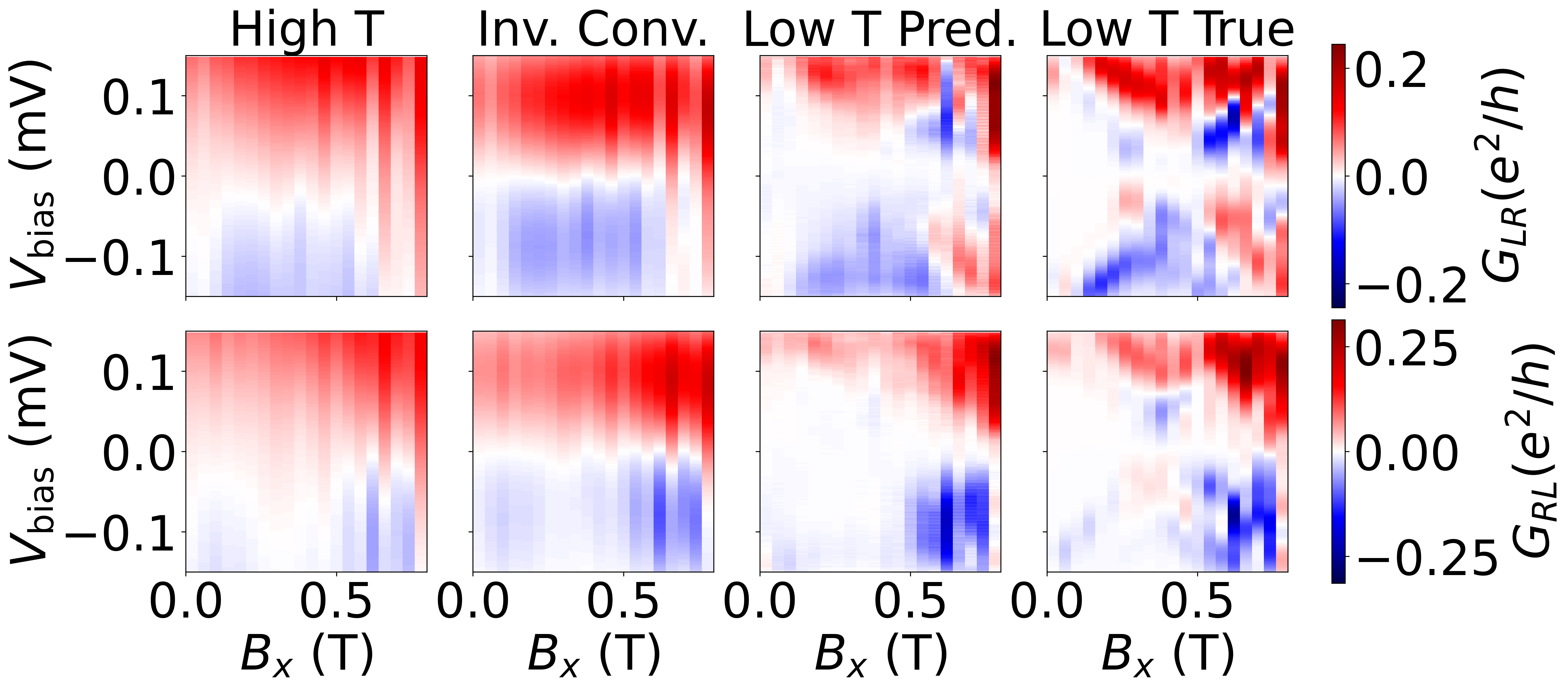}
        \caption{Comparison of nonlocal conductances. \HPadd{Columns show, from left to right, the high-temperature input, inverse-convolution result, neural-network prediction, and low-temperature ground truth, for $G_{LR}$ (top row) and $G_{RL}$ (bottom row).}}
        \label{fig:Extra1NonLocalCondCompare_a}
    \end{minipage}
    \hfill
    \begin{minipage}[t]{0.49\textwidth}
        \centering
        \includegraphics[width=\linewidth]{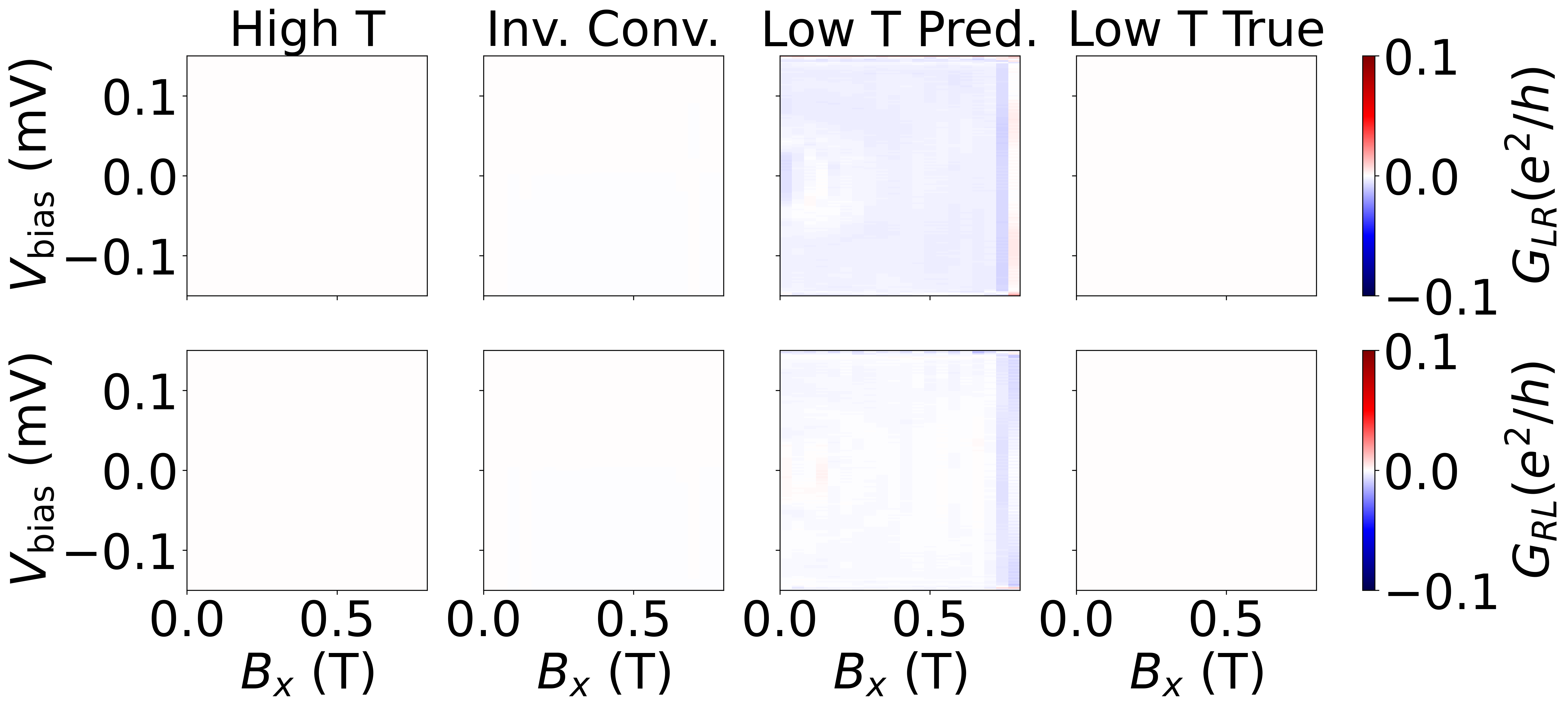}
        \caption{Comparison of nonlocal conductances. \HPadd{Columns show, from left to right, the high-temperature input, inverse-convolution result, neural-network prediction, and low-temperature ground truth, for $G_{LR}$ (top row) and $G_{RL}$ (bottom row).}}
        \label{fig:Extra1NonLocalCondCompare_b}
    \end{minipage}
\end{figure*}

\begin{figure*}[!t]
    \centering
    \begin{minipage}[t]{0.49\textwidth}
        \centering
        \includegraphics[width=\linewidth]{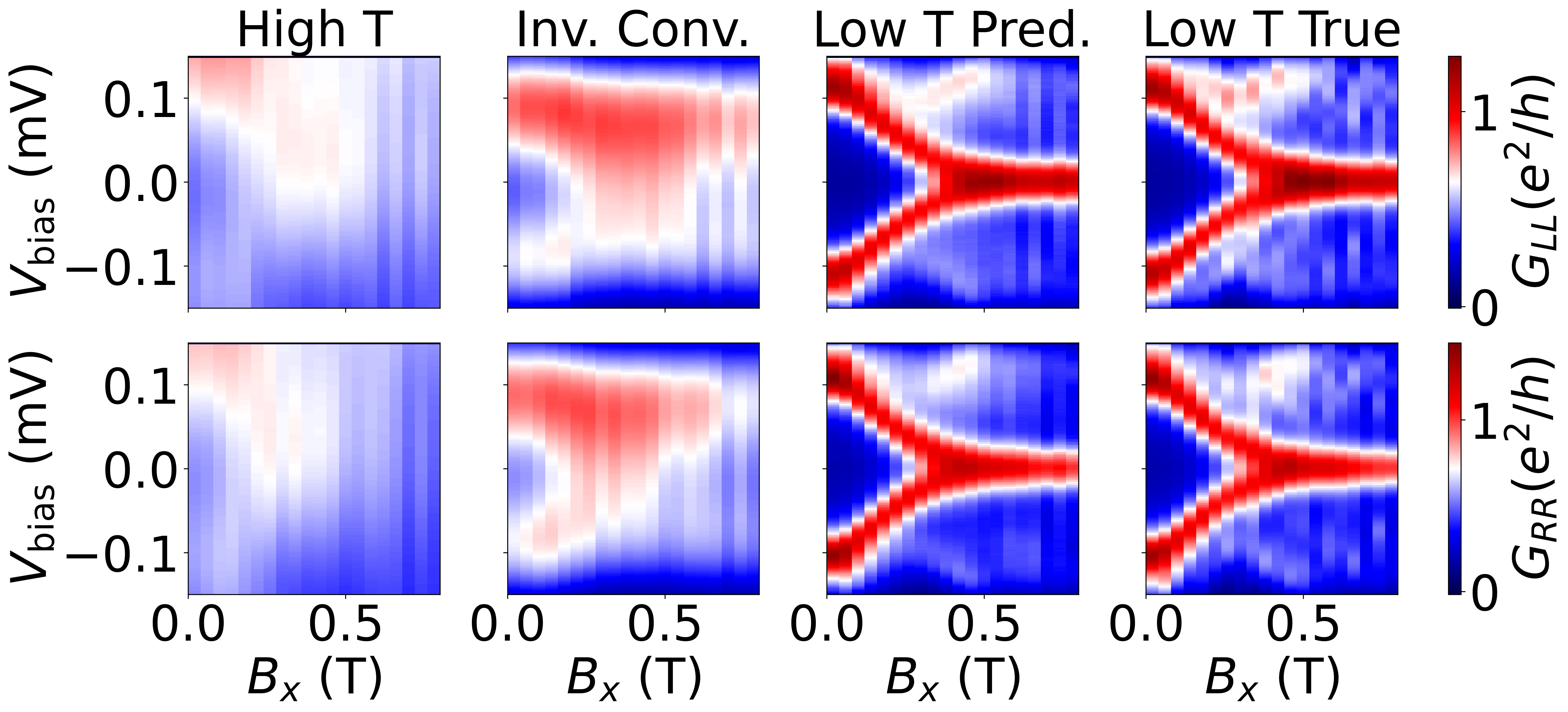}
        \caption{Comparison of local conductances. \HPadd{Columns show, from left to right, the high-temperature input, inverse-convolution result, neural-network prediction, and low-temperature ground truth, for $G_{LL}$ (top row) and $G_{RR}$ (bottom row).}}
        \label{fig:HighTE1LocalCondCompare_a}
    \end{minipage}
    \hfill
    \begin{minipage}[t]{0.49\textwidth}
        \centering
        \includegraphics[width=\linewidth]{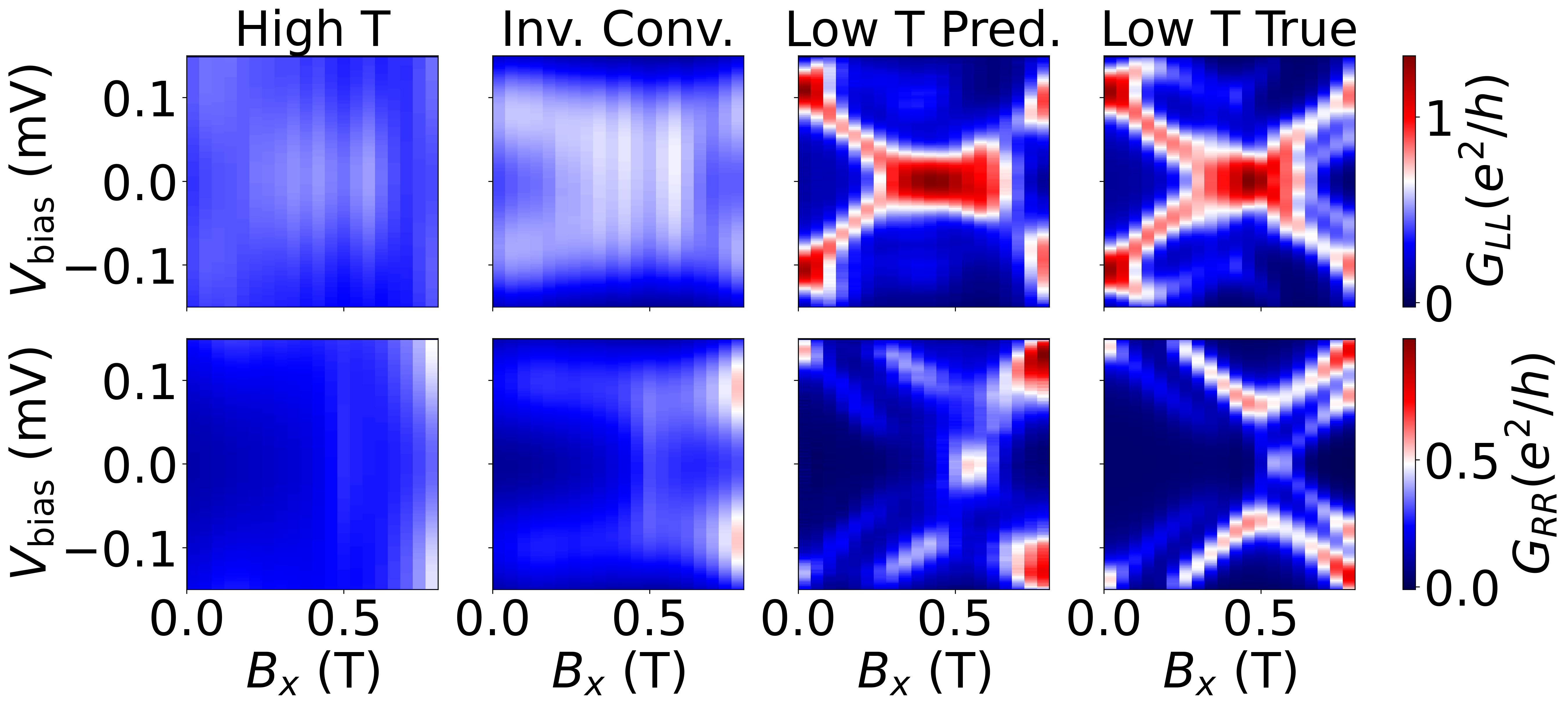}
        \caption{Comparison of local conductances. \HPadd{Columns show, from left to right, the high-temperature input, inverse-convolution result, neural-network prediction, and low-temperature ground truth, for $G_{LL}$ (top row) and $G_{RR}$ (bottom row).}}
        \label{fig:HighTE1LocalCondCompare_b}
    \end{minipage}
\end{figure*}

\begin{figure*}[!t]
    \centering
    \begin{minipage}[t]{0.49\textwidth}
        \centering
        \includegraphics[width=\linewidth]{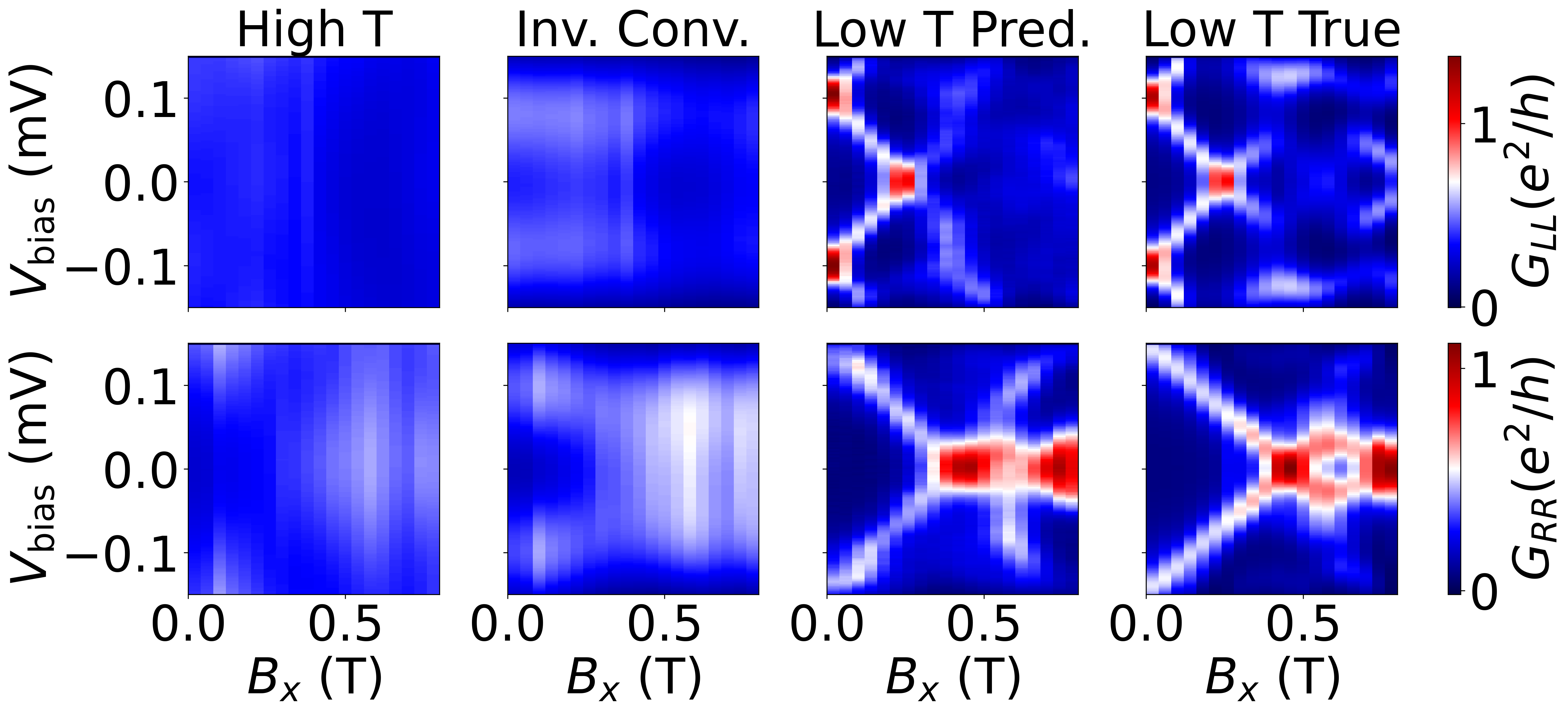}
        \caption{Comparison of local conductances. \HPadd{Columns show, from left to right, the high-temperature input, inverse-convolution result, neural-network prediction, and low-temperature ground truth, for $G_{LL}$ (top row) and $G_{RR}$ (bottom row).}}
        \label{fig:HighTE2LocalCondCompare_a}
    \end{minipage}
    \hfill
    \begin{minipage}[t]{0.49\textwidth}
        \centering
        \includegraphics[width=\linewidth]{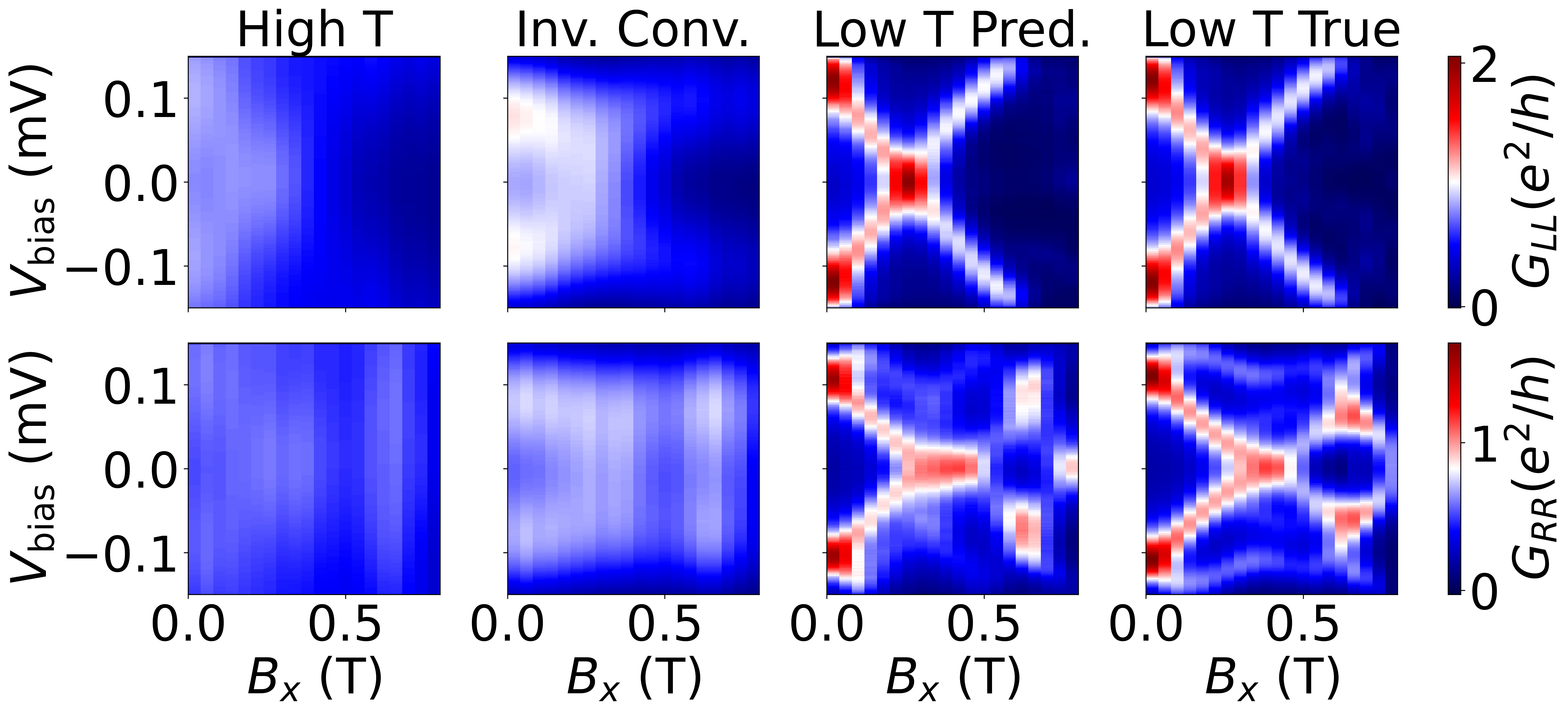}
        \caption{Comparison of local conductances. \HPadd{Columns show, from left to right, the high-temperature input, inverse-convolution result, neural-network prediction, and low-temperature ground truth, for $G_{LL}$ (top row) and $G_{RR}$ (bottom row).}}
        \label{fig:HighTE2LocalCondCompare_b}
    \end{minipage}
\end{figure*}

\begin{figure*}[!t]
    \centering
    \begin{minipage}[t]{0.49\textwidth}
        \centering
        \includegraphics[width=\linewidth]{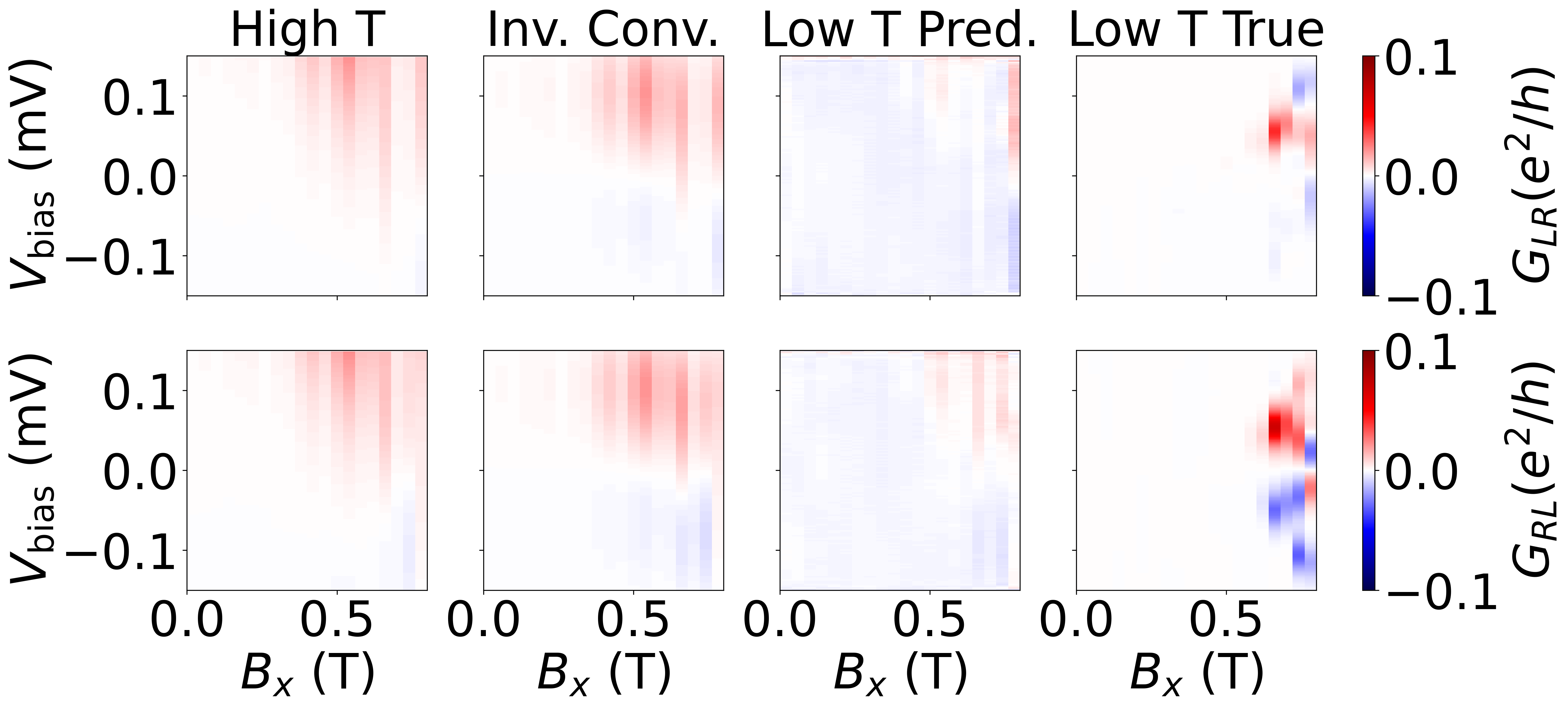}
        \caption{Comparison of nonlocal conductances. \HPadd{Columns show, from left to right, the high-temperature input, inverse-convolution result, neural-network prediction, and low-temperature ground truth, for $G_{LR}$ (top row) and $G_{RL}$ (bottom row).}}
        \label{fig:HighTE1NonLocalCondCompare_a}
    \end{minipage}
    \hfill
    \begin{minipage}[t]{0.49\textwidth}
        \centering
        \includegraphics[width=\linewidth]{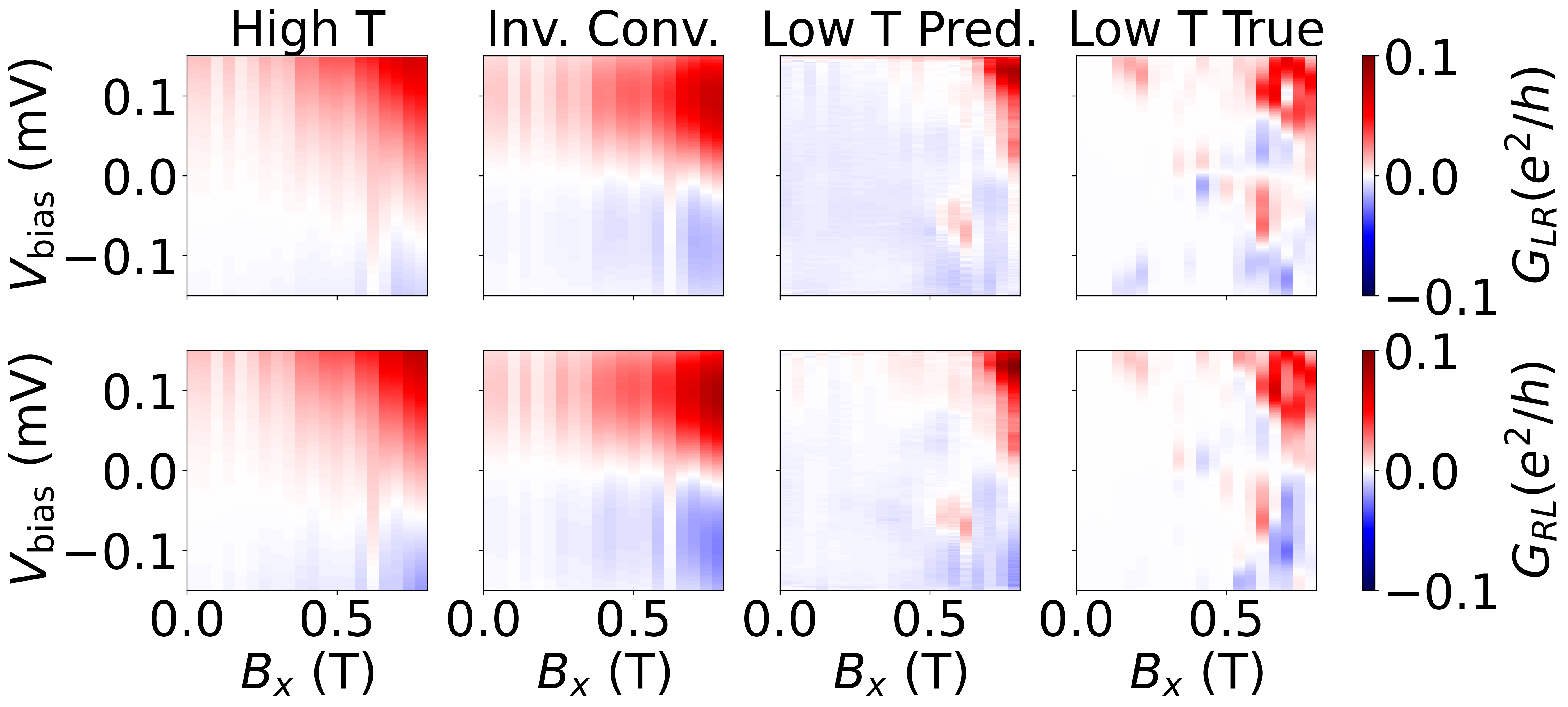}
        \caption{Comparison of nonlocal conductances. \HPadd{Columns show, from left to right, the high-temperature input, inverse-convolution result, neural-network prediction, and low-temperature ground truth, for $G_{LR}$ (top row) and $G_{RL}$ (bottom row).}}
        \label{fig:HighTE1NonLocalCondCompare_b}
    \end{minipage}
\end{figure*}

\begin{figure*}[!t]
    \centering
    \begin{minipage}[t]{0.47\textwidth}
        \centering
        \includegraphics[width=\linewidth]{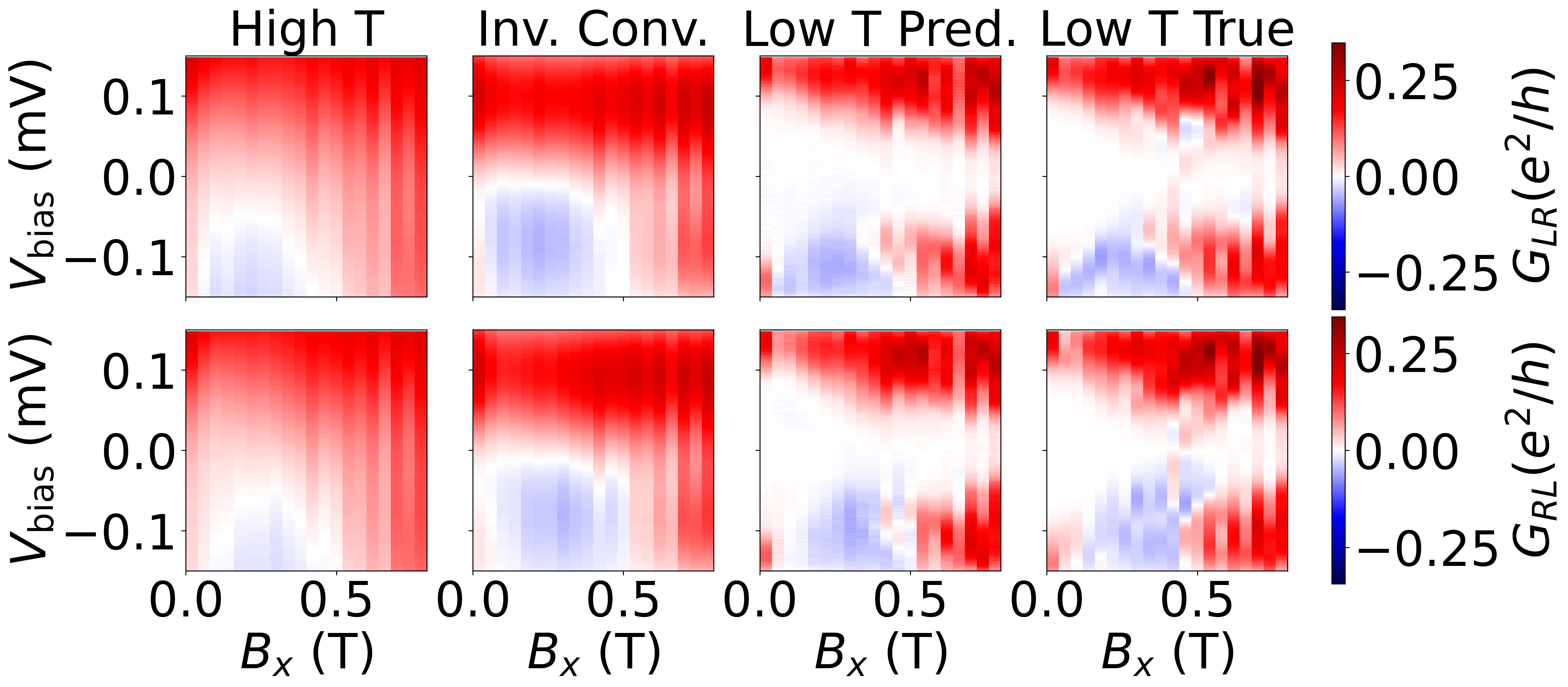}
        \caption{Comparison of nonlocal conductances. \HPadd{Columns show, from left to right, the high-temperature input, inverse-convolution result, neural-network prediction, and low-temperature ground truth, for $G_{LR}$ (top row) and $G_{RL}$ (bottom row).}}
        \label{fig:HighTE2NonLocalCondCompare_a}
    \end{minipage}
    \hfill
    \begin{minipage}[t]{0.47\textwidth}
        \centering
        \includegraphics[width=\linewidth]{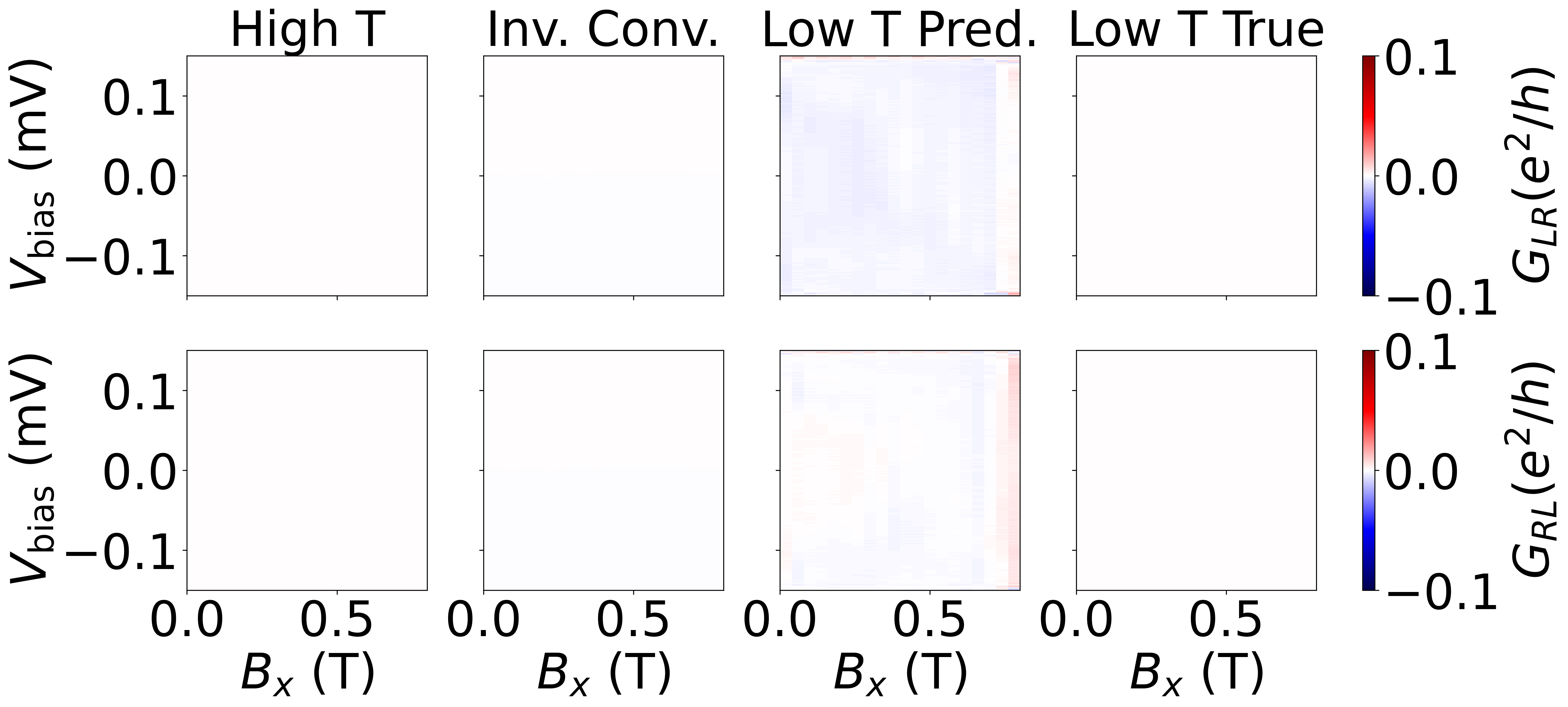}
        \caption{Comparison of nonlocal conductances. \HPadd{Columns show, from left to right, the high-temperature input, inverse-convolution result, neural-network prediction, and low-temperature ground truth, for $G_{LR}$ (top row) and $G_{RL}$ (bottom row).}}
        \label{fig:HighTE2NonLocalCondCompare_b}
    \end{minipage}
\end{figure*}

The current classical way to find low-temperature results from high-temperature measurements is to attempt an inverse convolutional operation. In general, this is only feasible with a resolution of about $k_B T$. Here we show a comparison between the high temperature input, the inverse Fourier transform process,  and our neural network prediction results. We find that effectively no meaningful features can be seen through the inverse convolutional process. See Fig.~\ref{fig:Extra1LocalCondCompare_a}--\ref{fig:HighTE2NonLocalCondCompare_b} for additional comparisons across different device realizations.

\end{document}